\documentclass[lettersize,journal]{IEEEtran}
\usepackage{amsmath,amsfonts}
\usepackage{algorithmic}
\usepackage{algorithm}
\usepackage{array}
\usepackage[caption=false,font=normalsize,labelfont=sf,textfont=sf]{subfig}
\usepackage{textcomp}
\usepackage{stfloats}
\usepackage{url}
\usepackage{verbatim}
\usepackage{graphicx}
\usepackage{cite}
\usepackage{makecell}
\usepackage{multirow}
\usepackage{threeparttable}
\usepackage[normalem]{ulem}
\usepackage{overpic}
\usepackage{hyperref}
\usepackage{footnote}
\makesavenoteenv{table}
\useunder{\uline}{\ul}{}
\begin{document}

\title{Cross-Modal Retrieval: A Systematic Review of Methods and Future Directions}

\author{Tianshi Wang$^{\dagger}$, Fengling Li$^{\dagger}$, Lei Zhu, Jingjing Li, Zheng Zhang and Heng Tao Shen\textit{, Fellow, IEEE}
\thanks{Tianshi Wang, Lei Zhu and Heng Tao Shen are with the School of Electronic and Information Engineering, Tongji University, Shanghai 201804, China. Tianshi Wang is also with the School of Information Science and Engineering, Shandong Normal University, Jinan 250358, China. Heng Tao Shen is also with the School of Computer Science and Engineering, University of Electronic Science and Technology of China, Chengdu 611731, China.}
\thanks{Fengling Li is with the Australian Artificial Intelligence Institute, University of Technology Sydney, Sydney, NSW 2007, Australia.}
\thanks{${\dagger}$ These authors contributed equally to this work.}
\thanks{Lei Zhu is the corresponding author (e-mail: leizhu0608@gmail.com).}
\thanks{Jingjing Li is with the School of Computer Science and Engineering, University of Electronic Science and Technology of China, Chengdu 611731, China.}
\thanks{Zheng Zhang is with Shenzhen Key Laboratory of Visual Object Detection and Recognition, Harbin Institute of Technology, Shenzhen, Shenzhen 518055, China.}}

\markboth{Proceedings of the IEEE}%
{Shell \MakeLowercase{\textit{et al.}}: A Sample Article Using IEEEtran.cls for IEEE Journals}


\maketitle

\begin{abstract}
With the exponential surge in diverse multi-modal data, traditional uni-modal retrieval methods struggle to meet the needs of users seeking access to data across various modalities. To address this, cross-modal retrieval has emerged, enabling interaction across modalities, facilitating semantic matching, and leveraging complementarity and consistency between heterogeneous data. Although prior literature has reviewed the field of cross-modal retrieval, it suffers from numerous deficiencies in terms of timeliness, taxonomy, and comprehensiveness. This paper conducts a comprehensive review of cross-modal retrieval's evolution, spanning from shallow statistical analysis techniques to vision-language pre-training models. Commencing with a comprehensive taxonomy grounded in machine learning paradigms, mechanisms, and models, the paper delves deeply into the principles and architectures underpinning existing cross-modal retrieval methods. Furthermore, it offers an overview of widely-used benchmarks, metrics, and performances. Lastly, the paper probes the prospects and challenges that confront contemporary cross-modal retrieval, while engaging in a discourse on potential directions for further progress in the field. To facilitate the ongoing research on cross-modal retrieval, we develop a user-friendly toolbox and an open-source repository at \url{https://cross-modal-retrieval.github.io}.
\end{abstract}


\section{Introduction}
\label{Sec:Introduction}
\begin{figure}[!t]
\centering
\hspace{-0.5mm}\includegraphics[width=0.49\textwidth]{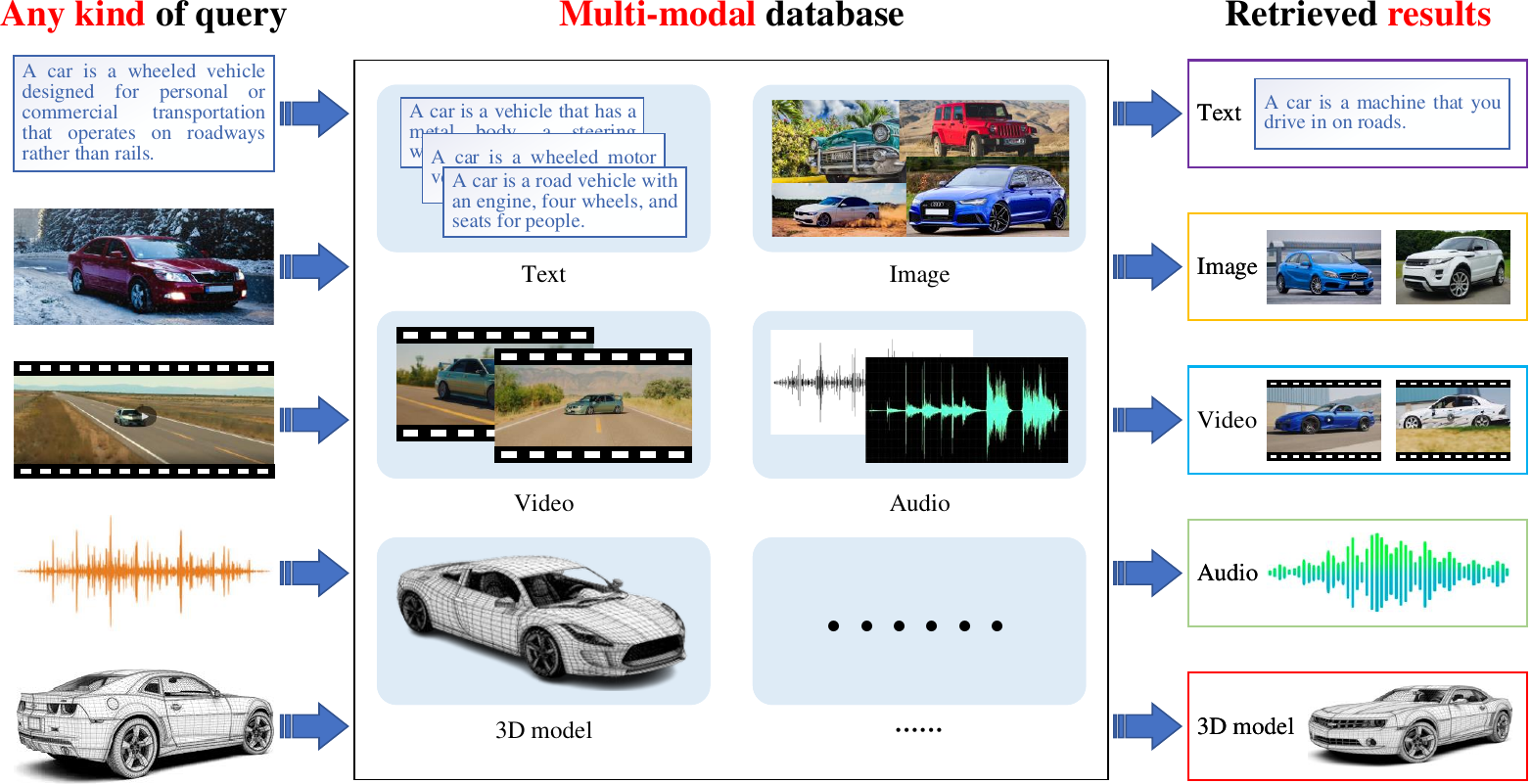}
\caption{Illustration of cross-modal retrieval. It involves retrieving information across different modalities, such as text, image, audio, and video, using a query from any one modality. For example, a user can enter a text query to retrieve relevant images or videos associated with that query.}
\label{Figure:intro}
\end{figure}

In recent decades, the Internet, smart devices, and sensors have undergone remarkable expansion, resulting in an exponential influx of diverse multi-modal data. This encompasses various forms such as image, text, audio, and video, often employed to portray the same event or topic. This surge in diversity has led to an increased demand from users for access to data spanning different modalities to achieve comprehensive insights\cite{SurveyMM}. However, conventional retrieval methods\cite{SurveyText, SurveyCBIR, SurveyHash} that concentrate on a single modality fall short in meeting these requirements due to the gap between heterogeneous modalities. Hence, the necessity for a retrieval strategy arises, one that fosters interaction among information sources and supports heterogeneous searches across modalities. Cross-modal retrieval, illustrated in Fig. \ref{Figure:intro}, has surfaced as a remedy, enhancing user experience and information assimilation by fostering semantic alignment and capitalizing on the synergy among multi-modal data. It empowers users to swiftly uncover captivating information, attain insights from varied perspectives, and pinpoint potential correlations and patterns.

Cross-modal retrieval has garnered significant attention and exploration from both academia and industry, resulting in the emergence of numerous learning-based methodologies in this dynamic research field. The trajectory of cross-modal retrieval research can be retraced to around 2010 when statistical analysis techniques predominantly prevailed, extracting features from multi-modal data and mapping them to a common space for relevance assessment\cite{CCA, CVH, SurveyML}. A notable example from this period is Canonical Correlation Analysis (CCA) \cite{CCA}, which focuses on understanding the correlation between two sets of variables by identifying linear combinations that maximize their relationship. Since 2014, the ascendancy of deep learning techniques has reverberated in the field of cross-modal retrieval, harnessing the potency of deep neural networks to autonomously learn high-level feature representations from multi-modal data\cite{Corr-AE, MMNN, SurveyDL}. Notably, Corr-AE \cite{Corr-AE} employs auto-encoders for feature learning, showcasing the transition from traditional statistical methods to neural network-based techniques. In recent years, a cascade of cross-modal retrieval approaches has been tailored to diverse open scenarios, harnessing the potential of Vision-Language Pre-training (VLP) models\cite{EI-CLIP, CSIC, SurveyVLP}. A significant example is EI-CLIP \cite{EI-CLIP}, which leverages VLP models as backbones and integrates contrastive learning with keyword enhancement, showcasing its effectiveness in E-commerce retrieval. These strides have notably bolstered the precision, robustness, and scalability of cross-modal retrieval systems by infusing sophisticated learning models and training strategies. Casting our gaze forward, cross-modal retrieval remains a challenging yet promising research frontier, poised to encompass a broader spectrum of data modalities, surmount intricate open retrieval scenarios, and necessitate efficient retrieval models.

To furnish researchers with a profound grasp of the research landscape, practical significance, and future prospects of cross-modal retrieval, this paper offers a methodical synopsis and analysis of prevailing representative methods, techniques, and frameworks. It also delves into experimental benchmarks, metrics, and performances, proffering novel ideas and recommendations for forthcoming research directions. Despite previous literature\cite{SurveyWang, SurveyPeng, SurveyKaur, SurveyNie} has reviewed cross-modal retrieval, it is riddled with significant deficiencies in timeliness, taxonomy, comprehensiveness, and more. Concretely, literature \cite{SurveyWang, SurveyPeng} furnish insights into the early studies of cross-modal retrieval, yet their portrayal of representative methods and contemporary advancements is hindered by temporal gaps. The impactful cross-modal retrieval techniques developed in the last five years, which have greatly influenced the field, have yet to be included. Notably, the recent emergence of Transformer architectures and vision-language pre-training models has wielded a profound impact on the domain of deep learning, fundamentally reshaping the cross-modal retrieval research landscape. While recent years have seen the publication of literature \cite{SurveyKaur, SurveyNie}, their scope and taxonomy fall markedly short. Within literature \cite{SurveyKaur}, the discourse on cross-modal retrieval methods rooted in Transformer architectures or large-scale pre-trained models is notably sparse. Conversely, literature \cite{SurveyNie} centers predominantly on the frontier of image-text matching and fails to comprehensively synthesize methodologies for supervised real-value and hashing-based cross-modal retrieval, as well as retrieval across broader modalities. Simultaneously, literature \cite{SurveyKaur, SurveyNie} lumps deep learning-based cross-modal retrieval methodologies into an overly simplistic category, an approach ill-suited for the present advanced state of deep learning. Within this realm, distinct deep techniques bear their core concepts, and consolidating them into a singular category obstructs a comprehensive grasp of the unique attributes inherent in divergent network architectures. Furthermore, neither of these works delves into the exploration of strategies to address pragmatic challenges encountered across an array of real-world scenarios. This omission is particularly noteworthy, given that resolving such practical concerns now stands as a focal point within cross-modal retrieval research. In light of this, we undertake a comprehensive review of over three hundred cross-modal retrieval articles, spanning from inception to the present, with the overarching goal of furnishing a holistic overview of this field. In summation, the primary contributions of this paper can be distilled as follows:

\begin{itemize}
\item{We introduce an exhaustive and nuanced taxonomy for cross-modal retrieval, dividing existing methods into five overarching categories and further into forty-four subcategories. By offering detailed elucidations of the principles and architectures employed in these techniques, this paper furnishes a comprehensive review of the entire cross-modal retrieval landscape, encompassing fundamental concepts and progressive innovation.}
\item{This study introduces a user-friendly cross-modal retrieval toolbox\footnote{\url{https://github.com/cross-modal-retrieval/cross-modal-retrieval}} that greatly enhances our work's practical utility, alongside a concise compilation of commonly used multi-modal datasets, evaluation metrics, and performance benchmarks. Together, these resources streamline cross-modal retrieval implementation and aid researchers in choosing suitable experimental paradigms for effective validation and evaluation.}
\item{Furthermore, in light of the current developmental status and application requisites, this paper probes into the opportunities and challenges confronting the realm of cross-modal retrieval. Drawing from this analysis, potential remedies and research directions are proposed, with a focal point on cutting-edge challenges and emergent trends within the discipline.}
\end{itemize}

The ensuing sections of this paper are organized as follows: Section \ref{Sec:Overview} furnishes a comprehensive overview of fundamental concepts and the classification criteria employed in cross-modal retrieval. Section \ref{Sec:Text-Image Cross-Modal Retrieval} delves deeply into various cross-modal retrieval methods centered around text-image, elucidating their intricate specifics and architectures. Section \ref{Sec:Cross-Modal Retrieval Beyond Text-Image} explores cross-modal retrieval methods beyond text-image, providing insights into the unique challenges and approaches for modalities. In Section \ref{Sec:Dataset and Evaluation}, an extensive compilation of widely used datasets, evaluation metrics, and performance comparisons within cross-modal retrieval research is presented. Section \ref{Sec:Application} highlights practical application scenarios where cross-modal retrieval techniques find relevance. Section \ref{Sec:Discussion and Outlook} undertakes a critical analysis of future developmental trends in the field. Ultimately, Section \ref{Sec:Conclusion} draws the paper closer, encapsulating key findings and contributions.

\begin{figure}[!t]
\centering
\includegraphics[width=0.485\textwidth]{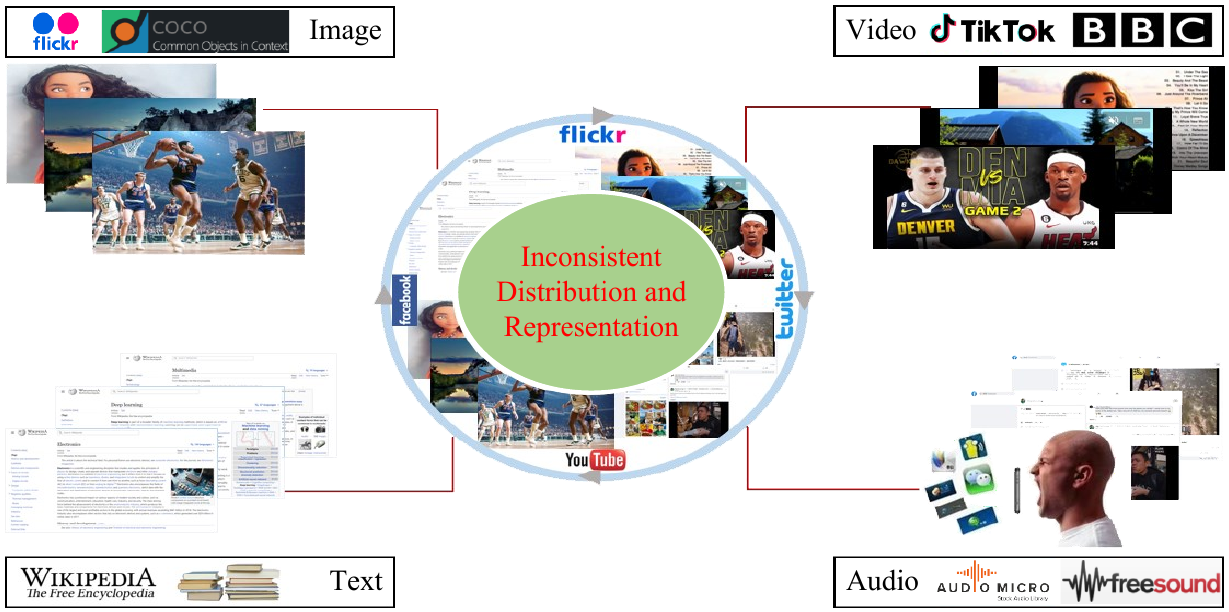}
\caption{Diagram of heterogeneous modality gap.}
\label{Figure:multimedia}
\end{figure}

\section{Overview}
\label{Sec:Overview}
Cross-modal retrieval stands as a pivotal domain within multimedia retrieval, poised with immense potential in the realm of artificial intelligence. Its purpose is to glean semantically pertinent information from disparate modalities, leveraging given modal cues like text, image, or video. Nonetheless, the landscape of cross-modal retrieval is rugged, the paramount among which is gauging content affinity amidst heterogeneous modal data---a conundrum often dubbed the heterogeneous modality gap, as shown in Fig. \ref{Figure:multimedia}. This quandary arises from divergences in data structures, feature spaces, and semantic portrayals across modalities, culminating in formidable obstacles for direct comparison and alignment. As a result, the crux of cross-modal retrieval research orbits around cultivating a shared framework for multi-modal data, thereby facilitating cross-modal similarity computation.

\begin{figure}[!t]
\centering
\includegraphics[width=0.47\textwidth]{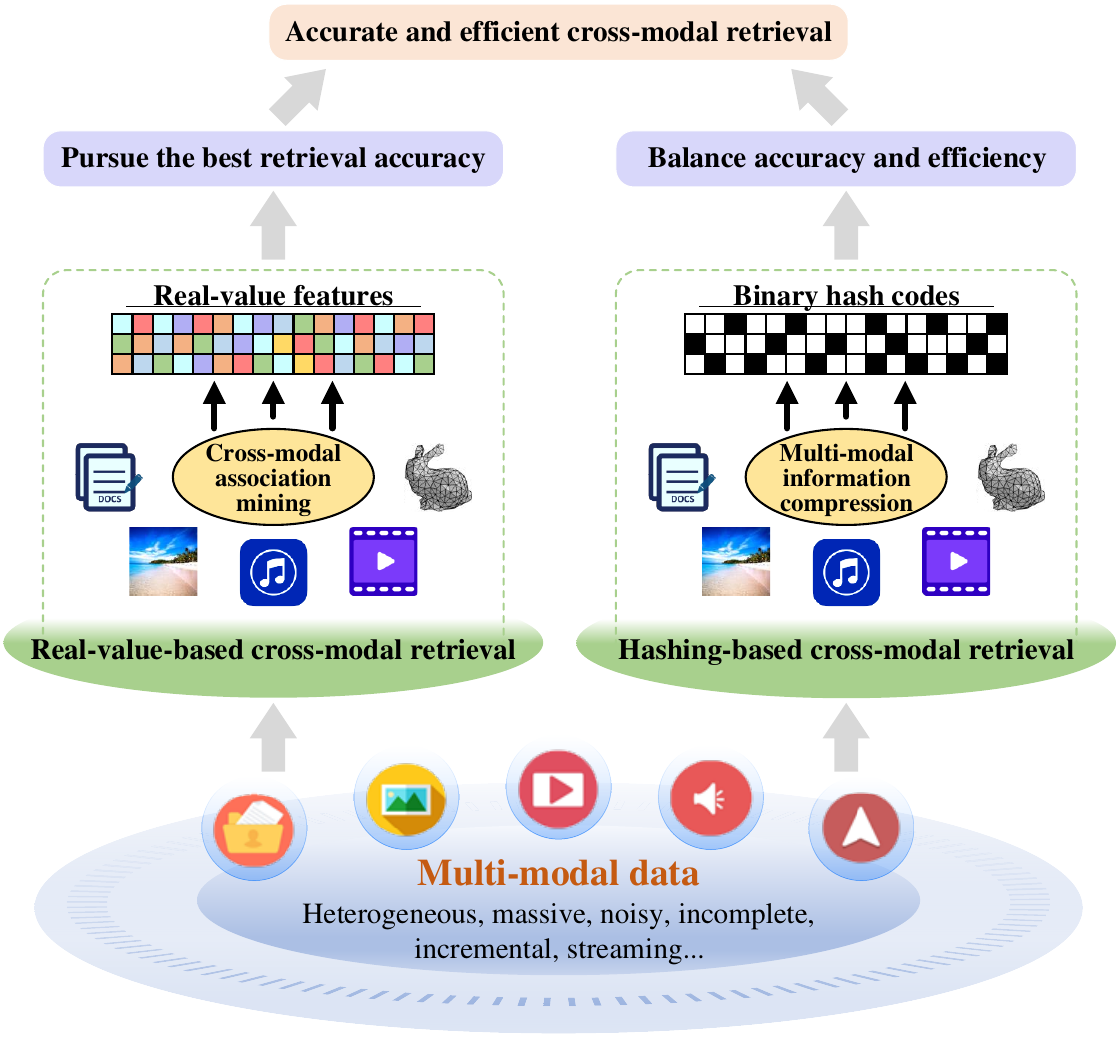}
\caption{A comparison of real-value and hashing-based cross-modal retrieval. While both can manage diverse multi-modal data, they differ significantly in modeling strategies, data representation, and core objectives.}
\label{Figure:comparison}
\end{figure}

In response to this challenge, researchers have proposed an array of common representation learning approaches. These methods endeavor to map diverse modal data into a shared, low-dimensional space. In this space, data with akin semantics cluster closely, while dissimilar data find themselves distanced. Broadly, cross-modal retrieval methods can be categorized into two archetypes based on their data encoding form: real-value retrieval and hashing retrieval. As shown in Fig. \ref{Figure:comparison}, real-value-based cross-modal retrieval \cite{CCA, CCL} strives to distill low-dimensional real-value features of multi-modal data. While this avenue preserves a richer tapestry of semantic information, it incurs elevated storage costs and computational demands. In contrast, hashing-based cross-modal retrieval \cite{CVH, MMNN} seeks to distill compressed binary representations of multi-modal data, fitting them into the Hamming space. This development enables more efficient retrieval. However, this efficiency comes at the cost of sacrificing some semantic information. Each possesses its unique set of advantages and drawbacks, and the decision between the two hinges on the particular requisites and limitations of practical applications.

Alongside the data coding form, supervisory information (i.e., manually annotated category labels) is a pivotal determinant for categorizing cross-modal retrieval methods. Supervised methods \cite{CCL, MMNN} rely on multi-modal data with labels to learn precise cross-modal associations, improving retrieval accuracy by leveraging explicit semantic guidance. However, they rely heavily on the availability and quality of multi-modal data labels, which can be expensive and time-consuming to annotate. In contrast, unsupervised methods \cite{CCA, CVH} operate without labels, making them more scalable and adaptable but often less effective in capturing fine-grained semantic relationships. Despite these differences, both approaches aim to map multi-modal data into a common representation space, with the choice depending on the availability of category labels and the trade-off between accuracy and scalability.

Beyond the two universal criteria for ideal retrieval scenarios, cross-modal retrieval under special scenarios should be treated as a separate category due to the unique challenges posed by real-world situations, such as noisy, incomplete, and incremental data\cite{WSJE, RUCMH, CCMR}. Traditional methods alone often fall short in addressing the complexity of these situations, necessitating customized solutions that account for practical constraints. Nonetheless, approaches in this category can still be classified by the data coding form and supervision type, adhering to real-value/hashing and supervised/unsupervised paradigms. This highlights both the specificity and versatility of cross-modal retrieval in these specialized contexts.

Combining these considerations, current cross-modal retrieval methods are categorized into five overarching categories: unsupervised real-value retrieval, supervised real-value retrieval, unsupervised hashing retrieval, supervised hashing retrieval, and cross-modal retrieval under special scenarios. Each of these overarching categories is subdivided based on specific technical architectures or scenarios:
\begin{itemize}
\item{The first four overarching categories undergo a meticulous refinement grounded in statistical analysis methods and deep network architectures. These encompass \emph{canonical correlation analysis}, \emph{topic learning}, \emph{dictionary learning}, \emph{matrix factorization}, \emph{spectral graph learning}, \emph{projection learning}, \emph{metric learning}, \emph{quantization learning}, \emph{self-reconstructing learning}, \emph{Convolutional Neural Network-Recurrent Neural Network (CNN-RNN)}, \emph{Generative Adversarial Network (GAN)}, \emph{Graph Neural Network (GNN)}, \emph{Transformer}, \emph{Vision-Language Pre-training model (VLP model)}, \emph{Cross-Modal Generation model (CMG model)}, \emph{distillation network}, \emph{memory network}, and \emph{quantization network}. Importantly, the last four subcategories differ from traditional network architectures, as they are tailored to advance the semantic representation learning of multi-modal data for cross-modal retrieval. Moreover, methods that integrate multiple techniques (e.g., GAN\cite{GAN}, GNN\cite{GNN}, and TransFormer\cite{TransFormer}) are classified into only one subcategory based on their key innovations.}
\begin{table*}[t]
\renewcommand\arraystretch{1.05}
\caption{A compilation of representative text-image cross-modal retrieval methods. Below, CNN-RNN stands for convolutional neural network and recurrent neural network, GAN stands for generative adversarial network, GNN stands for graph neural network, VLP model stands for vision-language pre-training model, and CMG model stands for cross-modal generation model.}
\label{Table:I}
\vspace{-3.93mm}
\begin{center}
    \resizebox{\linewidth}{!}{
        \begin{tabular}{|c|c|c|c|}
                \hline
                \multicolumn{3}{|c|}{Categories} & Representative methods \\
                \hline
                \hline
                \multirow{10}{*}{\makecell{Unsupervised \\Real-value Retrieval}} & \multirow{3}{*}{\makecell{Category-semantic\\ cross-modal retrieval}} &
                Canonical correlation analysis & \makecell{CCA\cite{CCA}, CR\cite{CR}, KCCA\cite{KCCA}, DCCA\cite{DCCA},  MCCA\cite{MCCA}, RCCA\cite{RCCA}} \\
                \cline{3-4}
                & & Topic learning & \makecell{Corr-LDA\cite{Corr-LDA}, tr-mmLDA\cite{tr-mmLDA}, MDRF\cite{MDRF}} \\
                \cline{3-4}
                & & Self-reconstructing learning & \makecell{Corr-AE\cite{Corr-AE}, CMDN\cite{CMDN}, MSAE\cite{MSAE}, CDPAE\cite{CDPAE}} \\
                \cline{2-4}
                & \multirow{6}{*}{\makecell{Image-text matching}} & CNN-RNN & \makecell{MSDS\cite{MSDS}, VSE++\cite{VSE++}, RRF-Net\cite{RRF-Net}, CHAIN-VSE\cite{CHAIN-VSE},
                 LDR\cite{LDR}, DRCE\cite{DRCE},\\ ESSE\cite{ESSE}, sm-LSTM\cite{sm-LSTM}, CRAN\cite{CRAN}, DSVEL\cite{DSVEL}, CAAN\cite{CAAN}, RANet\cite{RANet}}\\
                \cline{3-4}
                & & GNN & \makecell{KASCE\cite{KASCE}, VSRN\cite{VSRN}, GSMN\cite{GSMN}, SGRAF\cite{SGRAF}, CODER\cite{CODER}, CORA\cite{CORA}}\\
                \cline{3-4}
                & & Transformer & \makecell{PVSE\cite{PVSE}, PCME\cite{PCME}, DREN\cite{DREN}, TGDT\cite{TGDT}, HREM\cite{HREM}, TransTPS\cite{TransTPS}} \\
                \cline{3-4}
                & & VLP model & \makecell{COTS\cite{COTS}, TEAM\cite{TEAM}, EI-CLIP\cite{EI-CLIP},  AGREE\cite{AGREE}, CSIC\cite{CSIC}, IRRA\cite{IRRA}, \\ MAKE\cite{MAKE}, USER\cite{USER},  LAPS\cite{LAPS}}  \\
                \cline{3-4}
                & & CMG model & DVSA\cite{DVSA}, 2WayNet\cite{2WayNet}, GXN\cite{GXN}, LSCO\cite{LSCO}, X-MRS\cite{X-MRS}, GRACE\cite{GRACE}  \\
                \hline
                \multirow{9}{*}{\makecell{Supervised \\ Real-value Retrieval}} & \multirow{5}{*}{\makecell{Shallow supervised \\ real-value retrieval}}
                & Canonical correlation analysis & \makecell{GMA\cite{GMA}, ml-CCA\cite{ml-CCA}} \\
                \cline{3-4}
                & & Dictionary learning & \makecell{SliM$^2$\cite{SliM2}, DDL\cite{DDL}} \\
                \cline{3-4}
                & & Projection learning & \makecell{JRL\cite{JRL}, JFSSL\cite{JFSSL}, MDCR\cite{MDCR}, JLSLR\cite{JLSLR}} \\
                \cline{3-4}
                & & Topic learning & \makecell{M$^3$R\cite{M3R}, NPBUS\cite{NPBUS}} \\
                \cline{3-4}
                & & Metric learning & \makecell{PFAR\cite{PFAR}, PL-ranking\cite{PL-ranking}, CMOS\cite{CMOS1}} \\
                \cline{2-4}
                & \multirow{4}{*}{\makecell{Deep supervised \\ real-value retrieval}}
                & CNN-RNN & CCL\cite{CCL}, MCSM\cite{MCSM}, DSCMR\cite{DSCMR}, ED-Net\cite{ED-Net}, SDCMR\cite{SDCMR}\\
                \cline{3-4}
                & & GAN & \makecell{ACMR\cite{ACMR}, CM-GANs\cite{CM-GANs}, AACR\cite{AACR}, CAAL\cite{CAAL}, AAL\cite{AAL}} \\
                \cline{3-4}
                & & GNN & \makecell{SSPE\cite{SSPE}, DAGNN\cite{DAGNN}, ALGCN\cite{ALGCN}, HGE\cite{HGE}, WGSCN\cite{WGSCN}} \\
                \cline{3-4}
                & & Transformer & \makecell{RLCMR\cite{RLCMR}, CLIP4CMR\cite{CLIP4CMR}, IEFT\cite{IEFT}} \\
                \hline
                \multirow{9}{*}{\makecell{Unsupervised \\ Hashing Retrieval}} & \multirow{4}{*}{\makecell{Shallow unsupervised \\ hashing retrieval}}
                & Matrix factorization & \makecell{CMFH\cite{CMFH}, LSSH\cite{LSSH}, STMH\cite{STMH}, RFDH\cite{RFDH}, SUH\cite{SUH}} \\
                \cline{3-4}
                & & Spectral graph learning & \makecell{CVH\cite{CVH}, LCMH\cite{LCMH}, IMH\cite{IMH}, SM$^2$H\cite{SM2H}, HMR\cite{HMR}} \\
                \cline{3-4}
                & & Metric learning & \makecell{PDH\cite{PDH}, IMVH\cite{IMVH}, CRE\cite{CRE} UDC\cite{UDC}}\\
                \cline{3-4}
                & & Quantization learning & \makecell{ACQ\cite{ACQ}, CMCQ\cite{CMCQ}, CCQ\cite{CCQ}} \\
                \cline{2-4}
                & \multirow{5}{*}{\makecell{Deep unsupervised\\ hashing retrieval}}
                & CNN-RNN & \makecell{DMHOR\cite{DMHOR}, DBRC\cite{DBRC}, UDCMH\cite{UDCMH}, UCHM\cite{UCHM}, UCHSTM\cite{UCHSTM}} \\
                \cline{3-4}
                & & GAN & \makecell{UGACH\cite{UGACH}, CYC-DGH\cite{CYC-DGH}, UCH\cite{UCH}, MGAH\cite{MGAH}} \\
                \cline{3-4}
                & & GNN & \makecell{DJSRH\cite{DJSRH}, JDSH\cite{JDSH}, DGCPN\cite{DGCPN}, AGCH\cite{AGCH}, CIRH\cite{CIRH}, CMGCH\cite{CMGCH}} \\
                \cline{3-4}
                & & Transformer & HuggingHash\cite{HuggingHash}, UCMFH\cite{UCMFH} \\
                \cline{3-4}
                & & Distillation network & \makecell{UKD\cite{UKD}, KDCMH\cite{KDCMH}, JOG\cite{JOG}, DAEH\cite{DAEH}, CKDH\cite{CKDH}} \\
                \hline
                \multirow{10}{*}{\makecell{Supervised \\ Hashing Retrieval}} & \multirow{3}{*}{\makecell{Shallow supervised\\ hashing retrieval}}
                & Matrix factorization & \makecell{SMFH\cite{SMFH1}, DCH\cite{DCH}, LCMFH\cite{LCMFH}, BATCH\cite{BATCH}, ASFOH\cite{ASFOH}, \\ DASRH\cite{DASRH}, DCDH\cite{DCDH}, DLCMH\cite{DLCMH}} \\
                \cline{3-4}
                & & Projection learning & \makecell{MDBE\cite{MDBE}, ALECH\cite{ALECH}, QCH\cite{QCH}, HCCH\cite{HCCH}, SRDMH\cite{SRDMH1}, FDCH\cite{FDCH}, \\ MIEH\cite{MIEH}, CRH\cite{CRH}, RaHH\cite{RaHH}, SCM\cite{SCM}, LSRH\cite{LSRH}} \\
                \cline{2-4}
                & \multirow{6}{*}{\makecell{Deep supervised\\ hashing retrieval}}
                & CNN-RNN & \makecell{CMNNH\cite{CMNNH}, MMNN\cite{MMNN}, CAH\cite{CAH}, MCITR\cite{MCITR}, DVSH\cite{DVSH}, DSMHN\cite{DSMHN}, \\ DCHUC\cite{DCHUC}, MRDH\cite{MRDH}, PRDH\cite{PRDH}, DCMH\cite{DCMH}, DCH-SCR\cite{DCH-SCR},\\ TDH\cite{TDH}, RDCMH\cite{RDCMH}, SCH\cite{SCH}} \\
                \cline{3-4}
                & & GAN & \makecell{SSAH\cite{SSAH}, DADH\cite{DADH}, CPAH\cite{CPAH},
                HMAH\cite{HMAH}, SDAH\cite{SDAH}} \\
                \cline{3-4}
                & & GNN & \makecell{GCH\cite{GCH}, LGCNH\cite{LGCNH}, PGCH\cite{PGCH}, GCDH\cite{GCDH}} \\
                \cline{3-4}
                & & Transformer & \makecell{DCHMT\cite{DCHMT}, MITH\cite{MITH}, DSPH\cite{DSPH},  DHaPH\cite{DHaPH}} \\
                \cline{3-4}
                & & Memory network & \makecell{CMMN\cite{CMMN}, CMPD\cite{CMPD}} \\
                \cline{3-4}
                & & Quantization network & \makecell{CDQ\cite{CDQ}, ACQH\cite{ACQH}} \\
                \hline
                \multicolumn{2}{|c|}{\multirow{13}{*}{Cross-modal Retrieval under Special Scenarios}}
                &Noise-robust & \makecell{WSJE\cite{WSJE}, MRL\cite{MRL}, WASH\cite{WASH}, LCNME\cite{LCNME}, DECL\cite{DECL}, RCL\cite{RCL},\\ SREM\cite{SREM}, RDE\cite{RDE}} \\
                \cline{3-4}
                \multicolumn{2}{|c|}{}& Incomplete & \makecell{RUCMH\cite{RUCMH}, GSPH\cite{GSPH1}, AMSH\cite{AMSH}, TFNH\cite{TFNH}, PAN\cite{PAN}, \\ DAVAE\cite{DAVAE}, ICMR-DCT\cite{ICMR-DCT}, CICH\cite{CICH}} \\
                \cline{3-4}
                \multicolumn{2}{|c|}{}& \multicolumn{1}{c|}{Zero/few-shot} & \makecell{MASLN\cite{MASLN}, TANSS\cite{TANSS}, MDVAE\cite{MDVAE}, AG-MIH\cite{AG-MIH},\\ Self-Others\cite{Self-Others}, ACMM\cite{ACMM}} \\
                \cline{3-4}
                \multicolumn{2}{|c|}{}& Incremental & \makecell{CCMR\cite{CCMR}, ECMH\cite{ECMH}, CTP\cite{CTP}, CCMH-GAM\cite{CCMH-GAM}, SVHNs\cite{SVHNs},\\ MARS\cite{MARS}, SDML\cite{SDML}, Uni-Code\cite{Uni-Code}} \\
                \cline{3-4}
                \multicolumn{2}{|c|}{}& \multicolumn{1}{c|}{Online} & \makecell{OCMH\cite{OCMH}, OCMFH\cite{OCMFH}, ONION\cite{ONION}, ROH\cite{ROH}, POLISH\cite{POLISH}} \\
                \cline{3-4}
                \multicolumn{2}{|c|}{}& \multicolumn{1}{c|}{Cross-domain} & \makecell{DASG\cite{DASG}, ACP\cite{ACP}, MHTN\cite{MHTN}, CDTH\cite{CDTH}} \\
                \cline{3-4}
                \multicolumn{2}{|c|}{}& \multicolumn{1}{c|}{Federated} & \makecell{FedCMR\cite{FedCMR}, PT-FUCH\cite{PT-FUCH}, PEPFCH\cite{PEPFCH}, FedCAFE\cite{FedCAFE}} \\
                \cline{3-4}
                \multicolumn{2}{|c|}{}& \multicolumn{1}{c|}{Composed} & \makecell{CMJPA\cite{CMJPA}, TFUN\cite{TFUN}, Cola\cite{Cola}, AlRet\cite{AlRet}} \\
                \cline{3-4}
                \multicolumn{2}{|c|}{}& \multicolumn{1}{c|}{Hierarchical/fine-grained} & \makecell{HiCHNet\cite{HiCHNet}, HSSPH\cite{HSSPH}, FGCrossNet\cite{FGCrossNet}, PCMDA\cite{PCMDA}} \\
                \cline{3-4}
                \multicolumn{2}{|c|}{}& \multicolumn{1}{c|}{Adversary} & \makecell{CMLA\cite{CMLA}, TA-DCH\cite{TA-DCH}, AACH\cite{AACH}, EQB2A\cite{EQB2A}, BadCM\cite{BadCM}, \\ IB$^3$A\cite{IB3A}, PIP\cite{PIP}, PPCL\cite{PPCL}}\\
                \hline
        \end{tabular}}
\end{center}
\end{table*}
\item{The last overarching category, cross-modal retrieval under special scenarios, covers various constrained settings encountered in practical applications: \emph{noise-robust retrieval}, \emph{incomplete retrieval}, \emph{zero/few-shot retrieval}, \emph{incremental retrieval}, \emph{online retrieval}, \emph{cross-domain retrieval}, \emph{federated retrieval}, \emph{composed retrieval}, \emph{hierarchical/fine-grained retrieval}, and \emph{adversary against cross-modal retrieval}. As previously noted, methods in this category still adhere to the refinement criteria established by the first four overarching categories.}
\end{itemize}

The taxonomy outlined above is versatile and applicable to a wide range of modality combinations in cross-modal retrieval. Whether dealing with text-image, text-video, text-audio, image-audio, or other multi-modal combinations, the five overarching categories and forty-four subcategories offer a comprehensive framework for systematically evaluating and addressing the associated retrieval challenges.

\begin{figure*}[!b]
\centering
    \begin{overpic}[width=1\textwidth]{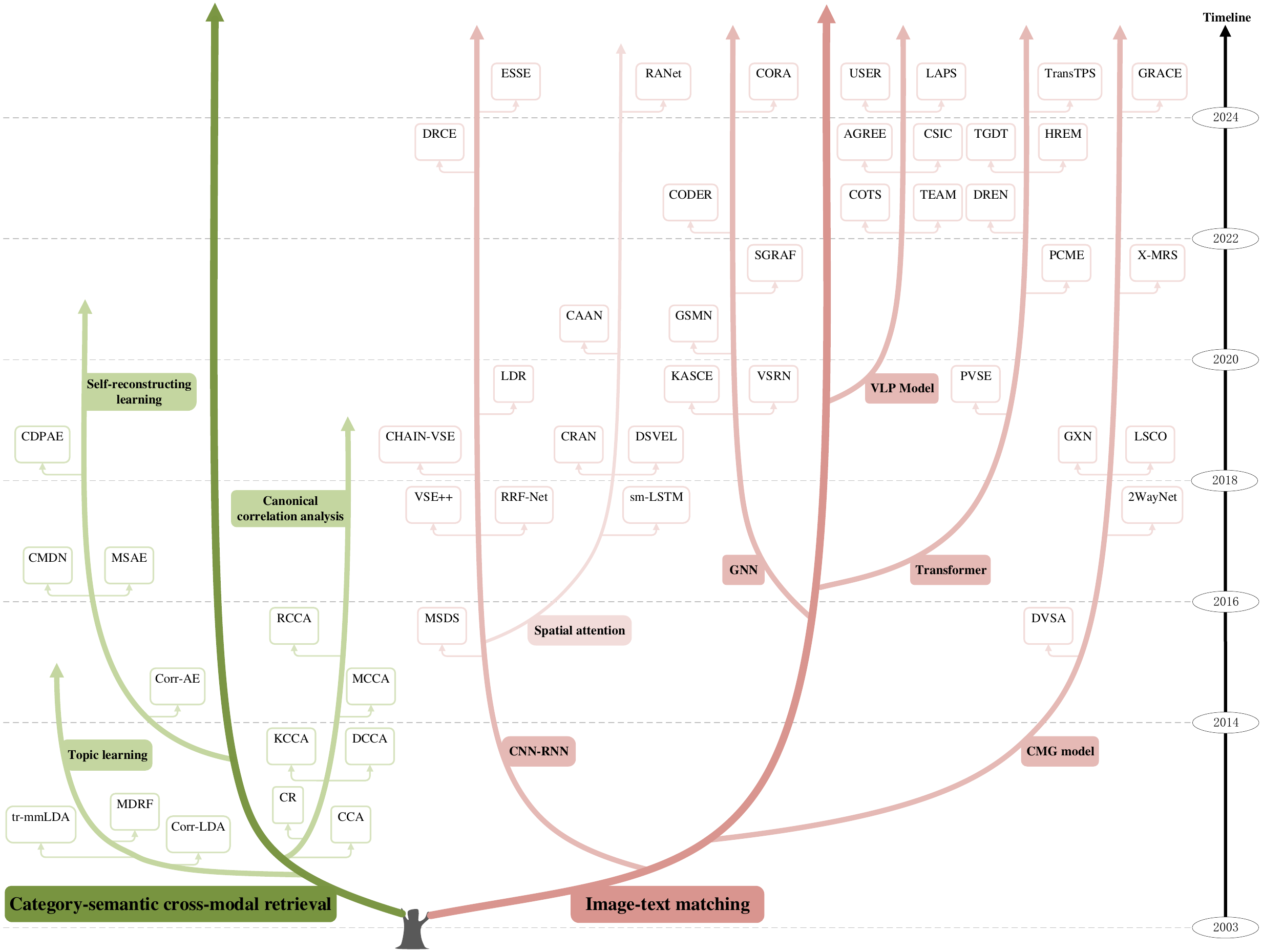}
        \put(2.01,9.2){\fontsize{6pt}{9pt}\selectfont \cite{tr-mmLDA}}
        \put(14.45,8.4){\fontsize{6pt}{9pt}\selectfont \cite{Corr-LDA}}
        \put(9.4,10.2){\fontsize{6pt}{9pt}\selectfont \cite{MDRF}}
        \put(13.1,20.15){\fontsize{6pt}{9pt}\selectfont \cite{Corr-AE}}
        \put(28.05,15.45){\fontsize{6pt}{9pt}\selectfont \cite{DCCA}}
        \put(21.55,10.8){\fontsize{6pt}{9pt}\selectfont \cite{CR}}
        \put(26.85,9.2){\fontsize{6pt}{9pt}\selectfont \cite{CCA}}
        \put(21.8,15.45){\fontsize{6pt}{9pt}\selectfont \cite{KCCA}}
        \put(22.05,25){\fontsize{6pt}{9pt}\selectfont \cite{RCCA}}
        \put(28.15,20.15){\fontsize{6pt}{9pt}\selectfont \cite{MCCA}}
        \put(33.8,25){\fontsize{6pt}{9pt}\selectfont \cite{MSDS}}
        \put(81.85,25){\fontsize{6pt}{9pt}\selectfont \cite{DVSA}}
        \put(2.5,29.8){\fontsize{6pt}{9pt}\selectfont \cite{CMDN}}
        \put(9.0,29.8){\fontsize{6pt}{9pt}\selectfont \cite{MSAE}}
        \put(2.1,39.35){\fontsize{6pt}{9pt}\selectfont \cite{CDPAE}}
        \put(32.,39.35){\fontsize{6pt}{9pt}\selectfont \cite{CHAIN-VSE}}
        \put(33.1,34.5){\fontsize{6pt}{9pt}\selectfont \cite{VSE++}}
        \put(44.6,39.35){\fontsize{6pt}{9pt}\selectfont \cite{CRAN}}
        \put(50.7,39.35){\fontsize{6pt}{9pt}\selectfont \cite{DSVEL}}
        \put(84.17,39.35){\fontsize{6pt}{9pt}\selectfont \cite{GXN}}
        \put(89.9,39.35){\fontsize{6pt}{9pt}\selectfont \cite{LSCO}}
        \put(90.1,34.5){\fontsize{6pt}{9pt}\selectfont \cite{2WayNet}}
        \put(40.3,34.5){\fontsize{6pt}{9pt}\selectfont \cite{RRF-Net}}
        \put(50.75,34.5){\fontsize{6pt}{9pt}\selectfont \cite{sm-LSTM}}
        \put(53.6,44.15){\fontsize{6pt}{9pt}\selectfont \cite{KASCE}}
        \put(39.46,44.15){\fontsize{6pt}{9pt}\selectfont \cite{LDR}}
        \put(60.05,44.15){\fontsize{6pt}{9pt}\selectfont \cite{VSRN}}
        \put(76.07,44.15){\fontsize{6pt}{9pt}\selectfont \cite{PVSE}}
        \put(45.05,48.88){\fontsize{6pt}{9pt}\selectfont \cite{CAAN}}
        \put(53.7,48.88){\fontsize{6pt}{9pt}\selectfont \cite{GSMN}}
        \put(60.17,53.68){\fontsize{6pt}{9pt}\selectfont \cite{SGRAF}}
        \put(83.25,53.68){\fontsize{6pt}{9pt}\selectfont \cite{PCME}}
        \put(90.5,53.68){\fontsize{6pt}{9pt}\selectfont \cite{X-MRS}}
        \put(53.5,58.5){\fontsize{6pt}{9pt}\selectfont \cite{CODER}}
        \put(67.35,58.5){\fontsize{6pt}{9pt}\selectfont \cite{COTS}}
        \put(73.15,58.5){\fontsize{6pt}{9pt}\selectfont \cite{TEAM}}
        \put(77.25,58.5){\fontsize{6pt}{9pt}\selectfont \cite{DREN}}
        \put(33.54,63.3){\fontsize{6pt}{9pt}\selectfont \cite{DRCE}}
        \put(67.3,63.3){\fontsize{6pt}{9pt}\selectfont \cite{AGREE}}
        \put(73.12,63.3){\fontsize{6pt}{9pt}\selectfont \cite{CSIC}}
        \put(77.3,63.3){\fontsize{6pt}{9pt}\selectfont \cite{TGDT}}
        \put(83.05,63.3){\fontsize{6pt}{9pt}\selectfont \cite{HREM}}
        \put(39.65,68.1){\fontsize{6pt}{9pt}\selectfont \cite{ESSE}}
        \put(51.3,68.1){\fontsize{6pt}{9pt}\selectfont \cite{RANet}}
        \put(60.1,68.1){\fontsize{6pt}{9pt}\selectfont \cite{CORA}}
        \put(67.35,68.1){\fontsize{6pt}{9pt}\selectfont \cite{USER}}
        \put(73.36,68.1){\fontsize{6pt}{9pt}\selectfont \cite{LAPS}}
        \put(83.5,68.1){\fontsize{6pt}{9pt}\selectfont \cite{TransTPS}}
        \put(90.65,68.1){\fontsize{6pt}{9pt}\selectfont \cite{GRACE}}
    \end{overpic}
\vspace{-5mm}
\caption{The evolutionary tree of representative unsupervised real-value retrieval methods.}
\label{Figure:UR}
\end{figure*}

\section{Text-Image Cross-Modal Retrieval}
\label{Sec:Text-Image Cross-Modal Retrieval}
In this section, we detail representative methods for cross-modal retrieval between text and image modalities. Table \ref{Table:I} compiles these methods and their respective categories, including unsupervised real-value retrieval, supervised real-value retrieval, unsupervised hashing retrieval, supervised hashing retrieval, and cross-modal retrieval under special scenarios. To make the literature review more uniform and accurate, we formalize the following notation and definition.

Consider a multi-modal dataset $\mathbf{\mathcal{O}}$ consists of $n$ instances, represented as $\mathcal{O}=\{V, T, L\}=\{v_i,t_i,l_i\}^n_{i=1}$, where $V, T, L$ are data matrices for images, texts, and labels respectively, and $v_i, t_i, l_i$ are the image feature vector, text feature vector, one-hot label vector for the $i$-th instance respectively. We define a similarity matrix $S=\{s_{ij}\}\in\mathbb{R}^{n\times n}$ to describe the relationship between multi-modal instances, where $s_{ij}=1$ if the $i$-th instance shares at least one category with the $j$-th instance, and $s_{ij}=0$ otherwise. Note that $L$ and $S$ are not available in unsupervised cross-modal retrieval. For the other symbols involved, we uniformly use capital letters (e.g., $A$, $B$, $C$, $\dots$) to represent matrices, lowercase letters (e.g., $a$, $b$, $c$, $\dots$) to represent vectors, and Greek letters (e.g., $\alpha$, $\beta$, $\gamma$, $\dots$) to represent scalars or hyper-parameters. Additionally, we define $\top$, $\|\cdot\|_1$, $\|\cdot\|_{2,1}$, $\|\cdot\|_F$, and ${\rm trace}(\cdot)$ as the matrix transposition, $l_1$-norm, $l_{2,1}$-norm, Frobenius-norm, and trace, respectively, while ${\rm sign}(\cdot)$ represents the sign function.

Below, we present the fundamental principles and development of cross-modal retrieval methods for each category between text and image modalities.

\subsection{Unsupervised Real-value Retrieval}
Unsupervised real-value retrieval aims to leverage the co-occurrence of multi-modal data, such as text and images appearing together, to capture their semantic correlation. Depending on the retrieval objectives, it can be categorized into two types: category-semantic cross-modal retrieval and image-text matching. The former focuses on retrieving results that belong to the same categories as queries. It typically employs statistical analysis methods or auto-encoders to learn sample-level co-occurrence in multi-modal data, modeling associations through one-to-one correspondence of paired heterogeneous samples. The retrieval accuracy is evaluated based on how well the results overlap with the query category (e.g., mean Average Precision, mAP). In contrast, the latter aims to match cross-modal samples that directly correspond to queries, rather than merely belonging to the same categories. It relies on various deep neural networks and optimizes performance using triplet-based ranking, prioritizing cross-modal samples that correspond to queries (e.g., Recall). The classification of the two categories is based on their differing retrieval objectives in unsupervised real-valued retrieval. In the other overarching categories, methods share a singular objective: retrieving results that belong to the same categories as the queries. Consequently, these overarching categories can be classified directly according to whether the methods utilize statistical analysis methods or deep neural networks. Additionally, based on the techniques used, further categorization is possible. As shown in Fig. \ref{Figure:UR}, category-semantic cross-modal retrieval can be subdivided into \emph{canonical correlation analysis methods}, \emph{topic learning methods}, and \emph{self-reconstructing learning methods}. Image-text matching, on the other hand, can be subdivided into \emph{CNN-RNN methods}, \emph{GNN methods}, \emph{Transformer methods}, \emph{VLP model methods}, and \emph{CMG model methods}.

\subsubsection{Canonical correlation analysis methods}
Canonical Correlation Analysis (CCA) is essentially a multi-variate statistical analysis method used to study the correlation between two sets of variables. CCA establishes a relationship model by finding linear combinations of each set that maximize their correlation. This allows the mapping of two sets of variables into the same low-dimensional statistical space, making similarity measurement easier. Formally, given an image feature matrix $V$ and a text feature matrix $T$ with pair-wise correspondence, by learning their transformation matrices $W_v$ and $W_t$, CCA-based methods \cite{CCA, CR, KCCA, DCCA, MCCA, RCCA} can map heterogeneous multi-modal data into a low-dimensional common space:
\begin{equation}
\small
\begin{aligned}
(W'_v, W'_t) &= \mathop{\arg\max}\limits_{W_v, W_t} \ {\rm corr}(VW_v, TW_t) \\
&=\mathop{\arg\max}\limits_{W_v, W_t} \ \frac{W_v^{\top}\sum_{VT}W_t}{\sqrt{W_v^{\top}\sum_{VV}W_v} \sqrt{W_t^{\top}\sum_{TT}W_t}},
\end{aligned}
\end{equation}
where $V\in\mathbb{R}^{n_{tr}\times d_v}$, $T\in\mathbb{R}^{n_{tr}\times d_t}$, $W_v\in\mathbb{R}^{d_v\times d_c}$, $W_t\in\mathbb{R}^{d_t\times d_c}$, $n_{tr}$ is the number of training samples, $d_v$ and $d_t$ are the feature dimensions of the images and texts respectively, and $d_c$ is the dimension of common representations. In addition, $\sum_{VV}\in\mathbb{R}^{d_v\times d_v}$ and $\sum_{TT}\in\mathbb{R}^{d_t\times d_t}$ are the covariance matrices of $V$ and $T$ respectively, while $\sum_{VT}\in\mathbb{R}^{d_v\times d_t}$ is their cross-covariance matrix. Among CCA-based methods, Rasiwasia et al. \cite{CCA} systematically prove the importance of cross-modal correlation in explicit modeling for the first time and realize semantic association matching by combining CCA and semantic modeling. Later, Kernel Canonical Correlation Analysis (KCCA)\cite{KCCA} and Deep Canonical Correlation Analysis (DCCA)\cite{DCCA} are proposed to overcome the limitation of CCA to model non-linear relationships. Multi-view Canonical Correlation Analysis (MCCA)\cite{MCCA} clusters several semantic centers and then integrates them with image-text views to realize multi-view learning.

\subsubsection{Topic learning methods}
Topic models, originally designed for textual data, assume that each document is composed of multiple topics, each of which is composed of multiple words. Latent Dirichlet Allocation (LDA) proposed by Blei et al., as a popular topic model, has been widely used in the field of text classification and information retrieval \cite{Corr-LDA, tr-mmLDA, MDRF}. LDA is able to represent multi-modal data as a combination of topic distributions and modality-specific distributions, from which it can infer the relationship between topics and modalities. The process can be defined through the following steps:
\begin{itemize}
\item{Sample a latent random variable $\theta\sim{\rm Dir}(\theta|\alpha)$ and generate its probability distribution ${\rm Multi}(\theta)$;}
\item{For each subregion of an image, sample the latent factor $v_n$, $n\in\{1,\dots,N\}$ that generates the $n$-th subregion, and generate the subregion $r_n$ conditioned on $v_n$;}
\item{For each word of a document, sample the latent factor $t_m$, $m\in\{1,\dots,M\}$ that generates the $m$-th word, and generate the word $w_m$ conditioned on $t_m$.}
\end{itemize}

From above, the joint distribution of subregions, words, and latent variables is as follows:
\begin{equation}
\small
\begin{aligned}
p({\bf r}, {\bf w},\theta, {\bf v}, {\bf t})=p(\theta|\alpha)&(\prod^N_{n=1}p(v_n|\theta)p(r_n|v_n,\mu,\sigma)) \\
&\cdot (\prod^M_{m=1}p(t_m|\theta)p(w_m|t_m,\beta)),
\end{aligned}
\end{equation}
where $\theta,\alpha,v_n,t_m\in \mathbb{R}^{n_k}$, $r_n\in\mathbb{R}^{d_v}$ and $w_m\in\mathbb{R}^{d_t}$ are the feature vectors of the $n$-th image subregion and the $m$-th word respectively. $\mu\in\mathbb{R}^{d_v}$ and $\sigma\in\mathbb{R}^{d_v}$ denote the conditional mean and variance of subregion feature vectors, respectively. $\beta\in\mathbb{R}^{n_k\times n_t}$ denotes the multi-dimensional distribution parameters of words, where $n_k$ is the number of latent factors, $d_v$ and $d_t$ are the feature dimensions of image subregions and words respectively, and $n_t$ is the size of the vocabulary. Based on the above principles, Correspondence Latent Dirichlet Allocation (Corr-LDA)\cite{Corr-LDA} assumes latent topics as the shared variables of multi-modal data, using cross-modal condition generation for modeling. In contrast, topic-regression multi-modal Latent Dirichlet Allocation (tr-mmLDA)\cite{tr-mmLDA} and Multi-modal Document Random Field (MDRF)\cite{MDRF} introduce a regression module and Markov random field to correlate latent topics across modalities, respectively, resulting in a more flexible form of topic modeling.

\subsubsection{Self-reconstructing methods}
Auto-encoders\cite{AE} are a type of neural network designed to compress data into a lower-dimensional form and then reconstruct it. Self-reconstructing methods\cite{Corr-AE, CMDN, MSAE, CDPAE} leverage this capability to learn similar representations from multi-modal data. One notable approach, Correspondence Auto-Encoder (Corr-AE) \cite{Corr-AE}, correlates latent representations of modality-specific auto-encoders by minimizing representation error within each modality and correlation error between different modalities. Subsequently, Cross-media Multiple Deep Network (CMDN) \cite{CMDN} and Multi-modal Stacked Auto-Encoders (MSAE) \cite{MSAE} employ stacked deep auto-encoders to learn intra- and inter-modal semantic relationships. Comprehensive Distance-Preserving Auto-Encoders (CDPAE) \cite{CDPAE} introduces denoising auto-encoders to reduce redundant noises and designs a distance-preserving common space and joint loss function to effectively coordinate intra- and inter-modal distance metrics.

\subsubsection{CNN-RNN methods}
\label{Naive network methods}
CNN-RNN methods \cite{MSDS, RRF-Net, VSE++, CHAIN-VSE, LDR, ESSE, DRCE, sm-LSTM, DSVEL, CRAN, CAAN, RANet} rely on Multi-Layer Perceptron (MLP), Convolutional Neural Network (CNN), and Recurrent Neural Network (RNN), and realize cross-modal constraint by extracting the features of each modal separately and constructing positive/negative samples according to paired information. With CNN as the backbone, Modality-Specific Deep Structure (MSDS)\cite{MSDS} adopts a maximum likelihood-based one-vs-more scheme to ensure the relevance of image-text pairs. Combining CNN and RNN to extract multi-modal features, Visual-Semantic Embeddings++ (VSE++)\cite{VSE++} integrates hard sample mining into the ranking loss and has been widely adopted in subsequent works. Subsequently, residual learning\cite{RRF-Net}, character-level convolution \cite{CHAIN-VSE}, and disentangled representation\cite{LDR} are introduced into image-text matching, aiming to achieve better retrieval performance by enhancing multi-modal feature mapping. Recent CNN-RNN methods focus on inherent challenges in image-text matching, such as Dual-path Rare Content Enhancement network (DRCE)\cite{DRCE} improves rare content representation and association using a dual-path structure, adaptive fusion, and a re-ranking strategy to handle the long-tail distribution issue, and Estimating the Semantics via Sector Embedding (ESSE)\cite{ESSE} addresses one-to-many correspondence by projecting data as sectors with uncertainty apertures.

Spatial attention \cite{sm-LSTM, DSVEL, CRAN, CAAN, RANet} is a widely adopted technique in CNN-RNN-based cross-modal retrieval due to its remarkable capability in mining fine-grained cross-modal associations. Its objective is to enhance feature representations of key regions by generating weighted masks for each local region (such as textual words or visual local images), thereby boosting the importance of target regions while downplaying irrelevant regions. Representatively, Deep Semantic-Visual Embedding with Localization (DSVEL)\cite{DSVEL} introduces the spatial-aware pooling mechanism to identify the image region corresponding to any given text. Cross-media Relation Attention Network (CRAN)\cite{CRAN} and Context-Aware Attention Network (CAAN)\cite{CAAN} build upon global-local alignment, introducing relation alignment and context-aware selection to uncover fine-grained alignments. One of the latest studies, Reference-aware Adaptive Network (RANet)\cite{RANet} improves attention mechanisms with reference attention that minimizes incorrect attention scores in cross-modal interaction and an adaptive aggregation that boosts useful information while reducing redundancy.

\begin{figure}[!t]
\centering
\includegraphics[width=0.49\textwidth]{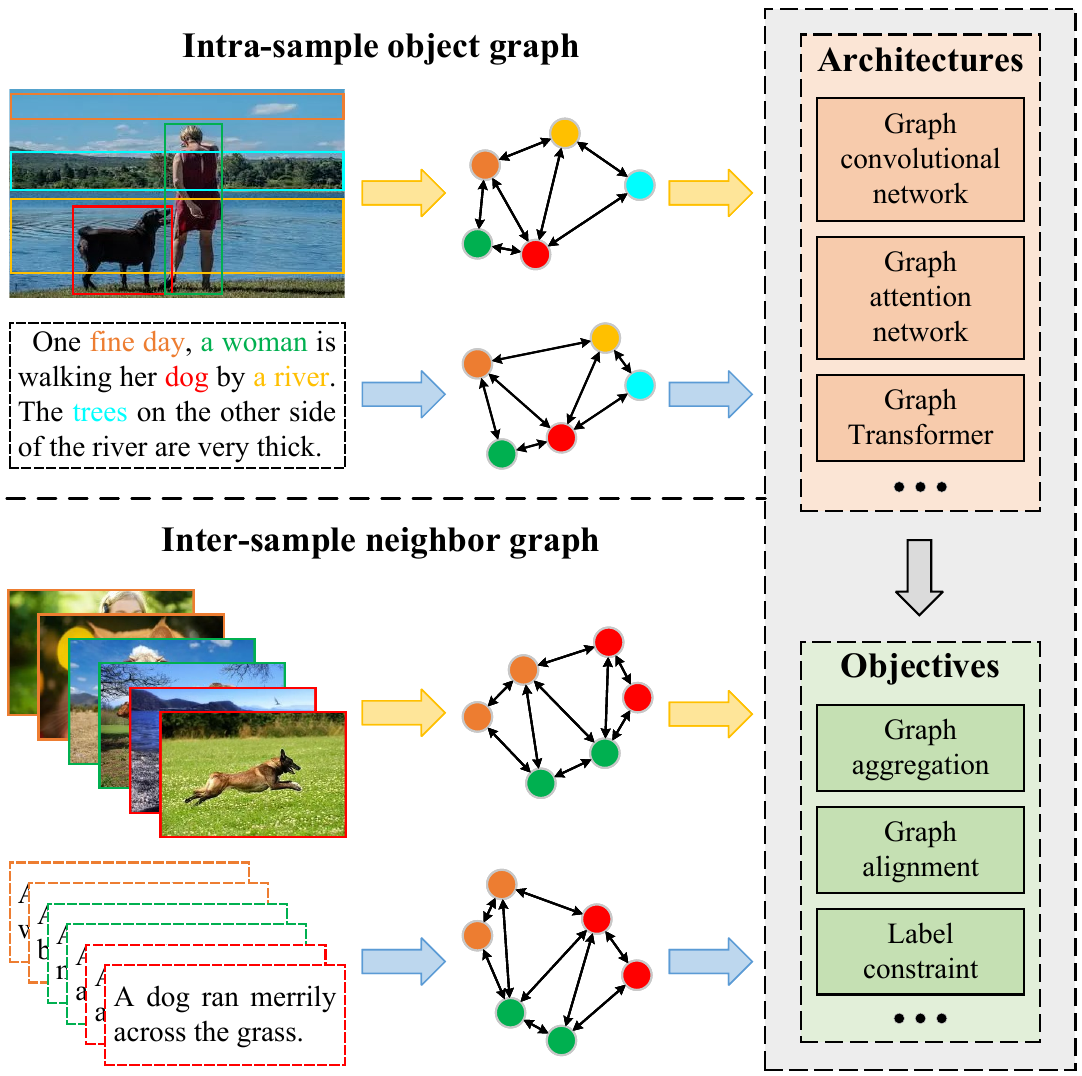}
\caption{Comparison of graph construction ways in different cross-modal retrieval tasks. Object-oriented image-text matching primarily builds object graphs within samples, whereas other cross-modal retrieval tasks focus on constructing neighbor graphs connecting samples.}
\label{Figure:Graph}
\end{figure}

\subsubsection{GNN methods}
GNN methods \cite{KASCE, VSRN, GSMN, SGRAF, CODER, CORA} construct object graphs for the basic components (e.g., words or local images) within each multi-modal instance, as illustrated in Fig. \ref{Figure:Graph}. This sets it apart from the graph-based studies in other cross-modal retrieval tasks. After constructing object graphs, they predominantly employ Graph Convolution Network (GCN)\cite{GCN} and Graph ATtention network (GAT)\cite{GAT} to align multi-modal data. Typically, Knowledge Aware Semantic Concept Expansion (KASCE)\cite{KASCE} enhances existing semantic concepts using a pre-constructed scene concept graph. Visual Semantic Reasoning Network (VSRN)\cite{VSRN} combines region-relational graphs with global semantic relations for better alignment with corresponding text. COupled DivERsity-sensitive momentum contrastive learning (CODER)\cite{CODER} enhances multi-modal graph embedding features through cross-attention gated fusion and coupled momentum contrastive learning. Recently, Composition model for Object Relations and Attributes (CORA)\cite{CORA} uses scene graphs to encode object-attribute and object-object relations, enhancing image-text matching through the relational inductive bias of graph networks.

\subsubsection{Transformer methods}
\label{Transformer methods}
Transformer methods \cite{PVSE, PCME, DREN, TGDT, HREM, TransTPS} harness the potency of multi-head self-attention mechanisms, enabling deep networks to focus concurrently on all incoming data. This advantageous attribute has been empirically substantiated for its heightened performance in multi-modal modeling, rendering it apt for cross-modal retrieval tasks. Notably, Polysemous Visual-Semantic Embedding (PVSE)\cite{PVSE} combines self-attention mechanisms with residual learning to create an enhanced Transformer architecture. Probabilistic Cross-Modal Embedding (PCME)\cite{PCME} employs probabilistic embeddings to adeptly model one-to-many correlations within image-text pairs, thereby delving deeper into the many-to-many associations inherent in multi-modal samples. Taking strides to refine multi-modal feature representation, Dual-level Representation Enhancement Network (DREN)\cite{DREN} introduces character-level augmentation for each instance and context-driven enhancement across instances. Subsequently, Token-Guided Dual Transformer (TGDT)\cite{TGDT} integrates coarse- and fine-grained representation learning into a unified framework, leveraging a multi-modal contrastive loss to align image and text features effectively. Hierarchical RElation Modeling (HREM)\cite{HREM} enhances image-text matching by capturing both intra- and inter-modal relationships at multiple levels, improving holistic embeddings and retrieval efficiency. Recently, TransTPS\cite{TransTPS} extends the Transformer architecture by incorporating cross-modal multi-granularity matching and contrastive loss with weakly positive pairs to improve feature distinction.

In conclusion, Transformer-based methods significantly advance cross-modal retrieval by improving Transformer architectures through techniques such as residual learning, multi-level alignment, and multi-grained integration.

\begin{table}[!b]
    \renewcommand\arraystretch{1.12}
	\centering
	\caption{The details of VLP model methods for image-text matching. The first six models \cite{Oscar, Uniter, Unicoder, SOHO, ALIGN, CLIP, Wukong} are themselves VLP models, so we do not document their backbones.}
	\label{Table:VLP}
    \resizebox{\linewidth}{!}{
	\begin{tabular}{|c|c|c|c|c|c|}
		\hline
		Model&Source&Stream&Main structure & Backbone & Core technology \\
		\hline
		\hline
		Oscar\cite{Oscar} & ECCV20 & Single & Transformer &---& Mask mechanism \\
		Uniter\cite{Uniter} & ECCV20  & Single & Transformer &---&Mask mechanism \\
		Unicoder\cite{Unicoder} & AAAI20  & Single & Transformer &---& Mask mechanism \\
        SOHO\cite{SOHO} & CVPR21 & Single & CNN, Transformer &---& Mask mechanism \\
        ALIGN\cite{ALIGN} & ICML21  & Dual & CNN, Transformer &---& Contrastive learning \\
        CLIP\cite{CLIP} & ICML21  & Dual & CNN, Transformer &---& Contrastive learning  \\
		\hline
        TEAM\cite{TEAM}&MM22&Single&Transformer&ViT\cite{ViT}, BERT\cite{Bert}&Contrastive learning\\
        CSIC\cite{CSIC}&TCSVT23&Single&Transformer & Uniter\cite{Uniter} & Hard-negative learning \\
        LAPS\cite{LAPS}&CVPR24 & Single & Transformer & ViT\cite{ViT}, BERT\cite{Bert} & Hard-negative learning \\
        COTS\cite{COTS}&CVPR22 & Dual & Transformer & ViT\cite{ViT}, BERT\cite{Bert} & Contrastive learning \\
        EI-CLIP\cite{EI-CLIP} & CVPR22 & Dual & Transformer & CLIP\cite{CLIP} & Contrastive learning \\
        AGREE\cite{AGREE}& WSDM23 & Dual & Transformer & CLIP\cite{CLIP}, Wukong\cite{Wukong} & Contrastive learning \\
        IRRA\cite{IRRA} & CVPR23 & Dual & Transformer & CLIP\cite{CLIP} & Mask mechanism \\
        MAKE\cite{MAKE} & WWW23 & Dual & Transformer & ALIGN\cite{ALIGN} & Mask mechanism \\
        USER\cite{USER} & TIP24 & Dual & Transformer & CLIP\cite{CLIP} & Contrastive learning \\
		\hline
	\end{tabular}}
\end{table}

\begin{figure}[!t]
\centering
\hspace{-0.5mm}\includegraphics[width=0.46\textwidth]{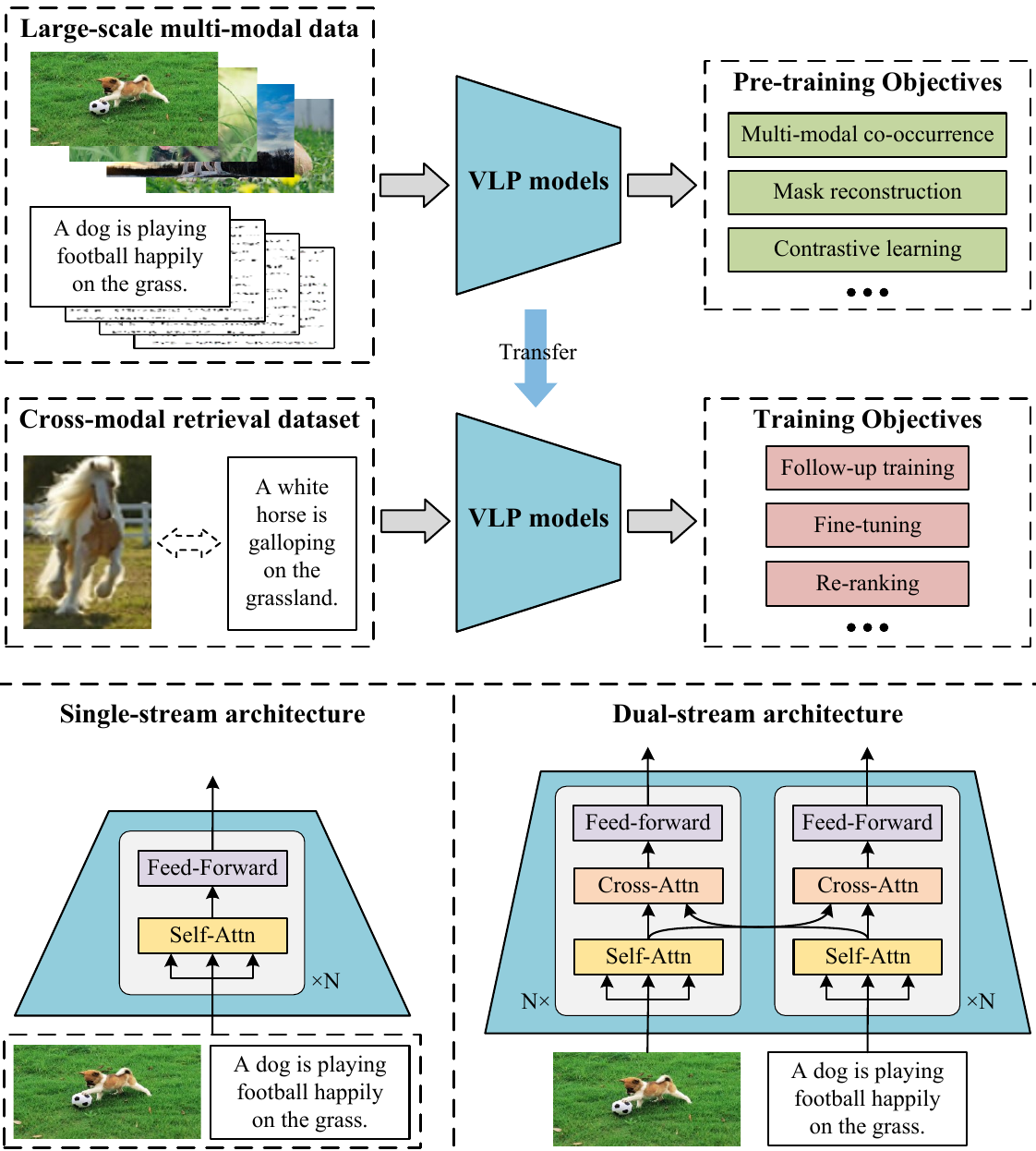}
\caption{The flow diagram of VLP model methods and its two architectures.}
\label{Figure:VLP}
\end{figure}

\subsubsection{VLP model methods}
Amidst rapid progress in artificial intelligence and computing hardware, Vision-Language Pre-training (VLP) models \cite{Oscar, Uniter, Unicoder, SOHO, ALIGN, CLIP, Wukong}, encompassing both single- and dual-stream architectures in Fig. \ref{Figure:VLP}, have harnessed expansive visual-linguistic datasets for joint pre-learning.

Researchers have harnessed the potent representational capabilities of VLP models to markedly augment cross-modal retrieval performance \cite{TEAM, CSIC, LAPS, COTS, AGREE, IRRA, USER, EI-CLIP, MAKE}, details are shown in Table \ref{Table:VLP}. As an instance, Token Embeddings AlignMent (TEAM)\cite{TEAM} trains a single-stream retrieval model on an expansive visual-language dataset they curated. This technique overtly aligns multi-modal token embeddings, culminating in the generation of token-level matching scores for input image-text pairs. In the realm of the Uniter-based backbone \cite{Uniter}, Cross-modal Semantic Importance Consistency (CSIC)\cite{CSIC} quantifies the semantic significance of both inter- and intra-modal instances. This is achieved while simultaneously realizing multi-modal data alignment, thereby fostering the acquisition of semantically cohesive image-text representations. Further, Linguistic-Aware Patch Slimming (LAPS)\cite{LAPS} framework enhances fine-grained cross-modal alignment by leveraging vision Transformers to align patch features with words, effectively addressing visual patch redundancy and ambiguity through linguistic supervision, thus improving semantic consistency in VLP models.

Departing from the previously mentioned single-stream architectures, COllaborative Two-Stream (COTS)\cite{COTS} embraces a dual-stream vision-language pre-training model. By combining instance-level contrastive learning with token- and task-level interaction, it enhances cross-modal collaboration for more effective image-text matching. Later, AliGn cRoss-modal EntitiEs (AGREE)\cite{AGREE} executes light-weight fine-tuning and re-ranking, thus harmonizing cross-modal entities atop VLP models. Advancing from the pre-trained CLIP\cite{CLIP}, Implicit Relation Reasoning and Aligning (IRRA)\cite{IRRA} designs a tailored text-specific mask mechanism that adeptly captures fine-grained relationships both within and across modalities. Also based on CLIP, USER\cite{USER} introduces a novel strategy to exploit and transfer knowledge from CLIP, thereby improving image-text retrieval by enriching the representation learning process. Furthermore, EI-CLIP\cite{EI-CLIP} and MAKE\cite{MAKE} leverage CLIP\cite{CLIP} and ALIGN\cite{ALIGN} as backbones, respectively. They merge contrastive learning and keyword enhancement mechanisms to empower E-commerce retrieval.

In summary, VLP models, whether single- or dual-stream, have become integral to cross-modal retrieval. Armed with the capacity to extrapolate representations gleaned from extensive datasets, these methods heighten cross-modal alignment and interaction in VLP models through strategies such as follow-up training, fine-tuning, re-ranking, and other avenues.

\subsubsection{CMG model methods}
In contrast to the methods previously discussed, which concentrate on refining model structures and optimization strategies to directly enhance object-oriented image-text matching performance, Cross-Modal Generation model (CMG model) methods\cite{DVSA, LSCO, X-MRS, 2WayNet, GXN, GRACE} represent a form of multi-task learning architecture. These methods incorporate cross-modal generation tasks into the retrieval learning process, indirectly optimizing cross-modal retrieval models by aligning generated cross-modal samples with retrieved results. Consequently, such methods can build upon existing object-oriented image-text matching methods, enhancing their representation capabilities through multi-task learning.

In these specific studies, Deep Visual-Semantic Alignment (DVSA)\cite{DVSA} optimizes image-text matching by predicting textual captions for image regions. Learning Semantic Concepts and Order (LSCO)\cite{LSCO} refines this with a multi-regional multi-label convolutional network for better text prediction. Leveraging text-to-image generation advances, Cross-Modal Retrieval and Synthesis (X-MRS)\cite{X-MRS} boosts image-text matching by generating images through conditional generation. Additionally, bi-directional cross-modal generation has been explored: 2WayNet\cite{2WayNet} employs two tied pipelines for bi-directional feature-level reconstruction, while Generative Cross-modal learning Network (GXN)\cite{GXN} integrates GAN-based bi-directional generation into cross-modal embedding learning. A recent study, GeneRAtive Cross-modal rEtrieval (GRACE)\cite{GRACE}, leverages large-model cross-modal generation techniques, embedding visual memory into multi-modal Large Language Models (LLMs) to transform image-text matching into an efficient generative task, eliminating the need for negative samples or retrieval indices.

\begin{figure*}[!b]
\centering
    \begin{overpic}[width=1\textwidth]{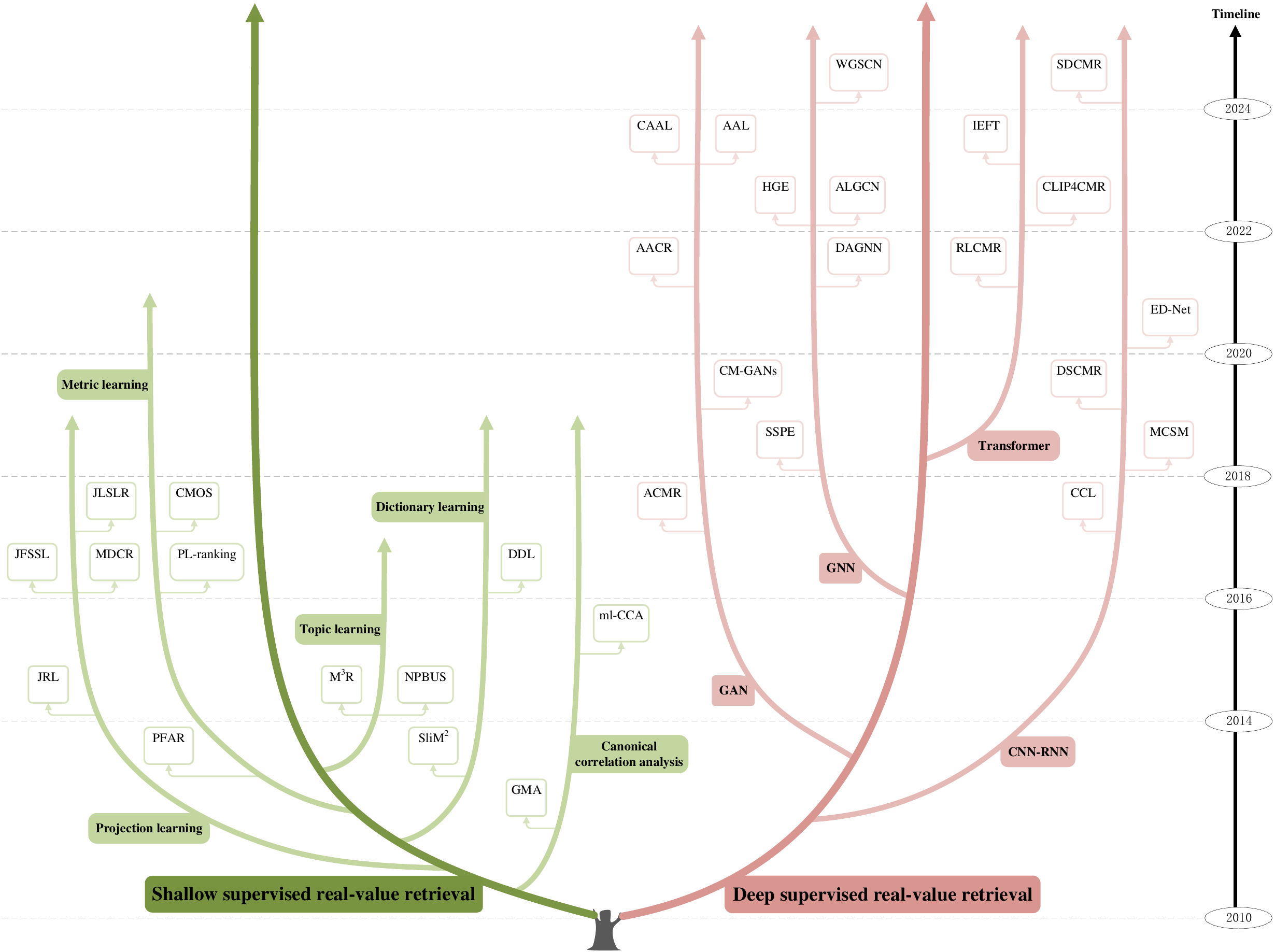}
        \put(40.05,11.2){\fontsize{6pt}{9pt}\selectfont \cite{GMA}}
        \put(12,15.3){\fontsize{6pt}{9pt}\selectfont \cite{PFAR}}
        \put(32.75,15.3){\fontsize{6pt}{9pt}\selectfont \cite{SliM2}}
        \put(2.5,20.1){\fontsize{6pt}{9pt}\selectfont \cite{JRL}}
        \put(25.55,20.1){\fontsize{6pt}{9pt}\selectfont \cite{M3R}}
        \put(32.15,20.1){\fontsize{6pt}{9pt}\selectfont \cite{NPBUS}}
        \put(47.48,24.9){\fontsize{6pt}{9pt}\selectfont \cite{ml-CCA}}
        \put(1.2,29.7){\fontsize{6pt}{9pt}\selectfont \cite{JFSSL}}
        \put(7.75,29.7){\fontsize{6pt}{9pt}\selectfont \cite{MDCR}}
        \put(14.95,29.7){\fontsize{6pt}{9pt}\selectfont \cite{PL-ranking}}
        \put(39.65,29.7){\fontsize{6pt}{9pt}\selectfont \cite{DDL}}
        \put(7.5,34.5){\fontsize{6pt}{9pt}\selectfont \cite{JLSLR}}
        \put(14.,34.5){\fontsize{6pt}{9pt}\selectfont \cite{CMOS1}}
        \put(50.7,34.5){\fontsize{6pt}{9pt}\selectfont \cite{ACMR}}
        \put(83.72,34.5){\fontsize{6pt}{9pt}\selectfont \cite{CCL}}
        \put(90.5,39.3){\fontsize{6pt}{9pt}\selectfont \cite{MCSM}}
        \put(59.90,39.3){\fontsize{6pt}{9pt}\selectfont \cite{SSPE}}
        \put(57.40,44.1){\fontsize{6pt}{9pt}\selectfont \cite{CM-GANs}}
        \put(83.40,44.1){\fontsize{6pt}{9pt}\selectfont \cite{DSCMR}}
        \put(90.5,48.9){\fontsize{6pt}{9pt}\selectfont \cite{ED-Net}}
        \put(75.2,53.68){\fontsize{6pt}{9pt}\selectfont \cite{RLCMR}}
        \put(66.15,53.68){\fontsize{6pt}{9pt}\selectfont \cite{DAGNN}}
        \put(50.05,53.68){\fontsize{6pt}{9pt}\selectfont \cite{AACR}}
        \put(59.6,58.48){\fontsize{6pt}{9pt}\selectfont \cite{ALGCN}}
        \put(66.0,58.48){\fontsize{6pt}{9pt}\selectfont \cite{HGE}}
        \put(82.7,58.48){\fontsize{6pt}{9pt}\selectfont \cite{CLIP4CMR}}
        \put(50.1,63.28){\fontsize{6pt}{9pt}\selectfont \cite{CAAL}}
        \put(56.5,63.28){\fontsize{6pt}{9pt}\selectfont \cite{AAL}}
        \put(75.83,63.28){\fontsize{6pt}{9pt}\selectfont \cite{IEFT}}
        \put(66.1,68.08){\fontsize{6pt}{9pt}\selectfont \cite{WGSCN}}
        \put(83.4,68.08){\fontsize{6pt}{9pt}\selectfont \cite{SDCMR}}
    \end{overpic}
\vspace{-6.mm}
\caption{The evolutionary tree of representative supervised real-value retrieval methods.}
\label{Figure:SR}
\end{figure*}
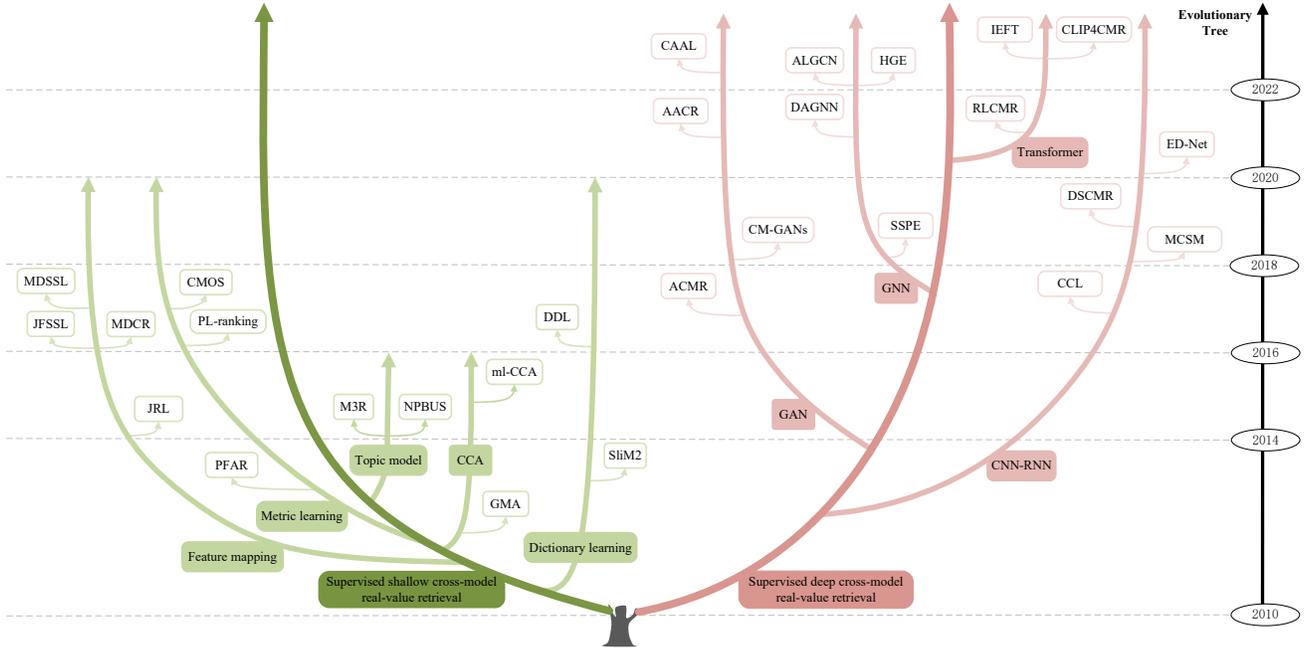

\subsection{Supervised Real-value Retrieval}
Supervised real-value retrieval, benefiting from manual annotation, explores semantic association and category discrimination in multi-modal data to achieve cross-modal retrieval. It involves two primary approaches: shallow and deep supervised real-value retrieval, based on different learning principles. Shallow methods use statistical analysis techniques such as CCA, dictionary learning, and projection learning to model multi-modal data associations. In contrast, deep methods utilize neural networks to capture complex semantic relationships. As shown in Fig. \ref{Figure:SR}, in shallow supervised real-value retrieval, methods are categorized into \emph{canonical correlation analysis methods}, \emph{dictionary learning methods}, \emph{projection learning methods}, \emph{topic learning methods}, and \emph{metric learning methods}. On the other hand, deep supervised real-value retrieval encompasses \emph{CNN-RNN methods}, \emph{GAN methods}, \emph{GNN methods}, and \emph{Transformer methods}.

\subsubsection{Canonical correlation analysis methods}
Early studies \cite{GMA, ml-CCA} implement supervised CCA to learn a latent space by maximizing the correlation between the projected features of different modalities. Formally, given the multi-modal feature matrices $V\in\mathbb{R}^{n_{tr}\times d_v}$ and $T\in\mathbb{R}^{n_{tr}\times d_t}$, supervised CCA learns their transformation matrices $W_v$ and $W_t$ under label supervision:
\begin{equation}
\small
\begin{aligned}
(W'_v, W'_t) &=\mathop{\arg\max}\limits_{W_v, W_t} \ \frac{W_v^{\top}\sum_{VT}W_t}{\sqrt{W_v^{\top}\sum_{VV}W_v} \sqrt{W_t^{\top}\sum_{TT}W_t}}.
\end{aligned}
\end{equation}
Similar to unsupervised CCA, here $W_v\in\mathbb{R}^{d_v\times d_c}$, $W_t\in\mathbb{R}^{d_t\times d_c}$, $\sum_{VV}\in\mathbb{R}^{d_v\times d_v}$, and $\sum_{TT}\in\mathbb{R}^{d_t\times d_t}$ are the transformation matrices and covariance matrices of $V$ and $T$, respectively. $n_{tr}$ is the number of training samples, $d_v$ and $d_t$ are the feature dimensions of images and texts respectively, and $d_c$ is the dimension of common representations. Particularly, $\sum_{VT}\in\mathbb{R}^{d_v\times d_t}$ is no longer the cross-covariance matrix but is instead equivalent to $V^{\top}LL^{\top}T$, where $L\in\mathbb{R}^{n_{tr}\times n_c}$ is the label matrix, $n_c$ is the number of multi-modal data categories. Based on this idea, Generalized Multi-view Analysis (GMA)\cite{GMA} solves a joint and relaxed quadratic constrained quadratic program on different feature spaces to obtain a non-linear subspace. Later, multi-label Canonical Correlation Analysis (ml-CCA) \cite{ml-CCA} extends CCA by learning shared subspaces and incorporating high-level semantics from multi-label annotations. Unlike traditional CCA, this method does not rely on explicit pairs between modalities but instead uses multi-label information to establish correspondences.

\subsubsection{Dictionary learning methods}
In early supervised retrieval studies, Dictionary Learning (DL) methods \cite{DDL, SliM2} have played a vital role. By employing dictionary learning, these methods effectively capture shared-modality structures and patterns in multi-modal data, resulting in a better understanding of inter-modal relationships. A notable example is Discriminative Dictionary Learning (DDL)\cite{DDL}, which concentrates on learning a discriminative dictionary capable of effectively representing each modality and mapping sparse codes to a unified label space. Formally, given the multi-modal feature matrices $V\in\mathbb{R}^{n_{tr}\times d_v}$ and $T\in\mathbb{R}^{n_{tr}\times d_t}$, it aims to find the sparse coefficients and dictionary matrices for each modality and align these coefficients in the label space:
\begin{equation}
\small
\begin{aligned}
&\min_{A_v,A_t,D_v,D_t,W_v,W_t}\|V-A_vD_v\|_F^2+\|T-A_tD_t\|_F^2\\ &\hspace{1cm}+\|C-A_vW_v^{\top}\|^2_F+\|C-A_tW_t^{\top}\|^2_F\\ &\hspace{1cm}+\alpha\|A_v\|_1+\beta\|A_t\|_1+\gamma(\|W_v\|^2_F+\|W_t\|^2_F),
\end{aligned}
\end{equation}
where $A_v\in\mathbb{R}^{n_{tr}\times n_k}$ and $D_v\in\mathbb{R}^{n_k\times d_v}$ are the coefficient and dictionary matrices of $V$, respectively, with analogous definitions for $A_t\in\mathbb{R}^{n_{tr}\times n_k}$ and $D_t\in\mathbb{R}^{n_k\times d_t}$. $C\in\mathbb{R}^{n_{tr}\times d_c}$ denotes a common representation derived from the label matrix $L\in\mathbb{R}^{n_{tr}\times n_c}$. Projection matrices for the sparse coefficients are $W_v\in\mathbb{R}^{d_c\times n_k}$ and $W_t\in\mathbb{R}^{d_c\times n_k}$. Here, $n_{tr}$, $n_{k}$, and $n_{c}$ are the number of training samples, dictionary atoms, and categories, respectively, while $d_v$, $d_t$, and $d_c$ are the feature dimensions of images, texts, and common representation. Finally, $\alpha$, $\beta$, $\gamma$ are trade-off hyper-parameters.

\subsubsection{Projection learning methods}
These methods \cite{JRL, JFSSL, MDCR, JLSLR} map multi-modal data to a shared low-dimensional subspace by learning projection matrices for each modality via linear regression, enabling cross-modal retrieval through direct comparisons. Typically, Joint Feature Selection and Subspace Learning (JFSSL)\cite{JFSSL} learns projection matrices $W_v$ and $W_t$ with label information, projecting multi-modal data $V$ and $T$ into a common space. During the learning process of projection matrices $W_v$ and $W_t$, an $l_{2,1}$-norm penalty is applied to realize feature selection on different feature spaces, and a multi-modal graph regularization term is also applied to preserve both inter- and intra-modality similarity. This idea can be formulated as
\begin{equation}
\small
\begin{aligned}
\min_{W_v,W_t}&\|VW_v-C\|_F^2+\|TW_t-C\|_F^2\\
+\alpha&(\|W_v\|_{2,1}+\|W_t\|_{2,1})+\beta\Omega(W_v,W_t),
\end{aligned}
\end{equation}
where $V\in\mathbb{R}^{n_{tr}\times d_v}$, $T\in\mathbb{R}^{n_{tr}\times d_t}$, $W_v\in\mathbb{R}^{d_v\times d_c}$, and $W_t\in\mathbb{R}^{d_t\times d_c}$. $C\in\mathbb{R}^{n_{tr}\times d_c}$ denotes a common latent representation, and $\Omega(\cdot)$ denotes a graph regularization function. $n_{tr}$ is the number of training samples, $d_v$, $d_t$, and $d_c$ are the feature dimensions of images, texts, and common representation, respectively, while $\alpha$ and $\beta$ are trade-off hyper-parameters. Modality-Dependent Cross-media Retrieval (MDCR)\cite{MDCR} improves upon common space mapping by learning two projection pairs for different retrieval tasks, mapping data into two distinct spaces.

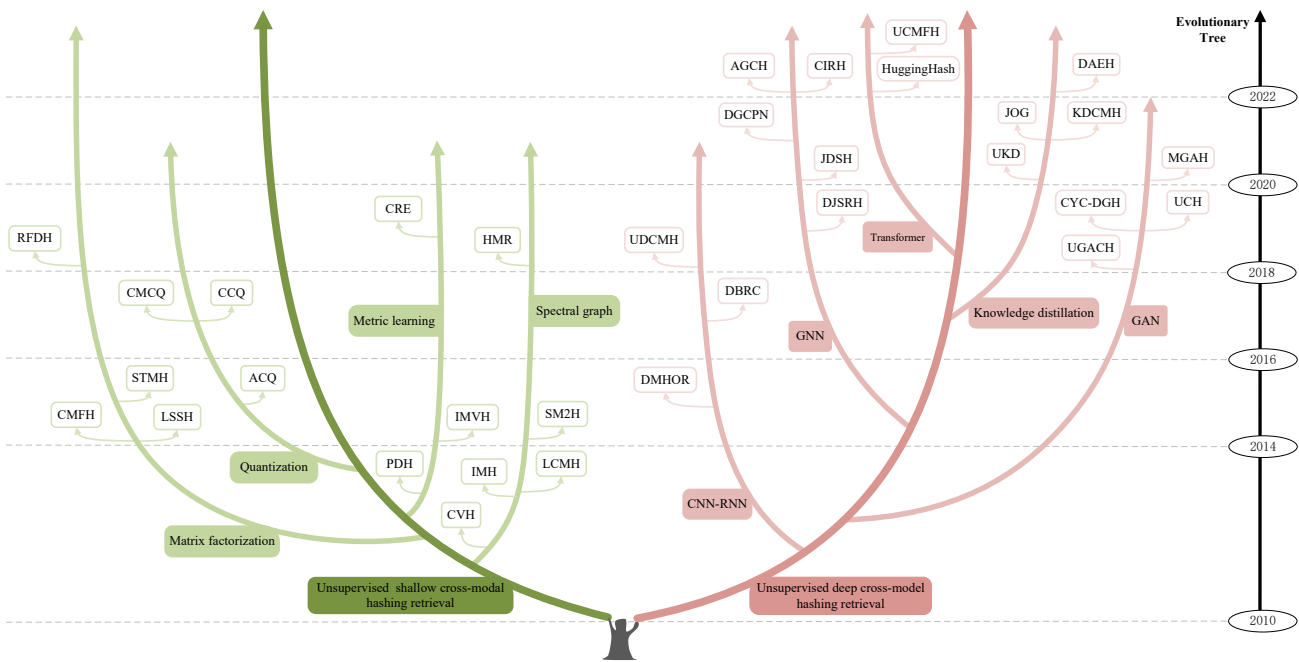
\begin{figure*}[!b]
\centering
    \begin{overpic}[width=1\textwidth]{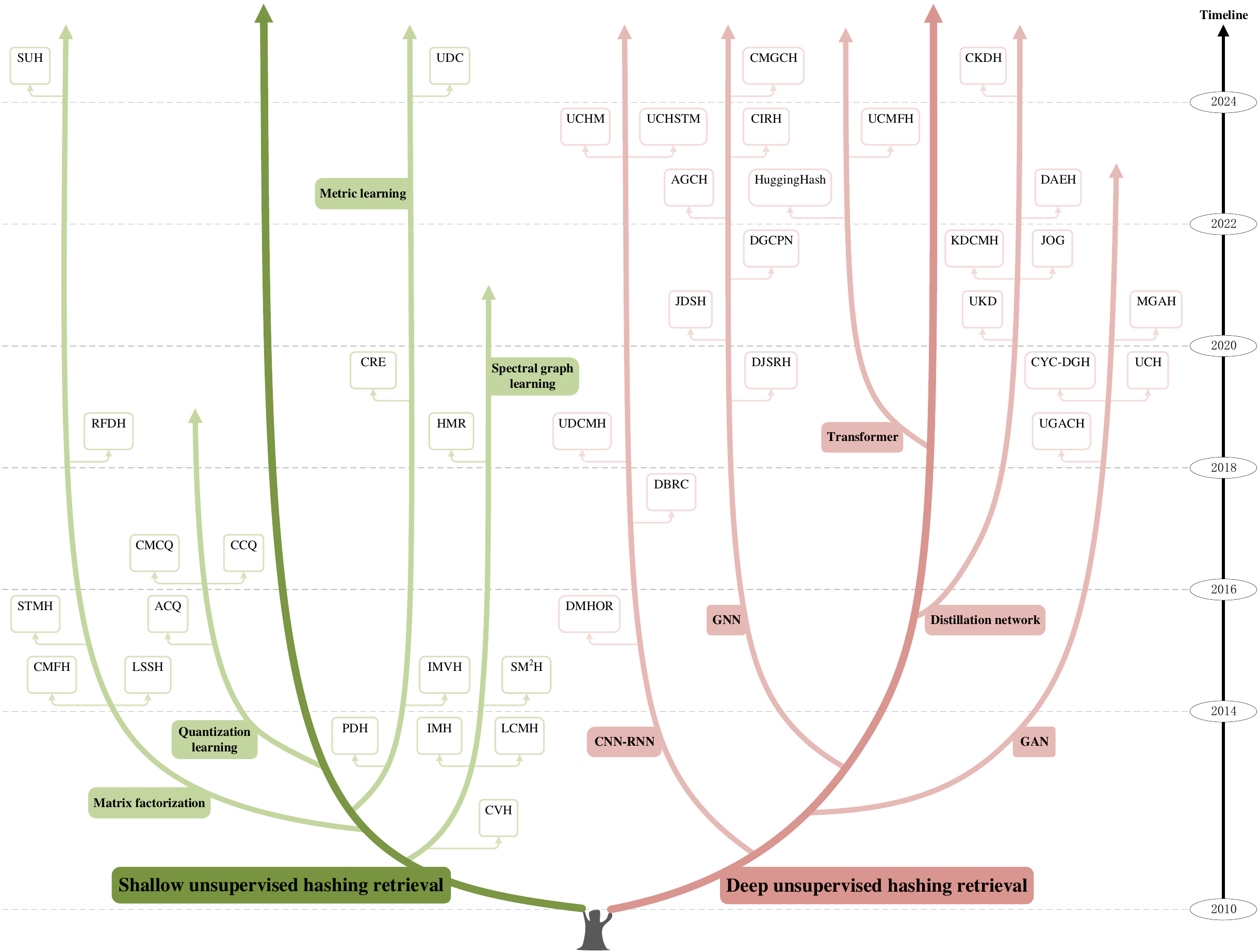}
        \put(1.26,25.95){\fontsize{6pt}{9pt}\selectfont \cite{STMH}}
        \put(11.8,25.95){\fontsize{6pt}{9pt}\selectfont \cite{ACQ}}
        \put(45.25,25.95){\fontsize{6pt}{9pt}\selectfont \cite{DMHOR}}
        \put(2.58,21.09){\fontsize{6pt}{9pt}\selectfont \cite{CMFH}}
        \put(10.18,21.09){\fontsize{6pt}{9pt}\selectfont \cite{LSSH}}
        \put(33.78,21.09){\fontsize{6pt}{9pt}\selectfont \cite{IMVH}}
        \put(40.25,21.09){\fontsize{6pt}{9pt}\selectfont \cite{SM2H}}
        \put(26.65,16.25){\fontsize{6pt}{9pt}\selectfont \cite{PDH}}
        \put(33.35,16.25){\fontsize{6pt}{9pt}\selectfont \cite{IMH}}
        \put(39.7,16.25){\fontsize{6pt}{9pt}\selectfont \cite{LCMH}}
        \put(38.64,9.73){\fontsize{6pt}{9pt}\selectfont \cite{CVH}}
        \put(10.7,30.77){\fontsize{6pt}{9pt}\selectfont \cite{CMCQ}}
        \put(17.8,30.77){\fontsize{6pt}{9pt}\selectfont \cite{CCQ}}
        \put(7.1,40.47){\fontsize{6pt}{9pt}\selectfont \cite{RFDH}}
        \put(34.3,40.47){\fontsize{6pt}{9pt}\selectfont \cite{HMR}}
        \put(44.65,40.47){\fontsize{6pt}{9pt}\selectfont \cite{UDCMH}}
        \put(82.7,40.47){\fontsize{6pt}{9pt}\selectfont \cite{UGACH}}
        \put(51.75,35.6){\fontsize{6pt}{9pt}\selectfont \cite{DBRC}}
        \put(28.1,45.3){\fontsize{6pt}{9pt}\selectfont \cite{CRE}}
        \put(59.65,45.3){\fontsize{6pt}{9pt}\selectfont \cite{DJSRH}}
        \put(82.55,45.3){\fontsize{6pt}{9pt}\selectfont \cite{CYC-DGH}}
        \put(89.6,45.3){\fontsize{6pt}{9pt}\selectfont \cite{UCH}}
        \put(90.25,50.1){\fontsize{6pt}{9pt}\selectfont \cite{MGAH}}
        \put(76.45,50.1){\fontsize{6pt}{9pt}\selectfont \cite{UKD}}
        \put(53.3,50.1){\fontsize{6pt}{9pt}\selectfont \cite{JDSH}}
        \put(82.,54.93){\fontsize{6pt}{9pt}\selectfont \cite{JOG}}
        \put(75.8,54.93){\fontsize{6pt}{9pt}\selectfont \cite{KDCMH}}
        \put(59.7,54.93){\fontsize{6pt}{9pt}\selectfont \cite{DGCPN}}
        \put(53.15,59.77){\fontsize{6pt}{9pt}\selectfont \cite{AGCH}}
        \put(82.48,59.77){\fontsize{6pt}{9pt}\selectfont \cite{DAEH}}
        \put(61.15,59.77){\fontsize{6pt}{9pt}\selectfont \cite{HuggingHash}}
        \put(44.95,64.57){\fontsize{6pt}{9pt}\selectfont \cite{UCHM}}
        \put(51.95,64.57){\fontsize{6pt}{9pt}\selectfont \cite{UCHSTM}}
        \put(59.32,64.57){\fontsize{6pt}{9pt}\selectfont \cite{CIRH}}
        \put(69.15,64.57){\fontsize{6pt}{9pt}\selectfont \cite{UCMFH}}
        \put(76.5,69.45){\fontsize{6pt}{9pt}\selectfont \cite{CKDH}}
        \put(59.85,69.45){\fontsize{6pt}{9pt}\selectfont \cite{CMGCH}}
        \put(.85,69.45){\fontsize{6pt}{9pt}\selectfont \cite{SUH}}
        \put(34.15,69.45){\fontsize{6pt}{9pt}\selectfont \cite{UDC}}
    \end{overpic}
\vspace{-8.5mm}
\caption{The evolutionary tree of representative unsupervised hashing retrieval methods.}
\label{Figure:UH}
\end{figure*}

\subsubsection{Topic learning methods}
Topic learning methods \cite{M3R, NPBUS} map features from various modalities into a unified semantic space by identifying latent topics or themes, enabling effective and meaningful data comparison. Multi-Modal Mutual topic Reinforce modeling (M$^3$R)\cite{M3R} constructs a joint cross-modal probabilistic graph model to explore mutually consistent topics through appropriate interactions between model factors. Non-Parametric Bayesian Upstream Supervised multi-modal topic model (NPBUS)\cite{NPBUS} utilizes a Gaussian process to capture the intra- and inter-modal correlation of multi-modal data. Additionally, it incorporates upstream supervised response variables shared by heterogeneous data into the normalized gamma representation of the hierarchical Dirichlet process, improving the method's prediction ability.

\subsubsection{Metric learning methods}
Metric learning methods \cite{PFAR, PL-ranking, CMOS1} aim to create a discriminative shared space by learning a linear projection matrix for features across different modalities, guided by label supervision. By optimizing pair-wise or triplet loss function, these methods encourage projected features from different modalities to be closer or farther apart based on their semantic similarity or dissimilarity. Typically, Parallel Field Alignment Retrieval (PFAR)\cite{PFAR} treats cross-modal retrieval as a manifold alignment problem, focusing on aligning parallel fields instead of directly fusing original or model output features. Pairwise-Listwise ranking (PL-ranking)\cite{PL-ranking} optimizes ranking through pair-wise ranking losses, reduces intra-neighbor variance, and maximizes inter-neighbor separability using class information.

\subsubsection{CNN-RNN methods}
CNN-RNN methods \cite{CCL, MCSM, DSCMR, ED-Net, SDCMR} use multi-layer perceptrons, convolutional neural networks, or recurrent neural networks to project multi-modal data into a unified retrieval space. These deep learning architectures capture complex relationships and patterns across different modalities effectively. For instance, Deep Supervised Cross-Modal Retrieval (DSCMR)\cite{DSCMR} uses modality-specific networks and fully connected layers to map multi-modal data into a shared representation space, learning discriminant features through label constraints and pair-wise loss with annotation supervision. Recently, Semantics Disentangling for Cross-Modal Retrieval (SDCMR)\cite{SDCMR} employs various CNNs to extract multi-modal features and a dual adversarial mechanism to isolate semantic-shared features from irrelevant ones, improving cross-modal retrieval by ensuring that only consistent semantic features contribute to the process.

\subsubsection{GAN methods}
GAN methods \cite{ACMR, CM-GANs, AACR, CAAL, AAL} leverage adversarial training\cite{GAN} to encourage the generation of more discriminative real-value features. These approaches explicitly consider modality and class information to improve alignment and separability in multi-modal data. Typically, Adversarial Cross-Modal Retrieval (ACMR)\cite{ACMR} is a pioneering approach that applies adversarial learning to cross-modal hashing, utilizing a feature projector and a modality classifier in a mini-max game with triplet constraints. The projector learns modality-invariant representation for multi-modal data, while the classifier distinguishes between them. Building on this, Adaptive Adversarial Learning (AAL)\cite{AAL} introduces an adaptive balance strategy that adjusts loss contributions based on data dispersion, thereby optimizing semantic and label utilization while preserving discriminability.

\subsubsection{GNN methods}
GNN methods \cite{SSPE, ALGCN, DAGNN, HGE, WGSCN} exploit graph topology to capture semantic information from neighboring nodes, improving multi-modal correlation understanding. Supervised approaches construct graphs using labels or integrate label information into sample features, allowing networks to harness both data relationships and label supervision. For instance, Semantic Structure Preserved Embedding (SSPE)\cite{SSPE} creates semantic graphs from label vectors, models multi-modal data with non-linear networks to preserve local graph structures, and reconstructs labels for global semantic retention. Conversely, Weighted Graph-structured Semantics Constraint Network (WGSCN)\cite{WGSCN} employs a GCN-based dual-branch encoder to generate semantic embeddings with modality-specific and shared correlations, boosting representation discriminability and modality invariance.

\subsubsection{Transformer methods}
Transformer methods \cite{RLCMR, CLIP4CMR, IEFT} draw on the Transformer architecture, utilizing multi-head attention to encode multi-modal relationships and optimize modality-specific encoders with supervised learning. Typically, Rethinking Label-wise Cross-Modal Retrieval (RLCMR)\cite{RLCMR} converts multi-modal data into individual tokens and combines them within a unified Transformer encoder, training the network to predict categories and capture semantic correlations across modalities. Interacting-Enhancing Feature Transformer (IEFT)\cite{IEFT} treats text-image pairs as a unified entity, modeling their intrinsic correlation and enhancing features to improve cross-modal retrieval accuracy.

\subsection{Unsupervised Hashing Retrieval}
Unsupervised hashing retrieval, akin to unsupervised real-value retrieval, leverages co-occurring multi-modal data (e.g., text-image pairs) to capture semantic correlations. As illustrated in Fig. \ref{Figure:UH}, it is divided into shallow and deep types based on their principles. Shallow methods primarily use matrix factorization or spectral graphs to uncover associations, while deep methods employ various neural networks. Further, shallow methods encompasses \emph{matrix factorization methods}, \emph{spectral graph learning methods}, \emph{metric learning methods}, and \emph{quantization learning methods}. In contrast, deep methods encompass \emph{CNN-RNN methods}, \emph{GAN methods}, \emph{GNN methods}, \emph{Transformer methods}, and \emph{distillation network methods}.

\subsubsection{Matrix factorization methods}
Early unsupervised hashing often uses matrix factorization\cite{CMFH, LSSH, STMH, RFDH, SUH} to derive hash codes from multi-modal features. One such method, Collective Matrix Factorization Hashing (CMFH)\cite{CMFH} decomposes multi-modal data $V\in\mathbb{R}^{n_{tr}\times d_v}$ and $T\in\mathbb{R}^{n_{tr}\times d_t}$ into common features and modality-specific factors, thus generating unified hash codes. The objective function is defined as
\begin{equation}
\small
\begin{aligned}
\min_{U_v,U_t,W_v,W_t,C}&\alpha(\|V-CU_v\|^2_F)+(1-\alpha)\|T-CU_t\|^2_F\\
+&\beta(\|C-VW_v\|^2_F+\|C-TW_t\|_F^2)\\
+&\gamma R(U_v,U_t,W_v,W_t,C),
\end{aligned}
\end{equation}
where $C\in\mathbb{R}^{n_{tr}\times d_c}$ is the common latent representation, $U_v\in\mathbb{R}^{d_c\times d_v}$ and $U_t\in\mathbb{R}^{d_c\times d_t}$ are the factor matrices of $V$ and $T$ respectively, $W_v\in\mathbb{R}^{d_{v}\times d_c}$ and $W_t\in\mathbb{R}^{d_{t}\times d_c}$ are the projection matrices, and $R(\cdot)=\|\cdot\|^2_F$ is a regularization term to prevent over-fitting. Here, $n_{tr}$ is the number of training samples, and $d_v$, $d_t$, and $d_c$ denote the feature dimensions of images, texts, and common representation respectively, with $\alpha$, $\beta$, and $\gamma$ as trade-off hyper-parameters. Another, Latent Semantic Sparse Hashing (LSSH)\cite{LSSH} maps texts and images into isomorphic latent semantic spaces via sparse coding and matrix factorization, then projects these spaces onto a joint high-level abstraction space to learn hash codes.

\subsubsection{Spectral graph learning methods}
Many unsupervised cross-modal hashing approaches \cite{CVH, LCMH, SM2H, IMH, HMR} use graphs to model cross-modal semantic correlations and facilitate hash learning. In these methods, multi-modal samples are depicted as vertices, with their relationships as edges. Typically, separate neighbor graphs are constructed for each modality, and similarity is evaluated within and between modalities using Laplacian operators. The core can be formalized as
\begin{equation}
\small
\begin{aligned}
\mathop{\arg\min}\limits_{W}\frac{1}{2}{\rm trace}(WX^{\top}L_{lap}XW^{\top}),
\end{aligned}
\end{equation}
where $X^{\top}=\left(\begin{array}{cc} V^{\top}&0\\0&T^{\top}\end{array}\right)$ is a hybrid feature matrix with visual data $V\in\mathbb{R}^{n_{tr}\times d_v}$ and text data $T\in\mathbb{R}^{n_{tr}\times d_t}$. The hybrid projection matrix $W=(W_v^{\top}, W_t^{\top})$ consists of modality-specific projections $W_v\in\mathbb{R}^{d_v\times d_c}$ and $W_t\in\mathbb{R}^{d_t\times d_c}$. The Laplacian matrix $L_{lap} = D - S$ is defined with degree matrix $D\in\mathbb{R}^{2n_{tr}\times 2n_{tr}}$ and similarity matrix $S\in\mathbb{R}^{2n_{tr}\times 2n_{tr}}$. Here, $n_{tr}$ is the number of training samples, and $d_v$, $d_t$, and $d_c$ are the feature dimensions for images, texts, and the common representation, respectively. Typically, Linear Cross-Modal Hashing (LCMH)\cite{LCMH} uses an anchor graph to preserve intra-modal similarity by clustering the training set into $k$ clusters and representing each data point by its distances to these cluster centroids. Sparse Multi-Modal Hashing (SM$^2$H)\cite{SM2H} employs a hyper-graph to capture both intra- and inter-modal similarities, using hyper-graph Laplacian sparse coding to learn multi-modal dictionaries and obtain sparse codes.

\subsubsection{Metric learning methods}
Metric learning methods for unsupervised hashing retrieval\cite{PDH, IMVH, CRE, UDC} focus on learning similarity metrics that preserve relationships between multi-modal data, facilitating the generation of meaningful hash codes. Typically, Predictable Dual-view Hashing (PDH)\cite{PDH} uses a dual-view mapping algorithm to learn hash codes from two distinct views, ensuring that similar samples receive identical binary codes. Iterative Multi-View Hashing (IMVH)\cite{IMVH} integrates intra-view similarity and inter-view correlation, using an iterative scheme for efficient joint optimization of distance measures. One notable recent work, Unsupervised Dual hashing Coding (UDC)\cite{UDC} encodes multi-modal data into dual hashing codes by aligning semantic properties and correlated content across modalities, improving cross-modal retrieval efficiency through content similarity in the tail code and semantic guidance from the head code.

\subsubsection{Quantization learning methods}
Quantization learning methods \cite{ACQ, CMCQ, CCQ} are essential for cross-modal retrieval because they directly learn binary hash codes, setting them apart from earlier hashing techniques. For example, Alternating Co-Quantization (ACQ)\cite{ACQ} iteratively seeks binary quantizers for each modality while minimizing the distances between inter-modal data pairs, their corresponding binary hash codes, and the vertices of the binary hypercube. Similarly, Composite Correlation Quantization (CCQ)\cite{CCQ} simultaneously identifies transformations that map different modalities into a correlated maximum mapping with an isomorphic potential space. The resulting composite quantizer then translates these isomorphic potential features into compact binary codes.

\subsubsection{CNN-RNN methods}
Early deep unsupervised hashing approaches \cite{DBRC, UDCMH, DMHOR, UCHM, UCHSTM} combine deep network structures with hash learning techniques in a straightforward manner. These approaches contribute significantly to the field by leveraging deep learning architectures and innovative regularization methods to generate effective hash codes from multi-modal data. Typically, Deep Binary ReConstruction (DBRC)\cite{DBRC} introduces the ATanh activation function, which adaptively binarizes the joint representation from two modality-specific networks into a hashing layer. The deep binary reconstruction network is trained to minimize reconstruction error based on the shared binary representation. Unsupervised Deep Cross-Modal Hashing (UDCMH)\cite{UDCMH} integrates deep learning and matrix factorization with binary latent factor models. By applying a Laplacian constraint to each modality and using the constructed graphs in the binary code learning process, UDCMH preserves the neighbor structures of the original data.
In recent CNN-RNN methods, Unsupervised Cross-modal Hashing with Modality-interaction (UCHM)\cite{UCHM} improves retrieval performance by generating a modality-interaction-enabled similarity matrix and high-quality unified binary codes through interactions among continuous codes from different modalities. Meanwhile, Unsupervised Cross-modal Hashing via Semantic Text Mining (UCHSTM)\cite{UCHSTM} enhances the construction of instance similarity matrices by mining correlations between text data, thereby providing more accurate guidance for training the hashing networks.

\subsubsection{GAN methods}
GAN methods \cite{UGACH, UCH, CYC-DGH, MGAH} reconstruct data in one modality using information from another modality, thereby enhancing the semantic association between different modalities. By incorporating the discriminator, these methods improve the category information of multi-modal features and the realism of reconstructed data, leading to the generation of robust hash codes. For example, Unsupervised Generative Adversarial Cross-modal Hashing (UGACH)\cite{UGACH} utilizes GANs to exploit the underlying manifold structure of multi-modal data. Its generator selects informative data from one modality to challenge the discriminator, which learns to differentiate between the generated data and true positive data sampled from the correlation graph. Similarly, Unsupervised coupled Cycle generative adversarial Hashing (UCH)\cite{UCH} utilizes a unified learning framework with two-cycle GANs. The outer-cycle network learns powerful common representations, and the inner-cycle network generates reliable hash codes.

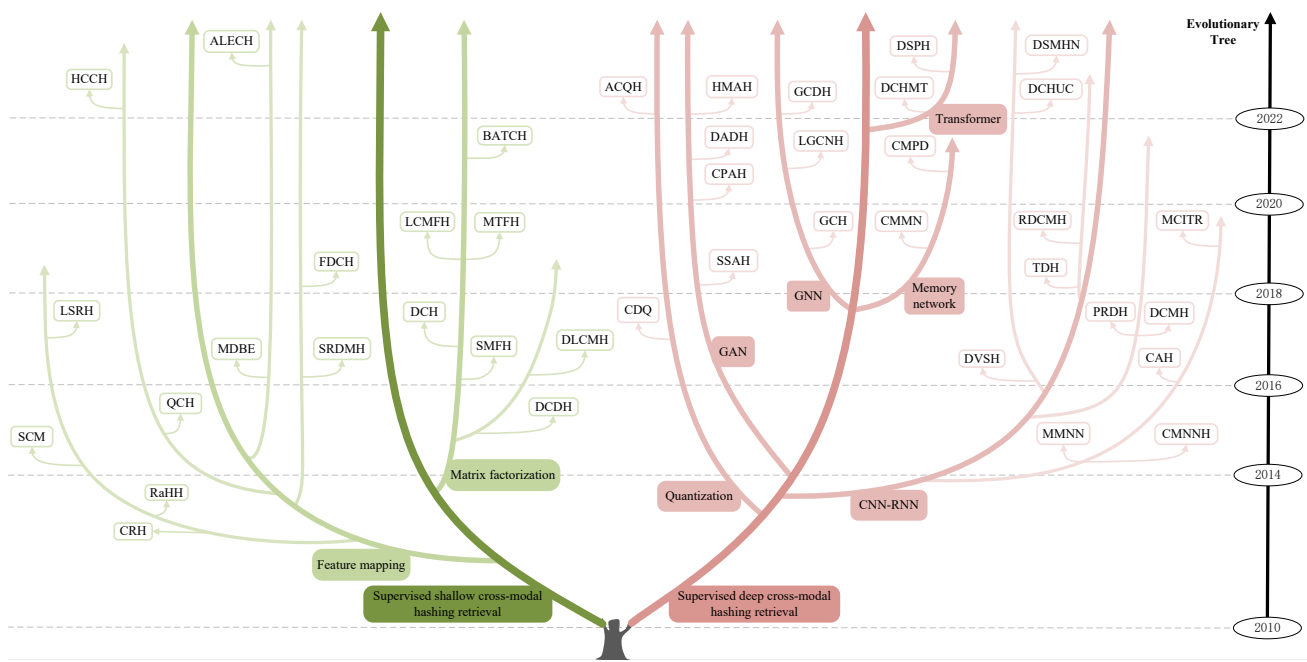
\begin{figure*}[!b]
\centering
    \begin{overpic}[width=\textwidth]{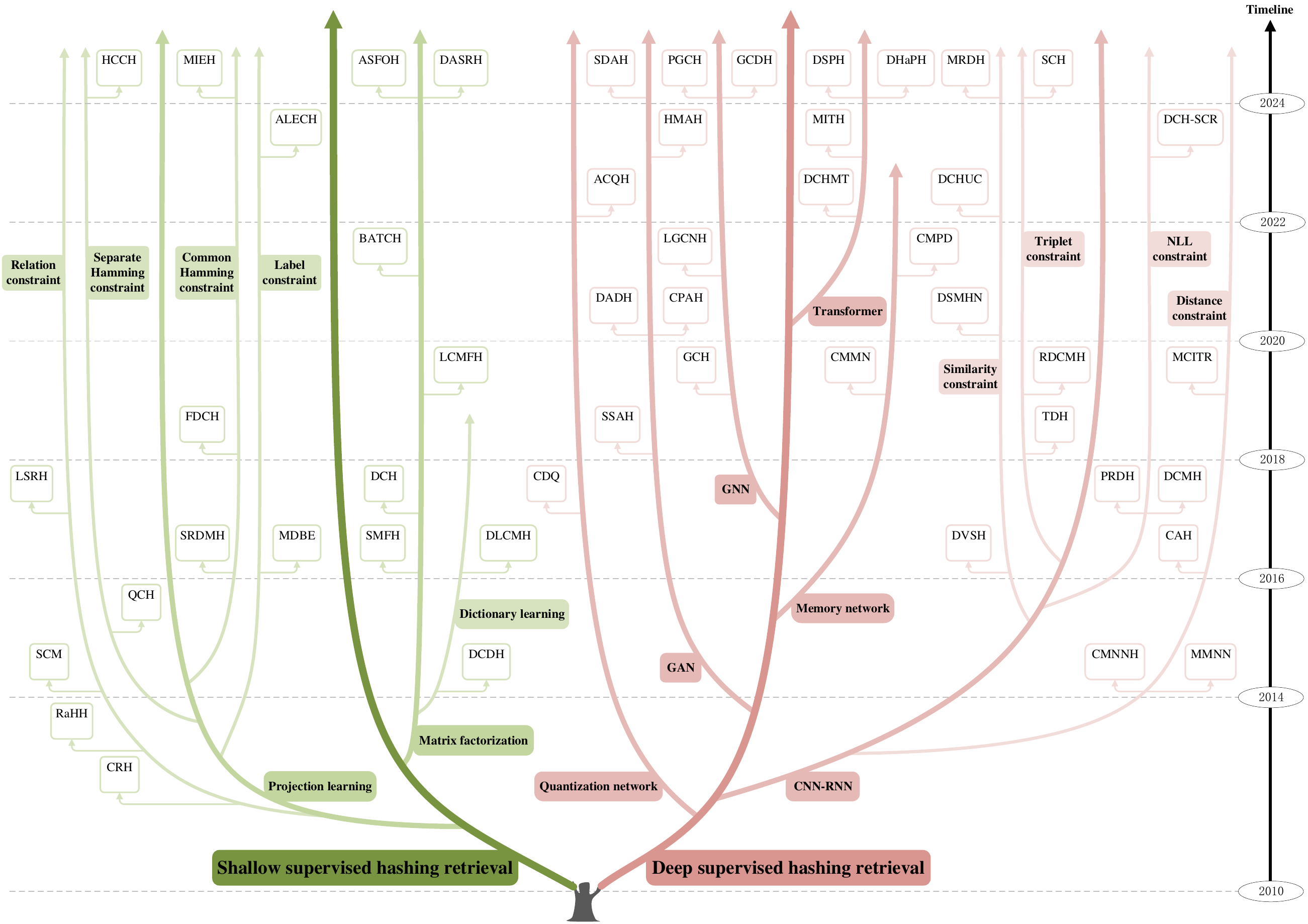}
        \put(2.21,19.25){\fontsize{6pt}{9pt}\selectfont \cite{SCM}}
        \put(35.65,19.25){\fontsize{6pt}{9pt}\selectfont \cite{DCDH}}
        \put(83.71,19.25){\fontsize{6pt}{9pt}\selectfont \cite{CMNNH}}
        \put(91.61,19.25){\fontsize{6pt}{9pt}\selectfont \cite{MMNN}}
        \put(3.97,14.65){\fontsize{6pt}{9pt}\selectfont \cite{RaHH}}
        \put(7.61,10.6){\fontsize{6pt}{9pt}\selectfont \cite{CRH}}
        \put(9.27,23.75){\fontsize{6pt}{9pt}\selectfont \cite{QCH}}
        \put(13.95,28.3){\fontsize{6pt}{9pt}\selectfont \cite{SRDMH1}}
        \put(21.17,28.3){\fontsize{6pt}{9pt}\selectfont \cite{MDBE}}
        \put(27.72,28.3){\fontsize{6pt}{9pt}\selectfont \cite{SMFH1}}
        \put(37.32,28.3){\fontsize{6pt}{9pt}\selectfont \cite{DLCMH}}
        \put(72.54,28.3){\fontsize{6pt}{9pt}\selectfont \cite{DVSH}}
        \put(88.56,28.3){\fontsize{6pt}{9pt}\selectfont \cite{CAH}}
        \put(88.9,32.82){\fontsize{6pt}{9pt}\selectfont \cite{DCMH}}
        \put(83.9,32.82){\fontsize{6pt}{9pt}\selectfont \cite{PRDH}}
        \put(40.24,32.82){\fontsize{6pt}{9pt}\selectfont \cite{CDQ}}
        \put(27.8,32.82){\fontsize{6pt}{9pt}\selectfont \cite{DCH}}
        \put(.88,32.82){\fontsize{6pt}{9pt}\selectfont \cite{LSRH}}
        \put(13.92,37.4){\fontsize{6pt}{9pt}\selectfont \cite{FDCH}}
        \put(79.12,37.4){\fontsize{6pt}{9pt}\selectfont \cite{TDH}}
        \put(89.62,41.9){\fontsize{6pt}{9pt}\selectfont \cite{MCITR}}
        \put(79.7,41.9){\fontsize{6pt}{9pt}\selectfont \cite{RDCMH}}
        \put(63.5,41.9){\fontsize{6pt}{9pt}\selectfont \cite{CMMN}}
        \put(51.72,41.9){\fontsize{6pt}{9pt}\selectfont \cite{GCH}}
        \put(33.72,41.9){\fontsize{6pt}{9pt}\selectfont \cite{LCMFH}}
        \put(45.68,37.4){\fontsize{6pt}{9pt}\selectfont \cite{SSAH}}
        \put(45.4,46.45){\fontsize{6pt}{9pt}\selectfont \cite{DADH}}
        \put(50.9,46.45){\fontsize{6pt}{9pt}\selectfont \cite{CPAH}}
        \put(71.85,46.45){\fontsize{6pt}{9pt}\selectfont \cite{DSMHN}}
        \put(69.9,51.0){\fontsize{6pt}{9pt}\selectfont \cite{CMPD}}
        \put(50.85,51.0){\fontsize{6pt}{9pt}\selectfont \cite{LGCNH}}
        \put(27.54,51.0){\fontsize{6pt}{9pt}\selectfont \cite{BATCH}}
        \put(71.84,55.55){\fontsize{6pt}{9pt}\selectfont \cite{DCHUC}}
        \put(61.6,55.55){\fontsize{6pt}{9pt}\selectfont \cite{DCHMT}}
        \put(45.2,55.55){\fontsize{6pt}{9pt}\selectfont \cite{ACQH}}
        \put(7.56,64.62){\fontsize{6pt}{9pt}\selectfont \cite{HCCH}}
        \put(21.1,60.08){\fontsize{6pt}{9pt}\selectfont \cite{ALECH}}
        \put(50.66,60.08){\fontsize{6pt}{9pt}\selectfont \cite{HMAH}}
        \put(61.81,60.08){\fontsize{6pt}{9pt}\selectfont \cite{MITH}}
        \put(89.5,60.08){\fontsize{6pt}{9pt}\selectfont \cite{DCH-SCR}}
        \put(79.01,64.62){\fontsize{6pt}{9pt}\selectfont \cite{SCH}}
        \put(72.31,64.62){\fontsize{6pt}{9pt}\selectfont \cite{MRDH}}
        \put(67.71,64.62){\fontsize{6pt}{9pt}\selectfont \cite{DHaPH}}
        \put(61.81,64.62){\fontsize{6pt}{9pt}\selectfont \cite{DSPH}}
        \put(56.15,64.62){\fontsize{6pt}{9pt}\selectfont \cite{GCDH}}
        \put(50.81,64.62){\fontsize{6pt}{9pt}\selectfont \cite{PGCH}}
        \put(45.2,64.62){\fontsize{6pt}{9pt}\selectfont \cite{SDAH}}
        \put(33.74,64.62){\fontsize{6pt}{9pt}\selectfont \cite{DASRH}}
        \put(27.41,64.62){\fontsize{6pt}{9pt}\selectfont \cite{ASFOH}}
        \put(13.7,64.62){\fontsize{6pt}{9pt}\selectfont \cite{MIEH}}
    \end{overpic}
\vspace{-7mm}
\caption{The evolutionary tree of representative supervised hashing retrieval methods.}
\label{Figure:SH}
\end{figure*}

\subsubsection{GNN methods}
GNN methods \cite{DJSRH, JDSH, DGCPN, AGCH, CIRH, CMGCH} construct separate graphs for each modality and leverage GCNs/GATs to enhance consistency between graphs, effectively aligning multi-modal data. Typically, Deep Joint Semantics Reconstructing Hashing (DJSRH)\cite{DJSRH} creates a joint-semantics affinity graph to reveal latent relations in batch-wise instances. It then learns semantically relevant binary codes by minimizing the reconstruction error between the target adjacency matrix and the to-be-learned hash code structure. Similarly, Aggregation-based Graph Convolutional Hashing (AGCH)\cite{AGCH} proposes various similarity measures, combines them to create a refined affinity graph, and designs a GCN-based unsupervised hashing network to capture neighbor relevance. One of the recent studies, Cross-Modal Graph Contrastive Hashing (CMGCH)\cite{CMGCH} advances the field by employing a multi-modal knowledge graph to capture fine-grained semantic associations. It uses cross-modal contrastive attention and multi-granularity embedding learning to enhance the discriminativeness and consistency of multi-modal representation.

\subsubsection{Transformer methods}
In the realm of unsupervised hashing retrieval, Transformer-based methods\cite{HuggingHash, UCMFH} are still emerging. These approaches leverage the formidable descriptive capabilities inherent in Transformer architectures, particularly their self-attention mechanisms, to capture complex interdependencies across different modalities. For example, HuggingHash\cite{HuggingHash} represents a groundbreaking effort in Transformer-based unsupervised hashing retrieval. It features the direct alignment of hash codes originating from global tokens as well as intricate fine-grained alignment from content token sequences, aiming to integrate global and fine-grained consistencies to support comprehensive cross-modal learning. Building upon the Transformer architecture\cite{CLIP}, Unsupervised Contrastive Multi-modal Fusion Hashing (UCMFH)\cite{UCMFH} propounds a multi-modal fusion Transformer to capture intricate correlations spanning multiple modalities. Concurrently, UCMFH employs contrastive learning to produce hash codes that are both informative and compact.

\subsubsection{Distillation network methods}
Distillation network methods \cite{UKD, JOG, KDCMH, DAEH, CKDH} focus on training a teacher model in an unsupervised manner and subsequently leveraging it to form a similarity matrix that guides the learning process of a student model. Thus, these methods play a crucial role in unsupervised hashing by effectively transferring valuable information from teacher to student models. Typically, Unsupervised Knowledge Distillation (UKD)\cite{UKD} employs two existing methods\cite{UGACH, SSAH} as the teacher and student models, respectively, to learn hash codes by transferring similarity matrices through knowledge distillation. The JOint-teachinG (JOG)\cite{JOG} approach aims to build a high-performing yet lightweight retrieval model by using a pre-trained model as the teacher to guide the student model's hashing learning. It also incorporates an online knowledge distillation strategy to eliminate noise in knowledge and enhance cross-task knowledge quality. Leveraging the success of CLIP\cite{CLIP}, CLIP-based Knowledge Distillation Hashing (CKDH)\cite{CKDH} transfers comprehensive similarity measurements and semantic relevance from the pre-trained CLIP to a lightweight student model, significantly improving retrieval accuracy and scalability.

\subsection{Supervised Hashing Retrieval}
Supervised hashing retrieval, benefiting from manual annotation, effectively maps multi-modal data into a low-dimensional Hamming space for efficient search by exploiting category discrimination and semantic associations. It can be divided into shallow and deep supervised hashing retrieval, based on different principles. The former employs matrix factorization and projection learning to generate common hash codes for multi-modal data, while the latter leverages neural networks to capture data associations and generate hash codes end-to-end. Further, as shown in Fig. \ref{Figure:SH}, shallow supervised hashing retrieval can be subdivided into \emph{matrix factorization methods} and \emph{projection learning methods}, while deep supervised hashing retrieval encompasses \emph{CNN-RNN methods}, \emph{GAN methods}, \emph{GNN methods}, \emph{Transformer methods}, \emph{memory network methods}, and \emph{quantization network methods}.

\subsubsection{Matrix factorization methods}
Matrix factorization methods\cite{SMFH1, DCH, LCMFH, BATCH, DASRH, ASFOH} decompose multi-modal data matrices into factor matrices and a common representation. This process involves optimizing both modality-specific factor matrices and modality-shared hash codes to uncover potential associations. An example is Supervised Matrix Factorization Hashing (SMFH)\cite{SMFH1}, which advances the unsupervised CMFH\cite{CMFH} by incorporating label information into a Laplacian matrix, thus enhancing cross-modal hashing learning with a constraint term formalized as
\begin{equation}
\small
\begin{aligned}
&\min_{U_v,U_t,W_v,W_t,C}\alpha\|V-CU_v\|^2_F+(1-\alpha)\|T-CU_t\|^2_F\\
&\hspace{1cm}+\beta(\|C-VW_v\|^2_F+\|C-TW_t\|^2_F)\\
&\hspace{1cm}+\gamma R(U_v,U_t,W_v,W_t,C)+\lambda{\rm trace}(C^{\top}L_{lap}C),
\end{aligned}
\end{equation}
where $V\in\mathbb{R}^{n_{tr}\times d_v}$, $T\in\mathbb{R}^{n_{tr}\times d_t}$, and $C\in\mathbb{R}^{n_{tr}\times d_c}$ are image data, text data, and common latent representation, respectively. The factor matrices $U_v\in\mathbb{R}^{d_c\times d_v}$ and $U_t\in\mathbb{R}^{d_c\times d_t}$ correspond to $V$ and $T$, respectively, while $W_v\in\mathbb{R}^{d_{v}\times d_c}$ and $W_t\in\mathbb{R}^{d_{t}\times d_c}$ are the projection matrices. $R(\cdot)=\|\cdot\|^2_F$ is a regularization term to prevent over-fitting, and $L_{lap}$ is a Laplacian matrix based on label similarity. Here, $n_{tr}$ denotes the number of training samples, $d_v$, $d_t$, and $d_c$ represent the feature dimensions of images, texts, and common representation, respectively, while $\alpha$, $\beta$, $\gamma$, and $\lambda$ are trade-off hyper-parameters. Recent advancements in Asymmetric Supervised Fusion-Oriented Hashing (ASFOH)\cite{ASFOH} further advance the field by explicitly formulating the matrix factorization problem into a common representation and transformation matrix. This approach enhances the discrimination and compactness of hash codes through an asymmetric hash learning framework.

Similar to the principles of matrix factorization, some methods \cite{DCDH, DLCMH} leverage supervised dictionary learning for cross-modal hashing retrieval. For example, Discriminative Coupled Dictionary Hashing (DCDH)\cite{DCDH} constructs modality-specific coupled dictionaries and unified multi-modal hash codes based on side information, such as category labels. This approach achieves more efficient and accurate cross-modal retrieval by minimizing quantization loss. Given an image feature matrix $V\in\mathbb{R}^{n_{tr}\times d_v}$ and text feature matrix $T\in\mathbb{R}^{n_{tr}\times d_t}$, its core can be formalized as
\begin{equation}
\small
\begin{aligned}
\min_{D_v,D_t,A_v,A_t}&\alpha\|V-A_vD_v\|^2_F+\beta\|T-A_tD_t\|^2_F\\
+&\gamma(\|A_v\|_1+\|A_t\|_1),
\end{aligned}
\end{equation}
where $D_v\in\mathbb{R}^{n_k\times d_v}$, $D_t\in\mathbb{R}^{n_k\times d_t}$, $A_v\in\mathbb{R}^{n_{tr}\times n_k}$, and $A_t\in\mathbb{R}^{n_{tr}\times n_k}$ represent the dictionary and sparse coefficient matrices, respectively. Here, $n_{tr}$ and $n_{k}$ denote the numbers of training samples and dictionary atoms respectively, $d_v$ and $d_t$ denote the feature dimensions of images and texts respectively, and $\alpha$, $\beta$, and $\gamma$ are trade-off hyper-parameters. Building on the above, Dictionary Learning Cross-Modal Hashing (DLCMH)\cite{DLCMH} imposes direct constraints on the similarity between hash codes using a label similarity matrix, thus enhancing the discriminative capability of multi-modal hash codes.

\subsubsection{projection learning methods}
projection learning methods \cite{MDBE, ALECH, QCH, HCCH, SRDMH1, FDCH, MIEH, RaHH, CRH, SCM, LSRH} focus on projecting multi-modal data into a lower-dimensional Hamming space by learning suitable mapping matrices. These methods are categorized based on the constraints on the mapping matrices: label, separate Hamming, common Hamming, and relation constraints.

Label constraint methods \cite{MDBE, ALECH} project multi-modal data $V\in\mathbb{R}^{n_{tr}\times d_v}$ and $T\in\mathbb{R}^{n_{tr}\times d_t}$ into a label space, with their core principles being formalized as
\begin{equation}
\small
\begin{aligned}
&\min_{W_v,W_t,U_v,U_t}\alpha\|L-{\rm sign}(VW_v)U_v\|^2_F\\
&\hspace{0.8cm}+\beta\|L-{\rm sign}(TW_t)U_t\|^2_F,
\end{aligned}
\end{equation}
where $W_v\in\mathbb{R}^{d_{v}\times d_h}$ and $W_t\in\mathbb{R}^{d_{t}\times d_h}$ are Hamming mapping matrices for $V$ and $T$, respectively. $U_v\in\mathbb{R}^{d_{h}\times n_c}$ and $U_t\in\mathbb{R}^{d_{h}\times n_c}$ bridge the label space and Hamming space, with $L\in\mathbb{R}^{n_{tr}\times n_c}$ as the label matrix. Here, $n_{tr}$ and $n_{c}$ are the numbers of training samples and categories, $d_v$ and $d_t$ are the feature dimensions of images and texts, respectively, and $d_h$ is the length of hash codes. The parameters $\alpha$ and $\beta$ control trade-offs. Typically, Multi-modal Discriminative Binary Embedding (MDBE)\cite{MDBE} learns hash functions from a classification perspective, using label information to reveal shared structures in heterogeneous data and ensuring that hash codes capture discriminative semantics.

Separate Hamming constraint methods \cite{QCH, HCCH} operate by projecting visual data $V\in\mathbb{R}^{n_{tr}\times d_v}$ and text data $T\in\mathbb{R}^{n_{tr}\times d_t}$ into modality-specific feature spaces individually, followed by the imposition of correlation constraints on multiple Hamming spaces:
\begin{equation}
\small
\begin{aligned}
\min_{W_v,W_t,B_v,B_t}\alpha\|&B_v-VW_v\|^2_F+\beta\|B_t-TW_t\|^2_F\\
+&\gamma \mathcal{R}(B_v,B_t),
\end{aligned}
\end{equation}
where $W_v\in\mathbb{R}^{d_{v}\times d_h}$, $W_t\in\mathbb{R}^{d_{t}\times d_h}$, $B_v\in\mathbb{R}^{n_{tr}\times d_h}$, and $B_t\in\mathbb{R}^{n_{tr}\times d_h}$ are the Hamming mapping matrices and binary codes for $V$ and $T$, respectively. $\mathcal{R}(B_v,B_t)$ represents the correlation constraint term across multiple Hamming spaces, such as $\|B_v-B_t\|^2_F$. Here, $n_{tr}$ is the number of training samples, $d_v$ and $d_t$ are the feature dimensions of images and texts, $d_h$ is the hash code length, and $\alpha$, $\beta$, and $\gamma$ are trade-off hyper-parameters. Typically, Quantized Correlation Hashing (QCH)\cite{QCH} maps multi-modal data to separate Hamming spaces and then aligns them through cosine similarity constraints.

Common Hamming constraint methods \cite{SRDMH1, FDCH, MIEH} project multi-modal data $V\in\mathbb{R}^{n_{tr}\times d_v}$ and $T\in\mathbb{R}^{n_{tr}\times d_t}$ into a common Hamming space or an intermediate space, incorporating label information to improve hash code discriminability:
\begin{equation}
\small
\begin{aligned}
&\min_{W_v,W_t,W_c,W_b,C,B}\alpha\|C-VW_v\|^2_F+\beta\|C-TW_t\|^2_F\\
&\hspace{1cm}+\gamma\|B-CW_c\|^2_F+\lambda \|L-BW_b\|^2_F,
\end{aligned}
\end{equation}
where $W_v\in\mathbb{R}^{d_{v}\times d_c}$ and $W_t\in\mathbb{R}^{d_{t}\times d_c}$ are the Hamming mapping matrices for multi-modal data. $C\in\mathbb{R}^{n_{tr}\times d_c}$, $B\in\mathbb{R}^{n_{tr}\times d_h}$, and $L\in\mathbb{R}^{n_{tr}\times n_c}$ are the intermediate representation, binary codes, and label matrix, respectively. $W_c\in\mathbb{R}^{d_{c}\times d_h}$ and $W_b\in\mathbb{R}^{d_{h}\times n_c}$ are the mapping matrices between them. Here, $n_{tr}$ and $n_c$ are the numbers of training samples and categories, respectively, $d_v$, $d_t$, and $d_c$ are the feature dimensions of images, texts, and common representation, respectively, $d_h$ is the length of hash codes, and $\alpha$, $\beta$, $\gamma$, and $\lambda$ are trade-off hyper-parameters. Typically, Supervised Robust Discrete Multi-modal Hashing (SRDMH)\cite{SRDMH1} projects multi-modal data into a shared intermediate space and learns hash functions and binary codes with label information. A recent method, Multiple Information Embedded Hashing (MIEH)\cite{MIEH}, enhances projection learning by extracting comprehensive semantic information across modality, category, and instance levels before refining it into binary hash codes.

Relation constraint methods \cite{RaHH, CRH, SCM, LSRH} specialize in learning multi-modal mapping functions that maintain original neighbor and ranking relations within projected multi-modal data. For image data $V\in\mathbb{R}^{n_{tr}\times d_v}$ and text data $T\in\mathbb{R}^{n_{tr}\times d_t}$, these methods involve two relation constraints: the distance measure constraint \cite{RaHH, CRH} in Eq. (\ref{distance measure}) and the similarity approximation constraint \cite{SCM, LSRH} in Eq. (\ref{similarity approximation}):
\begin{equation}
\label{distance measure}
\small
\begin{aligned}
&\min_{W_v,W_t}\alpha\sum_{v_i\in V}\sum_{t_j\in T} \xi^{vt}_{ij}+\beta\sum_{v_i\in V}\sum_{v_j\in V}\xi^{vv}_{ij}\\
&\hspace{1cm}+\gamma\sum_{t_i\in T}\sum_{t_j\in T}\xi^{tt}_{ij},
\end{aligned}
\end{equation}
\begin{equation}
\label{similarity approximation}
\small
\begin{aligned}
\min_{W_v,W_t} \|{\rm sign}(VW_v){\rm sign}(TW_t)^{\top}-\lambda S\|^2_F.
\end{aligned}
\end{equation}
In the above expression, $\xi^{vt}_{ij}=s_{ij}\cdot\|{\rm sign}(v_i\cdot W_v)-{\rm sign}(t_j\cdot W_t)\|^2$, and similarly for $\xi^{vv}_{ij}$ and $\xi^{tt}_{ij}$, $v_i\in\mathbb{R}^{d_v}$ and $t_j\in\mathbb{R}^{d_t}$ are feature vectors for the $i$-th image and $j$-th text, respectively. $W_v\in\mathbb{R}^{d_{v}\times d_h}$ and $W_t\in\mathbb{R}^{d_{t}\times d_h}$ are Hamming mapping matrices, and $S=\{s_{ij}\}^{n_{tr}\times n_{tr}}$ is a similarity matrix based on category labels. The parameters $\alpha, \beta, \gamma, \lambda$ are trade-off hyper-parameters. For distance measurement, Relation-aware Heterogeneous Hashing (RaHH)\cite{RaHH} employs multi-modal distance constraints to construct Hamming spaces and learn optimal mappings. In contrast, for similarity approximation, Semantic Correlation Maximization (SCM)\cite{SCM} integrates label information and similarity relationships for training.

\subsubsection{CNN-RNN methods}
CNN-RNN methods \cite{MMNN, CAH, CMNNH, MCITR, DVSH, DCHUC, DSMHN, MRDH, DCMH, PRDH, DCH-SCR, TDH, RDCMH, SCH} in supervised hashing use fundamental neural networks like MLP, CNN, and RNN to model multi-modal data. As illustrated in Table \ref{Constraint comparisons}, these methods can be categorized based on various training constraints, including distance, similarity, negative log-likelihood, and triplet constraints.

\begin{table}[!t]
    \renewcommand\arraystretch{1.1}
	\centering
	\caption{Constraint comparisons of CNN-RNN methods.}
	\label{Constraint comparisons}
    \resizebox{\linewidth}{!}{
    \begin{threeparttable}
	\begin{tabular}{|c|c|c|}
		\hline
		Constraints&\multicolumn{2}{c|}{Core formulations\footnote{Note that this presentation only covers the core loss for each constraint. In the above context, $H_v$ and $H_t$ denote modality-specific hashing networks, each consisting of feature mapping networks ($F_v$ or $F_t$) and the sign function ${\rm sign}(\cdot)$. $U_v$ and $U_t$ represent the reconstructed matrices. $t^+_i$, $t^-_i$, $v^+_j$, and $v^-_j$ are the positive and negative samples of $v_i$ and $t_j$, respectively. $\alpha$, $\beta$, and $\sigma$ are the preset margins for the hinge loss and triplet loss.}}\\
		\hline \hline
		\multirow{2}{*}{Distance}&Approximation&$\sum_{(v_i,t_i)\in (V,T)}\|H_v(v_i)-H_t(t_i)\|^2_2$\\[2pt]
		\cline{2-3}
		&Reconstruction&$\sum_{(v_i,t_i)\in (V,T)}\|v_i-U_tH_t(t_i)\|^2_2+\|t_i-U_vH_v(v_i)\|^2_2$\\[2pt]
		\hline
		Similarity&\multicolumn{2}{c|}{$\sum_{v_i\in V}\sum_{t_j\in T}\max(0,\sigma-s_{ij}\frac{F_v(v_i)\cdot F_t(t_j)}{\|F_v(v_i)\|\|F_t(t_j)\|})$}\\[2pt]
		\hline
		NLL&\multicolumn{2}{c|}{\makecell{$-\sum_{v_i\in V}\sum_{t_j\in T}(s_{ij}\theta_{ij}-\log(1+e^{\theta_{ij}}))$, \ $\theta_{ij}=\frac{1}{2}H_v(v_i)^\top H_t(t_j)$}}\\[2pt]
		\hline
		Triplet&\multicolumn{2}{c|}{\makecell{$\sum_{v_i\in V}\max(0,\alpha+\|H_v(v_i)-H_t(t_i^+)\|^2-\|H_v(v_i)-H_t(t_i^-)\|^2)$\\+$\sum_{t_j\in T}\max(0,\beta+\|H_t(t_j)-H_v(v_j^+)\|^2-\|H_t(t_j)-H_v(v_j^-)\|^2)$}}\\[3pt]
		\hline
	\end{tabular}
\end{threeparttable}
}
\end{table}

Deep supervised hashing methods with distance constraints \cite{MMNN, CAH, CMNNH, MCITR} involve training MLP-based deep networks for cross-modal approximation or feature reconstruction. For instance, Multi-Modal Neural Network (MM-NN)\cite{MMNN} employs a coupled Siamese network to map multi-modal data into a shared space, effectively managing both intra- and inter-modal similarity measures. Regarding feature reconstruction, Correlation Auto-encoder Hashing (CAH)\cite{CAH} maximizes inter-category separation and minimizes intra-category variance through intermediate feature constraints.

Deep supervised hashing with similarity constraints \cite{DVSH, DCHUC, DSMHN, MRDH} ensures that the generated hash codes closely approximate the similarity represented by the original category labels. For instance, Deep Visual-Semantic Hashing (DVSH)\cite{DVSH} captures intrinsic cross-modal correspondences between visual data and natural language by aligning similarity relationships across modalities. Similarly, Multi-Relational Deep Hashing (MRDH)\cite{MRDH} enforces cross-modal pair-wise similarity consistency and employs a global similarity metric to align similar hash codes while separating dissimilar ones, enhancing the discriminative capability of the hash codes.

Moreover, certain methods \cite{DCMH, PRDH, DCH-SCR} employ Negative Log-Likelihood (NLL) loss with similarity matrices to supervise deep cross-modal hashing models. Deep Cross-Modal Hashing (DCMH)\cite{DCMH} is the first end-to-end cross-modal hash learning framework based on NLL, which simultaneously extracts features and learns hash codes while preserving cross-modal similarity. Based on this, Deep Cross-modal Hashing based on Semantic Consistent Ranking (DCH-SCR)\cite{DCH-SCR} utilizes NLL to supervise the learning of both feature- and label-level ranking relationships, ensuring the model captures multi-level semantics and narrows the modality gap, resulting in improved cross-modal retrieval performance.

Finally, triple-based methods \cite{TDH, RDCMH, SCH} construct triples by selecting positive and negative samples based on semantic labels, and then compute ranking loss to optimize deep networks. Triplet-based Deep Hashing (TDH)\cite{TDH} leverages triplet-label supervision to capture sample relationships and transfers semantic similarity from the original feature space to the learned Hamming space using graph regularization. More recently, Semantic Channel Hashing (SCH)\cite{SCH} uses triplet-based supervision to classify sample pairs into distinct semantic groups, applying tailored constraints to better exploit the Hamming space and address the space gap issue.

\subsubsection{GAN methods}
GAN-based methods\cite{SSAH, CPAH, DADH, HMAH, SDAH} enhance network mapping by integrating discriminative constraints such as modality and category discrimination during cross-modal hashing learning. For instance, Self-Supervised Adversarial Hashing (SSAH)\cite{SSAH} employs adversarial networks to minimize discrepancies between multi-modal semantics and annotation information, supported by a self-supervised semantic network for feature learning. Consistency-Preserving Adversarial Hashing (CPAH)\cite{CPAH} integrates multi-modal information decoupling with multi-task adversarial learning to improve cross-modal interaction. The recent Semantic Disentanglement Adversarial Hashing (SDAH)\cite{SDAH} leverages GANs to decouple modality-shared features from modality-private ones, thus enhancing the discrimination and robustness of semantic embeddings. With a variational information bottleneck and structural similarity constraint, SDAH preserves more semantic information during feature compression.

\subsubsection{GNN methods}
GNN methods\cite{GCH, LGCNH, GCDH, PGCH} create modality-specific graphs for multi-modal data, utilizing label information and graph neighbor structures to capture intrinsic semantic relationships. For instance, Graph Convolutional Hashing (GCH)\cite{GCH} introduces a GCN to enhance correlation information and guide feature encoding networks more  effectively. Local Graph Convolutional Networks Hashing (LGCNH)\cite{LGCNH} employs GCNs to reconstruct the local graph of multi-modal data into various modality features, thereby preserving the underlying data structure. Recently, Proxy-based Graph Convolutional Hashing (PGCH)\cite{PGCH} improves cross-modal retrieval by constructing proxy hash codes to capture global similarities between modalities. By incorporating a multi-modal contrastive loss and applying graph convolution to a proxy hash matrix, PGCH effectively reduces modality differences and improves retrieval performance.

\subsubsection{Transformer methods}
Transformer-based approaches\cite{DCHMT, DSPH, MITH, DHaPH} utilize Transformer encoders to capture fine-grained semantic information effectively and incorporate label information to supervise the modeling of complex relationships within multi-modal data. For instance, Differentiable Cross-modal Hashing via Multi-modal Transformers (DCHMT)\cite{DCHMT} constructs a multi-modal Transformer to capture detailed cross-modal semantic information and introduces a micro-hashing module for mapping modal representations into hash codes. Similarly, Multi-granularity Interactive Transformer Hashing (MITH)\cite{MITH} employs Transformers to hierarchically capture semantic similarities across modalities, generating discriminative hash codes. Deep Hierarchy-aware Proxy Hashing (DHaPH)\cite{DHaPH} models semantic hierarchies with shared proxies and dynamically focuses on hard pairs to refine hash codes for better retrieval performance.

\subsubsection{Memory network methods}
Memory network methods\cite{CMMN, CMPD} leverage shared- or independent-modality memory banks to extract relevant information and update memory contents, thus significantly enhancing cross-modal representation learning. For example, Cross-Modal Memory Network (CMMN)\cite{CMMN} employs a memory mechanism to effectively capture relationships between different data modalities. Similarly, Cross Memory network with Pair Discrimination (CMPD)\cite{CMPD} incorporates a cross-memory unit to facilitate feature sharing between image and text generators, and is optimized with a joint distribution matching constraint to learn both inter-modal and inter-class relationships. Additionally, memory networks are widely used in special retrieval scenarios\cite{ACMM, CCMH-GAM} such as few-shot and incremental processing, continuously adapting to evolving data and making them valuable in dynamic environments.

\subsubsection{Quantization network methods}
Quantization network methods\cite{CDQ, ACQH} utilize deep networks and specialized loss functions to convert real-value features into compact binary codes. For instance, Collective Deep Quantization (CDQ)\cite{CDQ} integrates quantization operations within an end-to-end framework, jointly learning deep features and quantizers through well-designed hybrid networks and loss functions. Meanwhile, Asymmetric Correlation Quantization Hashing (ACQH)\cite{ACQH} learns low-dimensional embeddings using projection matrices and applies a compositional quantization strategy to generate hash codes.

\begin{table}[!b]
    \renewcommand\arraystretch{1.25}
	\centering
	\caption{A compilation of representative methods for cross-modal retrieval under special scenarios. Each method is accompanied by annotations on universal criteria, detailing its data coding format, type of supervision, and technical architecture. The following abbreviations are used: UR for unsupervised real-value, SR for supervised real-value, UH for unsupervised hashing, SH for supervised hashing, MF for matrix factorization, PL for projection learning, ML for metric learning, and MN for memory network.}
	\label{Table:CMRSS}
    \resizebox{\linewidth}{!}{
	\begin{tabular}{|c|c|c|}
		\hline
        Special Scenarios & Concrete issues & Representative methods  \\ \hline \hline
        \multirow{2}{*}{Noise-robust} & Noisy labels & \makecell{WSJE(SR, CNN-RNN)\cite{WSJE}, MRL(SR, CNN-RNN)\cite{MRL},\\ WASH(SH, MF\&PL)\cite{WASH}, LCNME(SR, CNN-RNN)\cite{LCNME}}  \\ \cline{2-3}
        ~ & Mismatched pairs & \makecell{DECL(UR, CNN-RNN)\cite{DECL}, RCL(UR, CNN-RNN)\cite{RCL},\\ SREM(UR, CNN-RNN)\cite{SREM}, RDE(UR, TransFormer)\cite{RDE}}  \\ \hline
        \multicolumn{2}{|c|}{Incomplete} & \makecell{RUCMH(UH, ML)\cite{RUCMH}, GSPH(SH, PL)\cite{GSPH1},\\ AMSH(SH, PL)\cite{AMSH}, TFNH(SH, CNN-RNN)\cite{TFNH},\\ PAN(SH, CNN-RNN)\cite{PAN}, DAVAE(SR, CNN-RNN)\cite{DAVAE},\\ ICMR-DCT(SR, GNN)\cite{ICMR-DCT}, CICH(SH, CNN-RNN)\cite{CICH}} \\ \hline
        \multirow{2}{*}{Zero/few-shot} & Zero-shot & \makecell{MASLN(SR, CNN-RNN)\cite{MASLN}, TANSS(SR, GAN)\cite{TANSS},\\ MDVAE(UR, CNN-RNN)\cite{MDVAE}, AG-MIH(SH, CNN-RNN)\cite{AG-MIH}}  \\ \cline{2-3}
        ~ & Few-shot & \makecell{Self-Others(SH, CNN-RNN)\cite{Self-Others}, ACMM(UR, CNN-RNN)\cite{ACMM}}  \\ \hline
        \multirow{2}{*}{Incremental} & Distribution shift & \makecell{CCMR(UR, CNN-RNN)\cite{CCMR}, ECMH(SH, CNN-RNN)\cite{ECMH},\\ CTP(UR, TransFormer)\cite{CTP}, CCMH-GAM(SH, CNN-RNN)\cite{CCMH-GAM}}  \\ \cline{2-3}
        ~ & Modality expansion & \makecell{SVHNs(SH, CNN-RNN)\cite{SVHNs}, MARS(SR, CNN-RNN)\cite{MARS},\\ SDML(SR, CNN-RNN)\cite{SDML}, Uni-Code(SR, Transformer)\cite{Uni-Code}}  \\ \hline
        \multicolumn{2}{|c|}{Online learning} & \makecell{OCMH(UH, ML)\cite{OCMH}, OCMFH(UH, MF)\cite{OCMFH},\\ ONION(SH, ML)\cite{ONION}, ROH(SH, PL)\cite{ROH},\\ POLISH(SH, PL)\cite{POLISH}}  \\ \hline
        \multicolumn{2}{|c|}{Cross-domain} & \makecell{DASG(UR, CNN-RNN)\cite{DASG}, ACP(UR, CNN-RNN)\cite{ACP},\\ MHTN(SR, GAN)\cite{MHTN}, CDTH(UH, TransFormer)\cite{CDTH}}  \\ \hline
        \multicolumn{2}{|c|}{Federated} & \makecell{FedCMR(SR, CNN-RNN)\cite{FedCMR}, PT-FUCH(UH, CNN-RNN)\cite{PT-FUCH},\\ PEPFCH(SH, CNN-RNN)\cite{PEPFCH}, FedCAFE(SH, MN)\cite{FedCAFE}}  \\ \hline
        \multicolumn{2}{|c|}{Composed} & \makecell{CMJPA(UR, CNN-RNN)\cite{CMJPA}, TFUN(UR, CNN-RNN)\cite{TFUN},\\ Cola(UR, TransFormer)\cite{Cola}, AlRet(UR, TransFormer)\cite{AlRet}}  \\ \hline
        \multirow{2}{*}{\makecell{Hierarchical\\ \& fine-grained}} & Hierarchical & \makecell{HiCHNet(SH, CNN-RNN)\cite{HiCHNet}, HSSPH(SH, CNN-RNN)\cite{HSSPH}}  \\ \cline{2-3}
        ~ & Fine-grained & \makecell{FGCrossNet(SR, CNN-RNN)\cite{FGCrossNet}, PCMDA(UR, CNN-RNN)\cite{PCMDA}}  \\ \hline
        \multirow{3}{*}{Adversary} & Evasion attack & \makecell{CMLA(SH, CNN-RNN)\cite{CMLA}, TA-DCH(SH, GAN)\cite{TA-DCH}, \\AACH(UH, CNN-RNN)\cite{AACH}, EQB$^2$A(UH, GAN)\cite{EQB2A}}  \\ \cline{2-3}
        ~ & Poisoning attack & \makecell{BadCM(SR, GAN)\cite{BadCM}, IB$^3$A(SH, GAN)\cite{IB3A}}  \\ \cline{2-3}
        ~ & Privacy protection & \makecell{PIP(UH, GAN)\cite{PIP}, PPCL(UH, GAN)\cite{PPCL}}  \\ \hline
	\end{tabular}}
\end{table}

\subsection{Cross-modal Retrieval under Special Scenarios}
The cross-modal retrieval methods mentioned earlier are based on ideal assumptions and are applicable to general retrieval scenarios. However, practical constraints such as incomplete data collection, annotation noise, and specific retrieval needs have led to the development of various cross-modal retrieval methods tailored to address issues encountered in special scenarios. Table \ref{Table:CMRSS} compiles representative cross-modal retrieval methods designed for these special cases, along with detailed annotations of their respective categories.

\subsubsection{Noise-robust cross-modal retrieval}
Cross-modal retrieval often grapples with challenges arising from noisy data, such as noisy labels \cite{WSJE, MRL, WASH, LCNME} or mismatched image-text pairs \cite{DECL, RCL, SREM, RDE}. To mitigate the impact of noisy annotations, Webly Supervised Joint Embedding (WSJE)\cite{WSJE} utilizes curriculum learning to prioritize clean data, thereby minimizing the impact of noisy labels and achieving better results. Multi-modal Robust Learning (MRL)\cite{MRL} introduces a robust clustering loss focused on clean samples, coupled with a multi-modal contrastive loss designed to reduce noise interference by comparing inter- and intra-modal correspondences. Recently, Label Correction using Network prediction based on Memorization Effects (LCNME)\cite{LCNME} improves label accuracy by employing network-predicted labels for corrections, leveraging memorization effects to identify and amend noisy labels based on changes in validation loss.

In addition, several strategies\cite{DECL, RCL, SREM, RDE} have been proposed to address the issue of mismatched image-text pairs. For instance, Deep Evidential Cross-modal Learning (DECL)\cite{DECL} employs an evidence extractor to estimate the uncertainty of pairs, enhancing the system's robustness against noisy correspondences. Robust Dual Embedding (RDE)\cite{RDE} tackles noisy correspondences through a confident consensus division module to identify clean training data and a triplet alignment loss to prevent model collapse and effectively manage hard-negative samples.

\subsubsection{Incomplete cross-modal retrieval}
Efforts in addressing the challenge of incomplete cross-modal retrieval focus on mitigating the issue of partial-missing or unpaired multi-modal samples in training data through the utilization of either shallow or deep techniques. In shallow methods \cite{RUCMH, GSPH1, AMSH}, such as Robust Unsupervised Cross-Modal Hashing (RUCMH)\cite{RUCMH}, heterogeneous spaces are associated with a common space for object reconstruction without relying on paired data. Advances in deep learning have led to Triplet Fusion Network Hashing (TFNH)\cite{TFNH}, which employs a triplet network with zero-padding and data classifiers to learn from unpaired data. There are also similar deep methods \cite{PAN, DAVAE, ICMR-DCT, CICH}, such as Prototype-based Adaptive Network (PAN)\cite{PAN} learns unified prototypes for each category across modalities and fuses them with nearest neighbors to address data incompleteness, while Contrastive Incomplete Cross-modal Hashing (CICH)\cite{CICH} reconstructs global similarities and aligns contextual correspondences through contrastive learning to handle missing or unpaired data.

\subsubsection{Zero/few-shot cross-modal retrieval}
Zero/few-shot cross-modal retrieval \cite{MASLN, TANSS, MDVAE, AG-MIH, Self-Others, ACMM} focuses on handling unseen classes during the learning process of cross-modal retrieval models. Typically, Ternary Adversarial Networks with Self-Supervision (TANSS)\cite{TANSS} achieves zero-shot cross-modal retrieval through a self-supervised semantic network and adversarial training. It also leverages the word vectors of both seen and unseen labels as side information to facilitate knowledge transfer to unseen classes. Attribute-Guided Multiple Instance Hashing (AG-MIH)\cite{AG-MIH} improves zero-shot cross-modal retrieval by learning instance-level hash codes for images using a novel 2-D category-attribute relation layer and multi-stream instance hashing refinement, thus enhancing retrieval performance with unseen categories and noisy labels.

In few-shot retrieval, Self-Others Net\cite{Self-Others} tackles the issue of insufficient samples for new classes by leveraging semantic information from new class samples and exploring their relationships with other samples. This method enhances the semantic features of new class samples, thereby improving retrieval performance in few-shot scenarios. Additionally, Aligned Cross-Modal Memory (ACMM)\cite{ACMM} advances few-shot retrieval by weakly supervising the alignment between few-shot regions and words. It continuously updates cross-modal prototypes to effectively measure the similarity between images and sentences containing few-shot content.

\subsubsection{Incremental cross-modal retrieval}
Incremental cross-modal retrieval focuses on integrating new training data while preserving existing knowledge. It tackles data distribution shift\cite{CCMR, ECMH, CTP, CCMH-GAM} and modality expansion\cite{SVHNs, MARS, SDML, Uni-Code}, ensuring the continued effectiveness and stability of retrieval systems. To address evolving data distributions, Continual Cross-Modal Retrieval (CCMR)\cite{CCMR} prevents catastrophic interference from new data by avoiding database re-indexing with updated networks. Subsequently, Compatible momentum contrast with Topology Preservation (CTP)\cite{CTP} enhances incremental cross-modal retrieval by employing compatible momentum contrast to integrate knowledge from both current and previous tasks, while topology preservation ensures the transfer of embedding knowledge across tasks.

In the context of modality expansion, Separated Variational Hashing Networks (SVHNs)\cite{SVHNs} and Modality-Agnostic Representation for Scalable cross-media retrieval (MARS)\cite{MARS} build label networks to learn discriminative representation through category annotations independent of multi-modal sample pairs. Scalable Deep Multi-modal Learning (SDML)\cite{SDML} utilizes modality-specific auto-encoders and a label projection network to manage modality increment. Amid recent progress, Uni-Code\cite{Uni-Code} addresses modality generalization by learning a unified discrete representation from paired multi-modal data, thereby enabling zero-shot generalization to new modalities with minimal labels.

\subsubsection{Online cross-modal retrieval}
Online cross-modal retrieval \cite{OCMH, OCMFH, ONION, ROH, POLISH} focuses on efficiently handling streaming multi-modal data using advanced online learning strategies, with a particular emphasis on shallow hashing. One notable method, Online Collective Matrix Factorization Hashing (OCMFH)\cite{OCMFH}, dynamically updates hash codes for existing data in response to evolving streaming data by using collective matrix factorization within an online scheme. Subsequently, Random Online Hashing (ROH)\cite{ROH} improves online cross-modal retrieval by simplifying pair-wise similarity factorization into a linear optimization problem, preserving semantic similarities in hash codes, and dynamically handling new data in an online manner. Building on this, adaPtive Online cLass-Incremental haSHing (POLISH)\cite{POLISH} advances online retrieval by addressing the challenge of class-incremental data, generating novel-category label representation, and leveraging stable global label embeddings to produce high-quality hash codes, even as new classes arrive in streaming data.

\subsubsection{Cross-domain cross-modal retrieval}
Cross-domain cross-modal retrieval \cite{DASG, ACP, MHTN, CDTH} aims to improve multi-modal data features in the target domain by utilizing labeled data from the source domain. One such approach, called Domain Adaptation with Scene Graph (DASG)\cite{DASG}, learns cross-modal correlation in the source domain and facilitates cross-modal retrieval through modality and distribution alignment. Another technique, Adaptive Cross-modal Prototypes (ACP)\cite{ACP}, generates concept prototypes by clustering pre-trained uni-modal features within each domain and preserves semantic correlation between these prototypes through regularization. It also maximizes mutual information between the multi-modal representations of the source and target domains, effectively aligning data distribution and semantic commonalities. A recent innovation, Cross-Domain Transfer Hashing (CDTH)\cite{CDTH} enhances cross-modal retrieval by using a semantically rich auxiliary domain to support the target domain's hashing process, addressing semantic deficiencies through both direct and indirect semantic transfer.

\subsubsection{Federated cross-modal retrieval}
Federated cross-modal retrieval\cite{FedCMR, PT-FUCH, PEPFCH, FedCAFE} leverages federated learning to train cross-modal retrieval models across decentralized multi-modal data sources, allowing multiple clients to collaboratively optimize models without sharing their local data. One representative method, Federated Cross-Modal Retrieval (FedCMR)\cite{FedCMR}, builds a common space by training local models upon each client and then aggregating them on a central server to update the global model. Prototype Transfer-based Federated Unsupervised Cross-modal Hashing (PT-FUCH)\cite{PT-FUCH}, Privacy-Enhanced Prototype-based Federated Cross-modal Hashing (PEPFCH)\cite{PEPFCH}, and Federated Cross-modal hashing with Adaptive Feature Enhancement (FedCAFE)\cite{FedCAFE} advance federated cross-modal hashing by incorporating global prototypes, encrypting prototype information, and enhancing feature representation to improve retrieval accuracy, address privacy concerns and boost model generalization.

\subsubsection{Composed cross-modal retrieval}
Composed cross-modal retrieval\cite{CMJPA, TFUN, Cola, AlRet} addresses the challenge of retrieving target images based on composed queries, which combine a reference image with modification text to refine search intentions. One notable method, Cross-Modal Joint Prediction and Alignment (CMJPA)\cite{CMJPA} enhances retrieval accuracy by treating the modification text as an implicit transformation, aligning the query image and modification text with the target image through a joint prediction module. Meanwhile, Trilinear FUsion Network (TFUN)\cite{TFUN} addresses ternary image-text retrieval by capturing high-level associations between three inputs using a trilinear fusion framework and bi-directional triplet loss, while reducing complexity through tensor decomposition. Further, Align and Retrieve (AlRet)\cite{AlRet} improves retrieval by learning multi-modal representation with a composition-and-decomposition approach, where a composition network aligns the reference image and modification text with the target image, and a decomposition network explores the correlations within each input-target triplet.

\subsubsection{Hierarchical/fine-grained cross-modal retrieval}
Fine-grained and hierarchical cross-modal retrieval \cite{FGCrossNet, PCMDA, HiCHNet, HSSPH} aims to retrieve multi-modal data with detailed and hierarchical semantic distributions. For hierarchical cross-modal retrieval, Hierarchical Cross-modal Hashing Network (HiCHNet)\cite{HiCHNet} integrates hierarchical semantic learning and regularized cross-modal hashing. Also, it uses hierarchical labels to supervise this integration effectively, enabling multi-grained cross-modal retrieval. A representative fine-grained cross-modal retrieval method, Fine-Grained Cross-media retrieval Network (FGCrossNet)\cite{FGCrossNet}, introduces a unified deep model for handling fine-grained multi-modal contexts and sets a new benchmark for fine-grained cross-modal retrieval.

\subsubsection{Adversary against cross-modal retrieval}
Despite the powerful capabilities of deep cross-modal retrieval using neural networks, it inherits vulnerability to adversarial perturbations. Well-designed subtle changes can easily mislead retrieval models, causing them to return irrelevant or manipulated results (that is, non-targeted and targeted attacks).

Current research on anti-interference in cross-modal retrieval primarily focuses on evasion attacks against hashing-based deep cross-modal retrieval, including white-box\cite{CMLA, TA-DCH} and black-box attacks\cite{AACH, EQB2A}. For instance, Cross-Modal correlation Learning with Adversarial samples (CMLA)\cite{CMLA} addresses adversarial samples in cross-modal retrieval and proposes a strategy to simultaneously maximize inter-modal similarity and minimize intra-modal similarity. For real scenes with agnostic-knowledge cross-modal hashing models, Adversarial Attack on Cross-modal Hamming (AACH)\cite{AACH} replaces the attacked model with a surrogate model for black-box transfer attacks.

Additionally, certain methods focus on poisoning attacks\cite{BadCM, IB3A} and privacy protection\cite{PIP, PPCL} in cross-modal retrieval tasks. Notably, Backdoor attack against Cross-Modal learning (BadCM)\cite{BadCM} and Invisible Black-Box Backdoor Attack (IB$^3$A)\cite{IB3A} contribute to adversaries against cross-modal retrieval by employing invisible backdoor attacks. Meanwhile, PrIvacy Protection (PIP)\cite{PIP} and Proactive Privacy-preserving Cross-modal Learning (PPCL)\cite{PPCL} achieve data privacy by inserting imperceptible perturbations into released data, making it escape automatic retrieval while remaining accessible to human users in its original form.

\begin{figure}[!b]
\centering
\includegraphics[width=0.49\textwidth]{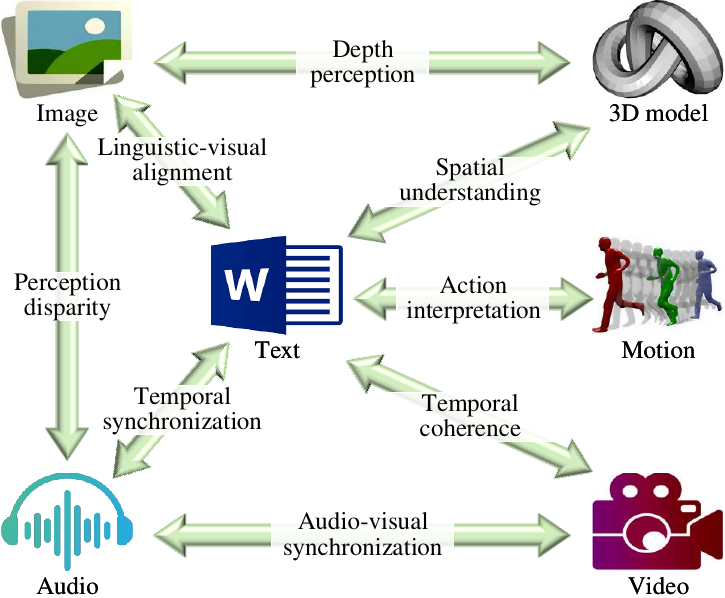}
\caption{The distinct challenges encountered in cross-modal retrieval across heterogeneous modalities.}
\label{Figure:modality_gap}
\end{figure}

\begin{table*}[!t]
\renewcommand\arraystretch{1.15}
\caption{A compilation of representative cross-modal retrieval methods beyond text-image. The abbreviations used remain consistent with the previous definitions.}
\label{Table:Cross-Modal Retrieval Beyond Text-Image}
\vspace{-3mm}
    \begin{center}
        \resizebox{\linewidth}{!}{
            \begin{tabular}{|c|c|c|c|}
                \hline
                Tasks & \multicolumn{2}{c|}{Categories} & Representative methods  \\ \hline \hline
                \multirow{8}{*}{Text-video retrieval} & \multirow{6}{*}{\makecell{Unsupervised \\real-value retrieval}} & Concept & Ad-hoc\_16A\cite{CBITR1}, Ad-hoc\_16B\cite{CBITR2}, Ad-hoc\_17A\cite{CBITR3}, QKR\cite{CBITR4} \\ \cline{3-4}
                ~ & ~ & CNN-RNN & \makecell{LLVE\cite{LLVE}, LJEMC\cite{LJEMC}, Howto100m\cite{Howto100m}, CT-SAN\cite{LSMDC17}, MEE\cite{MEE}, \\ Word2VisualVec\cite{Word2VisualVec}, CSL\cite{CSL}} \\ \cline{3-4}
                ~ & ~ & GNN & HGR\cite{HGR}, GHAN\cite{GHAN}, HCGC\cite{HCGC}, STGs\cite{STGs}, GVSI \cite{GVSI} \\ \cline{3-4}
                ~ & ~ & Transformer & MMT\cite{MMT}, HiT\cite{HiT}  \\ \cline{3-4}
                ~ & ~ & VLP model & \makecell{CLIP4Clip\cite{CLIP4Clip}, VoP\cite{VoP}, Cap4Video\cite{Cap4Video}, DGL\cite{DGL}, TeachCLIP\cite{TeachCLIP}, T-MASS\cite{T-MASS}}  \\ \cline{2-4}
                ~ & \multicolumn{2}{c|}{Other types of text-video retrieval} & \makecell{SVMR(SR, TransFormer)\cite{SVMR}, MCMN(SR, CNN-RNN)\cite{MCMN}, SpotVMR(SR, TransFormer)\cite{SpotVMR},\\ CLIP4Hashing(UH, TransFormer)\cite{CLIP4Hashing}, DSMHN(SH, CNN-RNN)\cite{DSMHN}}  \\ \hline
                \multirow{2}{*}{Text-audio retrieval} &  \multirow{2}{*}{\makecell{Unsupervised \\real-value retrieval}} & CNN-RNN & ATR\cite{ATR}, OML\cite{OML}, MRL2\cite{MRL2}  \\ \cline{3-4}
                ~ & ~ & Transformer & TAP-PMR\cite{TAP-PMR}, CMRF\cite{CMRF}, TTMR++\cite{TTMR++}  \\ \hline
                \multicolumn{3}{|c|}{Image-audio retrieval} & \makecell{ALAC(UR, CNN-RNN)\cite{ALAC}, SCFEM(UR, Transformer)\cite{SCFEM}, SSIA(UR, Transformer)\cite{SSIA},\\ SCRL(SR, CNN-RNN)\cite{SCRL}, CSIR(SR, Transformer)\cite{CSIR}, DIVR(SH, CNN-RNN)\cite{DIVR},\\ DQH(SH, CNN-RNN)\cite{DQH}, FADH(SH, TransFormer)\cite{FADH}}  \\ \hline
                \multicolumn{3}{|c|}{Video-audio retrieval} & \makecell{CDN(UR, CNN-RNN)\cite{CDN}, MAC(UR, TransFormer)\cite{MAC}, ACIENet(UR, TransFormer)\cite{ACIENet}, \\ LEIDCS(SR, CNN-RNN)\cite{LEIDCS}, VAE-CCA(SR, CNN-RNN)\cite{VAE-CCA}}  \\ \hline
                \multicolumn{3}{|c|}{Image-3D retrieval} & \makecell{CMCL(SR, CNN-RNN)\cite{CMCL}, CMI-CCL(SR, CNN-RNN)\cite{CMI-CCL}, RONO(SR, CNN-RNN)\cite{RONO}, \\FIVE(SR, CNN-RNN)\cite{FIVE}, HOPE(SR, CNN-RNN)\cite{HOPE}}  \\ \hline
                \multicolumn{3}{|c|}{Text-3D retrieval} & xM\_Match(SR, TransFormer)\cite{xM_Match}, RMARN(UR, TransFormer)\cite{RMARN}  \\ \hline
                \multicolumn{3}{|c|}{Text-motion retrieval} & TMR(UR, TransFormer)\cite{TMR}, TMR++(UR, TransFormer)\cite{TMR++}  \\ \hline
                \multicolumn{3}{|c|}{Cross-modal 3D retrieval} & HSR(UR, CNN-RNN)\cite{HSR}, CMCL(SR, CNN-RNN)\cite{CMCL}  \\ \hline
                \multicolumn{3}{|c|}{Cross-modal image retrieval} & SYRER(SR, TransFormer)\cite{SYRER}, CGAE(SR, GNN)\cite{CGAE}  \\ \hline
                \multicolumn{3}{|c|}{Image-video retrieval} & LSVR(UR, Fisher\_vector)\cite{LSVR}, APIVR(SR, CNN-RNN)\cite{APIVR}  \\ \hline
            \end{tabular}}
    \end{center}
\end{table*}

\section{Cross-Modal Retrieval Beyond Text-Image}
\label{Sec:Cross-Modal Retrieval Beyond Text-Image}
As cross-modal retrieval has advanced, it now extends beyond traditional text-image retrieval to encompass a broader array of data modalities and retrieval tasks. This section offers a comprehensive review of cross-modal retrieval methods that involve modalities beyond text-image, as summarized in Table \ref{Table:Cross-Modal Retrieval Beyond Text-Image}, including text-video, text-audio, image-audio, image-3D, and more. We explore the unique challenges associated with these extended modalities, illustrated in Fig. \ref{Figure:modality_gap}, offering insights into the specific techniques and architectures designed to handle the complexities of each combination.

While this section focuses on cross-modal retrieval beyond text-image, it employs the same taxonomy as text-image retrieval for consistency. We leverage this taxonomy to analyze each retrieval task, emphasizing the unique challenges and the corresponding solutions crafted for each.

\subsection{Text-video Cross-modal Retrieval}
Text-video retrieval involves searching for corresponding videos in a database based on a given text description, and vice versa. It introduces unique challenges due to the temporal nature of video data. Unlike text-image retrieval, which aligns static visual features with texts, text-video retrieval manages the evolving content of video frames over time. This requires methods that capture temporal coherence and synchronize text with specific video moments, making text-video retrieval inherently more complex.
Current research in text-video retrieval primarily falls under the overarching category of unsupervised real-value retrieval. Only a few studies explore supervised real-value video moment retrieval \cite{SVMR, MCMN, SpotVMR} (discussed in Section \ref{Sec:Application}) and hashing-based retrieval \cite{CLIP4Hashing, DSMHN}. Further, unsupervised real-value retrieval can be subdivided based on the matching strategies and modeling techniques employed, including \emph{concept methods}, \emph{CNN-RNN methods}, \emph{GNN methods}, \emph{Transformer methods}, and \emph{VLP model methods}.

\subsubsection{Concept methods}
Concept-based methods \cite{CBITR1, CBITR2, CBITR3, CBITR4} rely on semantic concept similarity between texts and videos to enable their mutual retrieval. For example, some studies \cite{CBITR1, CBITR2} use pre-trained classifiers to identify key objects and scenes in videos and assign them to relevant concept categories, while applying complex linguistic rules to extract corresponding concepts from texts. Subsequent improvements mainly focus on expanding the set of semantic concepts \cite{CBITR3} and employing more powerful concept classifiers \cite{CBITR4}. While concept-based representations provide some interpretability in text-video retrieval, a significant drawback is the difficulty of capturing the rich temporal information in texts and videos with a limited set of selected concepts.

\subsubsection{CNN-RNN methods}
CNN-RNN methods \cite{LLVE, LJEMC, Howto100m, LSMDC17, MEE, Word2VisualVec, CSL} encode videos and texts separately, mapping them into a common latent space for direct similarity measurement. An early exploration, Learning Language-Visual Embedding (LLVE) \cite{LLVE}, uses CNN and LSTM to capture latent features of frame images and texts, with LSTM further modeling temporal relationships between video frames. Some works \cite{LJEMC, Howto100m} use mean/max-pooling on video frame sequences to aggregate them into video-level representations, trading precision for efficiency. Later, additional modality cues like audio and motion are integrated to enrich video semantic representation \cite{MEE}. Text processing, being simpler, primarily relies on Word2Vec, LSTM, or GRU for encoding \cite{LJEMC, Howto100m, LSMDC17}, with some studies \cite{Word2VisualVec} showing that combining multiple text encoding strategies improves retrieval performance. Recently, Complementarity-aware Space Learning (CSL) \cite{CSL} introduces concept-based video representation, leveraging complementary latent and symbolic spaces to enhance text-video retrieval.

\subsubsection{GNN methods}
GNN methods \cite{HGR, GHAN, HCGC, STGs, GVSI} for text-video retrieval model the relationships between video components and text elements by constructing hierarchical or spatial-temporal graphs. For instance, Hierarchical Graph Reasoning (HGR) \cite{HGR} and Graph-based Hierarchical Aggregation Network (GHAN) \cite{GHAN} focus on decomposing text-video matching into global-to-local levels, creating hierarchical graphs to capture fine-grained details. Concurrently, Hierarchical Cross-modal Graph Consistency learning (HCGC) \cite{HCGC} introduces multi-level graph consistency, addressing both global and local alignment between videos and texts. Other approaches, like Spatial-Temporal Graphs (STGs) \cite{STGs} and Graph convolutional networks and Visual Semantic Inference (GVSI) \cite{GVSI}, further enrich video representations by modeling object interactions across time, employing GCNs to capture hidden semantic relationships within video frames. 

\subsubsection{Transformer methods}
Transformer methods \cite{MMT, HiT} leverage self-attention mechanisms to jointly encode videos and texts, enabling each modality to attend to the other. Typically, Multi-Modal Transformer (MMT) \cite{MMT} achieves this by enabling mutual attention between video and text modalities and incorporating temporal information for better feature representation. Hierarchical Transformer (HiT) \cite{HiT} introduces hierarchical cross-modal contrastive matching at both feature and semantic levels, offering comprehensive retrieval results. It also employs cross-modal momentum contrast, inspired by MoCo \cite{MoCo}, to facilitate extensive negative sample interactions. Additionally, these methods \cite{MMT, HiT} encode various modalities in video data, such as audio and motion, further enriching video representation.

\subsubsection{VLP model methods}
Recent advances in text-video retrieval also harness the capability of Vision-Language Pre-training (VLP) models \cite{CLIP, GPT-2} to improve cross-modal understanding. Such methods \cite{CLIP4Clip, VoP, Cap4Video, DGL, TeachCLIP, T-MASS} primarily leverage pre-training models like CLIP \cite{CLIP} to directly transfer knowledge from text-image tasks to text-video tasks, and enhance text-video retrieval by exploiting cross-modal cues and incorporating temporal dependencies. Typically, CLIP4Clip\cite{CLIP4Clip} transfers CLIP's text-image pre-training to text-video retrieval and captioning, examining factors like temporal dependencies and post-pre-training effects. Later, text-Video co-operative Prompt tuning (VoP)\cite{VoP} introduces a text-video cooperative prompt tuning mechanism that efficiently tunes the CLIP model, minimizing the trainable parameters while modeling spatio-temporal aspects of video data. Cap4Video\cite{Cap4Video} explores the role of auxiliary captions generated from videos using zero-shot captioning based on VLP models like CLIP\cite{CLIP} and GPT-2\cite{GPT-2}, integrating these captions into text-video retrieval at various stages to boost performance. Dynamic prompt tuning with Global-Local video attention (DGL)\cite{DGL} proposes dynamic global-local prompt tuning for CLIP, emphasizing inter-modal interaction by generating local-level text and frame prompts in a shared latent space and capturing global video information through a global-local attention mechanism. TeachCLIP\cite{TeachCLIP} enhances CLIP4Clip\cite{CLIP4Clip} by incorporating fine-grained cross-modal knowledge from more advanced models, using an attentional frame-feature aggregation block to refine text-video similarity without adding computational overhead. Lastly, T-MASS\cite{T-MASS} introduces stochastic text modeling in VLP models to address the limitations of short, concise text content in datasets, enriching text embeddings with a flexible semantic range and optimizing them for accurate text-video retrieval.

\subsection{Text-audio Cross-modal Retrieval}
Text-audio retrieval involves the challenge of finding matching audio content based on text queries, and vice versa. It deals with temporal sequences of auditory data, requiring the alignment of textual semantics with varying acoustic patterns, such as speech, music, or environmental sounds. Early text-audio retrieval \cite{LCAR} used pre-defined category labels and individual words for retrieval. Recently, with the advent of audio caption datasets \cite{Clotho, SoundDecs}, language-based text-audio retrieval \cite{ATR, OML, MRL2, TAP-PMR, CMRF, TTMR++} has gained attention, exploring aggregation methods, objective functions, and training strategies. According to our taxonomy, they all fall under unsupervised real-value retrieval and can be further categorized into \emph{CNN-RNN methods} and \emph{Transformer methods}.

\subsubsection{CNN-RNN methods}
CNN-RNN methods \cite{ATR, OML, MRL2} focus on encoding text and audio separately, aligning them in a common space for similarity measurement. For instance, Audio-Text Retrieval (ATR) \cite{ATR} leverages pre-trained audio networks with CNNs to extract audio features from a large sound event dataset, combined with NetRVLAD \cite{NetRVLAD} pooling to aggregate these features into comprehensive representation. On Metric Learning (OML) \cite{OML} leverages CNNs to extract robust audio features while employing metric learning to enhance alignment between audio and text embeddings. Recently, Multi-grained Representation Learning (MRL) \cite{MRL2} utilizes CNNs to extract localized audio features and applies an adaptive aggregation strategy to handle varying granularity levels in text-audio associations.

\subsubsection{Transformer methods}
Transformer methods \cite{TAP-PMR, CMRF, TTMR++} leverage multi-head attention mechanisms and fine-tuning to improve cross-modal interactions. Notably, Text-aware Attention Pooling and Prior Matrix Revised loss (TAP-PMR) \cite{TAP-PMR} introduces a scaled dot-product attention mechanism, allowing text to selectively focus on semantically similar audio frames, thus mitigating misleading information. Additionally, the prior matrix revised loss optimizes dual matching by addressing discrepancies in text-to-audio and audio-to-text similarities. Recently, Cross-Modal interaction via Reinforcement Feedback (CMRF) \cite{CMRF} enhances audio-lyrics retrieval by employing directional cross-modal attention and reinforcement learning to refine multi-modal embeddings and interactions. Also, improved Text-To-Music Retrieval (TTMR++) \cite{TTMR++} integrates a fine-tuned large language model and rich meta-data to generate detailed text descriptions, improving retrieval effectiveness by addressing both specific musical attributes and user preferences.

\subsection{Image-audio Cross-modal Retrieval}
Image-audio retrieval uniquely focuses on aligning visual features with auditory cues, requiring methods to bridge the gap between the static nature of images and the dynamic, temporal characteristics of audio, which differs from the more straightforward alignment in text-based retrieval tasks. Current image-audio retrieval research primarily focuses on cross-modal Remote Sensing (RS) retrieval and scene matching, with limited research but broad coverage, including both supervised and unsupervised methods, real-value and hashing techniques, as well as diverse deep neural networks\footnote{Since there is little research on cross-modal retrieval beyond text-image/video/audio, we no longer separate subcategories into distinct sections. Instead, we clarify the subcategory for each method within its description.}.

Unsupervised real-value retrieval spans both CNN-RNN \cite{ALAC} and Transformer methods \cite{SCFEM, SSIA}. CNN-RNN methods, such as Adaptive Learning for Aligning Correlation (ALAC) \cite{ALAC}, address the challenge of imbalanced information between RS images and audio by integrating region-level learning into image annotation, providing comprehensive visual feature representation. Transformer methods, such as Self- and Cross-modal Feature Embedding Memory (SCFEM) \cite{SCFEM} and Self-Supervision Interactive Alignment (SSIA) \cite{SSIA}, enhance retrieval by employing advanced cross-modal alignment strategies. SCFEM \cite{SCFEM} uses metric learning and embedding memory, while SSIA \cite{SSIA} leverages interactive alignment and audio-guided de-redundant modules to capture salient information, ensuring accurate image-audio retrieval.

In supervised real-value retrieval, CNN-RNN methods include Semantics-Consistent Representation Learning (SCRL) \cite{SCRL}, which integrates pair-wise, intra-modality, and non-paired inter-modality relationships to improve semantic consistency in RS image-audio retrieval. Also, it employs CNNs for image encoding and dilated convolutions for audio encoding, refining the consistency of multi-modal representation. Transformer methods such as Contrastive Speech Image Retrieval (CSIR) \cite{CSIR} utilize Transformer architectures for space-aware representation learning and supervised contrastive learning to improve cross-modal similarity.

Current hashing-based research primarily focuses on supervised RS image-audio retrieval \cite{DIVR, DQH, FADH}. Among these approaches, Deep Image-Voice Retrieval (DIVR)\cite{DIVR} and Deep Quadruple-based Hashing (DQH)\cite{DQH} utilize CNN-RNN architectures to capture latent features and learn relative semantic similarities through hashing.
DIVR\cite{DIVR} employs an inception dilated convolution module to capture multi-scale contextual information, whereas DQH\cite{DQH} uses a quadruple-based hashing network to enhance semantic correlations and minimize quantization errors.
Fine Aligned Discriminative Hashing (FADH)\cite{FADH}, a Transformer-based method, advances this field by learning detailed discriminative information and fine alignment between image regions and audio, with a novel objective function to preserve semantic information and reduce cross-modal discrepancies.

\subsection{Video-audio Cross-modal Retrieval}
Video-audio retrieval involves the unique challenge of synchronizing temporal audio features with dynamic video content. This requires aligning audio cues with the visual context and the temporal progression of video frames. Unlike static or static-dynamic matching seen in other retrieval tasks, this process demands modeling the intricate interactions between audio and video sequences over time. Current research primarily focuses on real-value retrieval, encompassing both unsupervised and supervised methods.

In unsupervised real-value retrieval, CNN-RNN methods include Collaborative Decision Network (CDN)\cite{CDN}, which models the task as a sequential decision process, allowing event-relevant segments to collaborate and achieve better localization results. In contrast, Transformer methods such as Multi-modal Aggregation and Co-attention network (MAC)\cite{MAC} enhance video-audio retrieval by leveraging co-attentional interactions and self-attention mechanisms to better capture mutual relationships between video and audio modalities. Similarly, the Attribute-guided Cross-modal Interaction and Enhancement Network (ACIENet)\cite{ACIENet}, also a Transformer-based approach, focuses on refining cross-modal matching by incorporating attribute-guided interaction and enhancement, thus improving the correlation between subtle local features in audio-visual matching.

Currently, supervised real-value retrieval methods\cite{LEIDCS, VAE-CCA} are predominantly based on CNN-RNN architectures. For instance, Learning Explicit and Implicit Dual Common Subspaces (LEIDCS) \cite{LEIDCS} employs an end-to-end architecture that projects audio-visual data into explicit and implicit dual common subspaces. It learns modality-common features in the explicit subspace to bridge the modality gap while using the implicit subspace to maintain modality-specific distinctions. Variational Auto-Encoder with CCA (VAE-CCA) \cite{VAE-CCA} leverages canonical correlation analysis to learn paired audio-visual correlation embeddings and category correlation embeddings, enhancing feature extraction and reducing cross-modal discrepancies through probabilistic modeling.

\subsection{Image-3D Cross-modal Retrieval}
Image-3D retrieval uniquely involves aligning 2D image data with 3D object representations, which requires bridging the gap between flat image features and volumetric 3D structures. This challenge necessitates specialized methods to map and correlate complex 3D spatial information with 2D visual content. Current research in this field focuses on CNN-RNN-based supervised real-value methods \cite{CMCL, CMI-CCL, RONO, FIVE, HOPE}, with 3D shapes predominantly represented as point clouds.

Supervised real-value retrieval often employs metric learning or contrastive techniques to map both modalities into a common space, facilitating accurate retrieval based on semantic similarity. For instance, Cross-Modal Center Loss (CMCL) \cite{CMCL} reduces modality discrepancies using a cross-modal center loss, combined with cross-entropy and mean-square-error losses to enhance feature alignment. Cross-Modal Instance and Category Contrastive Learning (CMI-CCL) \cite{CMI-CCL} employs contrastive learning with multi-view grayscale images and a category-level contrastive loss to improve retrieval accuracy from single 2D images. RONO \cite{RONO} addresses noisy labels by integrating robust discriminative center learning and shared space consistency learning, enhancing retrieval performance despite data imperfections. Both FIVE \cite{FIVE} and HOPE \cite{HOPE} address label scarcity and heterogeneous data by incorporating advanced techniques like fine-grained prototypes, heterogeneous mixup, hierarchical alignment, and curriculum learning to enhance retrieval accuracy and robustness.

\subsection{Broader Scope of Cross-modal Retrieval}
In addition to the cross-modal retrieval tasks previously discussed, there is ongoing research that explores retrieval across a broader range of data modalities \cite{xM_Match, RMARN, TMR, TMR++, HSR, CMCL, SYRER, CGAE, LSVR, APIVR}. In this context, we examine the distinctive features and representative methods of several emerging cross-modal retrieval tasks, including:

\subsubsection{Text-3D retrieval}
Text-3D retrieval \cite{xM_Match, RMARN} challenges the mapping of textual descriptions to 3D objects, requiring the alignment of abstract language with multi-dimensional spatial structures. Typically, xM\_Match \cite{xM_Match} uses a Bird's-Eye-View (BEV)-based information embedding module to match local visual features with text descriptions. It employs a visual transformer to model long-range dependencies and cross-modal consistency learning to align feature spaces, enhancing matching accuracy.

\subsubsection{Text-motion retrieval}
Text-motion retrieval \cite{TMR, TMR++} specifically focuses on matching textual descriptions to dynamic sequences of motion data, such as skeletal movements or animations, rather than visual frames as in text-video retrieval. Typically, Text-to-Motion Retrieval (TMR) \cite{TMR} extends text-to-motion synthesis models with a contrastive loss to structure the cross-modal latent space, significantly improving retrieval performance by maintaining the motion generation loss alongside contrastive training.

\subsubsection{Cross-modal 3D retrieval}
Cross-modal 3D retrieval \cite{HSR, CMCL} handles complex 3D objects such as voxels, meshes, and point clouds, requiring effective alignment across diverse spatial forms. Methods like Hierarchical Set-to-set Representation (HSR) \cite{HSR} address limitations of global feature reliance by incorporating global-to-global and local-to-local similarity metrics. It uses bi-linear pooling for compact-set features and joint loss functions to optimize hierarchical similarity measurement and cross-modal discrepancies.

\subsubsection{Cross-modal image retrieval}
Cross-modal image retrieval\cite{SYRER, CGAE} involves identifying relevant images across different imaging types, such as matching RGB images with depth images, infrared images, or sketches. The goal is to bridge the gap between the visual characteristics of RGB images and the unique features of these other image types. Typically, SYnergistic RElational Reasoning (SYRER)\cite{SYRER} introduces synergistic relational reasoning by focusing on both intra- and inter-class cross-modal relationships. It employs a heterogeneous relationship contrast branch and a point-wise depth extractor to tackle modality discrepancies.

\subsubsection{Image-video retrieval}
Image-video retrieval \cite{LSVR, APIVR} aligns static images with dynamic video sequences by integrating stable image features with frame-level details from videos, leveraging both temporal and spatial cues to bridge the modality gap. Typically, Large-Scale Video Retrieval (LSVR) \cite{LSVR} enhances video search by directly comparing image queries with videos in the database. It employs an asymmetric comparison technique using Fisher vectors and novel video descriptors, which significantly improves compression and accuracy compared to frame-based methods.

\begin{table}[!t]
    \renewcommand\arraystretch{1.1}
	\centering
	\caption{Detailed statistics for cross-modal retrieval datasets. Modality: T is text, I is image, V is video, A is audio, and 3D is 3D model. Label: S is single-label and M is multi-label.}
	\label{Dataset}
    \resizebox{\linewidth}{!}{
	\begin{tabular}{|c|c|c|c|c|c|}
		\hline
		Dataset&Instance&Category&\makecell{Modality\\(T, I, V, A, 3D)}&\makecell{Label\\(S,M)}&Main  scene \\
		\hline
		\hline
		Wikipedia\cite{CCA} & 2,866 & 10 & T, I & S & Scenery, Humanity \\
		\hline
		Pascal-Sentence\cite{Pascal} & 1,000 & 20 & T, I & S & Scenery, Humanity \\
		\hline
		INRIA-Websearch\cite{INRIA-Websearch}& 71,478 & 353 & T, I & S & Scenery, Humanity \\
		\hline
		MIR Flickr\cite{flickr}& 20,015 & 24 & T, I & M & Scenery \\
		\hline
		Flickr-8K\cite{flickr-8k}& 8,092 & --- & T, I & None & Scenery \\
		\hline
		Flickr-30K\cite{flickr-30k}& 31,783 & --- & T, I & None & Scenery \\
		\hline
		MS COCO\cite{COCO}& 123,287 & 80 & T, I & M & Scenery, Humanity \\
		\hline
		MS COCO-5K\cite{COCO}& 5,000 & --- & T, I & None & Scenery, Humanity \\
		\hline
		NUS-WIDE\cite{NUS}& 195,834 & 21 & T, I & M & Scenery, Humanity \\
		\hline
        NUS-WIDE-10K\cite{NUS}& 10,000 & 10 & T, I & S & Scenery, Humanity \\
		\hline
		IAPR TC-12\cite{IAPR}& 20,000 & 255 & T, I & M & Scenery, Humanity \\
		\hline
		Xmedia\cite{Xmedia}& 5,000 & 20 & T, I, V, A, 3D & S & Animal, Object \\
		\hline
		Fashion-Gen\cite{Fashion-Gen}& 67,666 & --- & T, I & None & Clothing \\
		\hline
		Recipe1M\cite{Recipe1M}& 1,029,720 & --- & T, I & None & Recipe \\
		\hline
		CUB 200-2011\cite{CUB}& 11,788 & 200 & T, I & S & Bird \\
		\hline
		Caltech-256\cite{Caltech-256}& 30,607 & 256 & T, I & S & Scenery, Humanity \\
		\hline
		LabelMe\cite{LabelMe}& 30,369 & 183 & T, I & M & Animal, Object \\
		\hline
		Visual Genome\cite{Visual-Genome}& 108,077 & --- & T, I & None & Humanity \\
		\hline
		CC3M\cite{CC3M}& 3,369,218 & --- & T, I & None & Scenery, Object \\
		\hline
		MSR-VTT\cite{MSR-VTT}& 10,000 & 20 & T, V & S & Humanity \\
		\hline
		ActivityNet\cite{ActivityNet}& 20,000 & --- & T, V & None & Humanity, Object \\
		\hline
		LSMDC\cite{LSMDC}& 128,085 & --- & T, V & None & Movie \\
		\hline
		Vatex\cite{Vatex}& 41,250 & 600 & T, V & S & Humanity \\
		\hline
		HowTo100M\cite{Howto100m}& 136M & 12 & T, V & S & Teaching \\
		\hline
		Clotho\cite{Clotho}& 4,981 & --- & T, A & None & Show, Wild \\
		\hline
		AudioCaps\cite{Audiocaps}& 39,597 & --- & T, A & None & Wild \\
		\hline
		UCM\cite{UCM_Sydney}& 2,100 & 21 & I, A & S & Remote sensing \\
		\hline
		Sydney\cite{UCM_Sydney}& 613 & 7 & I, A & S & Remote sensing \\
		\hline
		RSITMD\cite{RSITMD}& 4,743 & 32 & I, A & S & Remote sensing \\
		\hline
		RSICD\cite{RSICD}& 10,921 & 30 & I, A & S & Remote sensing \\
		\hline
		AVE\cite{AVE}& 4,143 & 28 & V, A & S & Humanity, Animal \\
		\hline
        VEGAS\cite{VEGAS}& 28,109 & 10 & V, A & S & Humanity, Animal \\
		\hline
        3D MNIST\cite{3D_MNIST}& 6,000 & 10 & I, 3D & S & Object \\
		\hline
        ModelNet10\cite{ModelNet}& 4,900 & 10 & I, 3D & S & Object \\
		\hline
        ModelNet40\cite{ModelNet}& 12,311 & 40 & I, 3D & S & Object \\
		\hline
        Pix3D\cite{Pix3D}& 10,069 & 9 & I, 3D & S & Object \\
		\hline
        T3DR-HIT\cite{T3DR-HIT}& 3,380 & --- & T, 3D & None & Indoor, Object \\
		\hline
	\end{tabular}}
\end{table}

\section{Evaluation Dataset, Metric, and Result}
\label{Sec:Dataset and Evaluation}
To provide researchers with a comprehensive understanding of data sources and characteristics in cross-modal retrieval, as well as performance evaluation methods, this section introduces widely used evaluation datasets, metrics, and representative results for cross-modal retrieval.

\subsection{Evaluation Dataset}
Cross-modal retrieval datasets typically comprise multiple modalities with unique feature representations and semantic annotations. The data from different modalities are grouped or paired based on their semantic relevance or similarity. We compile commonly used datasets in various cross-modal retrieval tasks, and their statistics are shown in Table \ref{Dataset}.
In previous cross-modal retrieval studies, the commonly used datasets to evaluate the cross-modal retrieval performance between text and image include Wikipedia \cite{CCA}, Pascal Sentence \cite{Pascal}, INRIA-Websearch \cite{INRIA-Websearch}, Flickr \cite{flickr, flickr-8k, flickr-30k}, MS COCO \cite{COCO}, NUS-WIDE \cite{NUS}, IAPR TC-12 \cite{IAPR}, Xmedia \cite{Xmedia}, Fashion-Gen \cite{Fashion-Gen}, Recipe1M \cite{Recipe1M}, CUB 200-2011 \cite{CUB}, Caltech-256 \cite{Caltech-256}, LabelMe \cite{LabelMe}, Visual Genome \cite{Visual-Genome}, CC3M \cite{CC3M}. Most of the samples in these datasets are sampled from online sites, such as Wikipedia\footnote{\url{https://www.wikipedia.org}} and Flickr\footnote{\url{http://www.flickr.com}}. Also, many public repositories have been established to make them available to researchers, such as MS COCO\footnote{\url{https://cocodataset.org}}, NUS-WIDE\footnote{\url{https://lms.comp.nus.edu.sg/wp-content/uploads/2019/research/nuswide/NUS-WIDE.html}}, and Fashion-Gen\footnote{\url{https://fashion-gen.com}}. Additionally, several datasets have been used to evaluate broader cross-modal retrieval tasks, such as text-video and text-audio retrieval. These include MSR-VTT \cite{MSR-VTT}, ActivityNet Captions \cite{ActivityNet}, LSMDC \cite{LSMDC}, Vatex \cite{Vatex}, HowTo100M \cite{Howto100m}, Clotho \cite{Clotho}, AudioCaps \cite{Audiocaps}, UCM\cite{UCM_Sydney}, Sydney\cite{UCM_Sydney}, RSITMD\cite{RSITMD}, RSICD\cite{RSICD}, AVE\cite{AVE}, VEGAS\cite{VEGAS}, 3D MNIST\cite{3D_MNIST}, ModelNet10\cite{ModelNet}, ModelNet40\cite{ModelNet}, Pix3D\cite{Pix3D}, and T3DR-HIT\cite{T3DR-HIT}.

\subsection{Evaluation Metric}
In cross-modal retrieval tasks, several metrics such as Precision, Recall, mean Average Precision (mAP), and Normalized Discounted Cumulative Gain (NDCG) are used to assess performance. Since these metrics are calculated based on the top-$k$ retrieved results, they are typically represented as \textit{Precision}@$k$, \textit{Recall}@$k$, \textit{mAP}@$k$, and \textit{NDCG}@$k$. Here, the symbol @ serves as a placeholder indicating that the value following it represents the number of top-$k$ retrieved results considered\cite{CCMH-GAM, APIVR, ESSE}. Their details are as follows:

\subsubsection{Precision@$k$} This metric computes the ratio of relevant items among the top-$k$ retrieved items for each query and then averages over all queries. For each query,
\begin{equation}
\small
\textit{Precision}\text{@}k=\frac{\textit{TP}}{\textit{TP+FP}},
\end{equation}
where \textit{TP} and \textit{FP} are the sample number of true positives and false positives in the top-$k$ retrieved results, respectively.

\subsubsection{Recall@$k$} This metric computes the ratio of relevant items among the top-$k$ retrieved items to the total number of relevant items for each query and then averages over all queries. For each query,
\begin{equation}
\small
\textit{Recall}\text{@}k=\frac{\textit{TP}}{\textit{TP+FN}},
\end{equation}
where \textit{TP} and \textit{FN} represent the sample number of true positives and false negatives in the top-$k$ retrieved results, respectively. It is worth noting that \textit{Recall}@$k$ is a standard evaluation metric for unsupervised cross-modal retrieval across various modalities, such as text-image\cite{MSRM, P2RM, R2RM}, text-video\cite{CLIP4Clip, VoP, Cap4Video}, and image-audio\cite{ALAC, SCFEM, SSIA}. Additionally, its variants, Median Rank (\textit{MedR}) and Mean Rank (\textit{MeanR}), which measure the median and mean positions of cross-modal samples corresponding to queries in the retrieved results, are also widely used\cite{PVSE, CLIP4Clip, ALAC}.

\subsubsection{mAP@$k$} This metric computes the average precision for each query and then averages over all queries. First, the top-$k$ Average Precision (\textit{AP}) of a query $q$ is defined as
\begin{equation}
\small
\textit{AP}(q)=\frac{{\textstyle\sum_{i=1}^{k}}(P(i)*rel(i))}{{\textstyle \sum_{i=1}^{k}}rel(i)},
\end{equation}
where $rel(i)$ is an indicator function, indicating whether the $i$-th retrieved result is related to the query $q$, if so $rel(i)=1$, and vice versa $rel(i)=0$. $P(i)$ denotes the \textit{Precision}@$k$ metric. Further, \textit{mAP}@$k$ can be represented as
\begin{equation}
\small
\textit{mAP}\text{@}k = \frac{1}{n}\sum_{j=1}^{n}\textit{AP}(q_j),
\end{equation}
where $n$ is the sample number of queries.

\begin{table}[!b]
\caption{The \textit{Recall}@$k$ results of unsupervised real-value retrieval methods. The experimental results are from ViSTA\cite{ViSTA} and AGREE\cite{AGREE}. '---' indicates that these results are not recorded in the original literature\cite{ViSTA, AGREE}.}
\label{tab:URR}
\renewcommand\arraystretch{1.15}
\centering
\resizebox{\linewidth}{!}{
\begin{tabular}{|c|c|c|c|ccc|ccc|}
\hline
\multicolumn{1}{|c|}{\multirow{2}{*}{Task}} & \multicolumn{1}{c|}{\multirow{2}{*}{Method}} & \multicolumn{1}{c|}{\multirow{2}{*}{Category}} & \multicolumn{1}{c|}{\multirow{2}{*}{Source}} & \multicolumn{3}{c|}{Flickr-30K}                 & \multicolumn{3}{c|}{MS COCO-5K}   \\ \cline{5-10}
\multicolumn{1}{|c|}{}                      & \multicolumn{1}{c|}{}                        & \multicolumn{1}{l|}{}                          & \multicolumn{1}{c|}{}                        & R@1           & \multicolumn{1}{c}{R@5}    & \multicolumn{1}{c|}{R@10}        & \multicolumn{1}{c}{R@1}           & \multicolumn{1}{c}{R@5}  & \multicolumn{1}{c|}{R@10}  \\  \hline \hline
\multicolumn{1}{|c|}{\multirow{11}{*}{I$\to$T}}
& SCAN\cite{SCAN} & CNN-RNN & ECCV18  & 0.674 & 0.903 & 0.958  & 0.504 & 0.822 & 0.900 \\
& VSRN\cite{VSRN} & GNN & ICCV19 & 0.713   & 0.906 & 0.960 & 0.530 & 0.811 & 0.894 \\
& GSMN\cite{GSMN} & GNN & CVPR20 & 0.764 & 0.943 & 0.973 & --- & --- & --- \\
& IMRAM\cite{IMRAM} & CNN-RNN & CVPR20 & 0.741 & 0.930 & 0.966 & 0.537 & 0.832 & 0.910 \\
& Unicoder\cite{Unicoder} & VLP model & AAAI20 & 0.862 & 0.963 & 0.990 & 0.623 & 0.871 & 0.928 \\
& Uniter\cite{Uniter} & VLP model & ECCV20 & 0.859 & 0.971 & 0.988 & 0.644 & 0.874 & 0.931 \\
& SGRAF\cite{SGRAF} & GNN & AAAI21 & 0.778 & 0.941 & 0.974 & 0.578 & --- & 0.916 \\
& PCME\cite{PCME} & Transformer & CVPR21 & --- & --- & --- & 0.442 & 0.738 & 0.836 \\
& SOHO\cite{SOHO} & VLP model & CVPR21 & 0.865 & 0.981 & 0.993 & 0.664 & 0.882 & 0.938 \\
& ViSTA\cite{ViSTA} & Transformer & CVPR22 & 0.895 & 0.984 & 0.996 & 0.689 & 0.901 & 0.954 \\
& AGREE\cite{AGREE} & VLP model & WSDM23 & 0.921 & 0.987 & 0.992 & --- & --- & --- \\ \hline
\multicolumn{1}{|c|}{\multirow{11}{*}{T$\to$I}}
& SCAN\cite{SCAN} & CNN-RNN & ECCV18 & 0.486 & 0.777 & 0.852 & 0.386 & 0.693 & 0.804 \\
& VSRN\cite{VSRN} & GNN & ICCV19 & 0.547 & 0.818 & 0.882 & 0.405 & 0.706 & 0.811 \\
& GSMN\cite{GSMN} & GNN & CVPR20 & 0.574 & 0.823 & 0.890 & --- & --- & --- \\
& IMRAM\cite{IMRAM} & CNN-RNN & CVPR20 & 0.539 & 0.794 & 0.872 & 0.397 & 0.691 & 0.798 \\
& Unicoder\cite{Unicoder} & VLP model & AAAI20 & 0.715 & 0.912 & 0.952 & 0.484 & 0.767 & 0.859 \\
& Uniter\cite{Uniter} & VLP model & ECCV20 & 0.725 & 0.924 & 0.961 & 0.503 & 0.785 & 0.872  \\
& SGRAF\cite{SGRAF} & GNN & AAAI21 & 0.585 & 0.830 & 0.888 & 0.419 & --- & 0.813 \\
& PCME\cite{PCME} & Transformer & CVPR21  & --- & --- & --- & 0.319 & 0.621 & 0.745 \\
& SOHO\cite{SOHO} & VLP model & CVPR21 & 0.725 & 0.927 & 0.961 & 0.506 & 0.780 & 0.867 \\
& ViSTA\cite{ViSTA} & Transformer & CVPR22 & 0.758 & 0.942 & 0.969 & 0.526 & 0.796 & 0.876 \\
& AGREE\cite{AGREE} & VLP model & WSDM23 & 0.828 & 0.959 & 0.978 & --- & --- & --- \\ \hline
\end{tabular}}
\end{table}

\begin{table}[!b]
\caption{The \textit{mAP}@\textit{All} results of supervised real-value retrieval methods. The experimental results are from CLIP4CMR\cite{CLIP4CMR}.}
\label{tab:SRR}
\renewcommand\arraystretch{1.15}
\centering
\resizebox{\linewidth}{!}{
\begin{tabular}{|c|c|c|c|c|c|c|c|}
\hline
Task                 & Methods         & Category & Source & Wikipedia      & Pascal-Sentence & NUS-WIDE& Xmedia\\ \hline \hline
\multirow{9}{*}{I$\to$T}
                     & JRL\cite{JRL} & PL & TCSVT14 & 0.479 & 0.563 & 0.466 & 0.488 \\
                     & JFSSL\cite{JFSSL} & PL & TPAMI16 & 0.458 & 0.553 & 0.514 & 0.525 \\
                     & JLSLR\cite{JLSLR} & PL & SIGIR17 & 0.473 & 0.568 & 0.536 & 0.544 \\
                     & CCL\cite{CCL} & CNN-RNN & TMM17 & 0.505 & 0.576 & 0.506 & 0.537 \\
                     & ACMR\cite{ACMR} & GAN & MM17 & 0.468 & 0.538 & 0.519 & 0.536 \\
                     & MCSM\cite{MCSM} & CNN-RNN & TIP18 & 0.516 & 0.598 & 0.522 & 0.540 \\
                     & CM-GANS\cite{CM-GANs} & GAN & TOMM19 & 0.521 & 0.603 & 0.536 & 0.567 \\
                     & DSCMR\cite{DSCMR} & CNN-RNN & CVPR19 & 0.521 & 0.674 & 0.611 & 0.697 \\
                     & CLIP4CMR\cite{CLIP4CMR} & Transformer & ARXIV22 & 0.592 & 0.698 & 0.609 & 0.746 \\ \hline
\multirow{9}{*}{T$\to$I}
                     & JRL\cite{JRL} & PL & TCSVT14 & 0.428 & 0.505 & 0.499 & 0.405 \\
                     & JFSSL\cite{JFSSL}  & PL & TPAMI16 & 0.426 & 0.542 & 0.523 & 0.518 \\
                     & JLSLR\cite{JLSLR} & PL & SIGIR17 & 0.440 & 0.551 & 0.531 & 0.553 \\
                     & CCL\cite{CCL} & CNN-RNN & TMM17 & 0.457 & 0.561 & 0.535 & 0.528 \\
                     & ACMR\cite{ACMR} & GAN & MM17 & 0.412 & 0.544 & 0.542 & 0.519 \\
                     & MCSM\cite{MCSM} & CNN-RNN & TIP18 & 0.458 & 0.598 & 0.546 & 0.550 \\
                     & CM-GANS\cite{CM-GANs} & GAN & TOMM19 & 0.466 & 0.604 & 0.551 & 0.551 \\
                     & DSCMR\cite{DSCMR} & CNN-RNN & CVPR19 & 0.478 & 0.682 & 0.615 & 0.693 \\
                     & CLIP4CMR\cite{CLIP4CMR} & Transformer & ARXIV22 & 0.574 & 0.692 & 0.621 & 0.758 \\ \hline
\end{tabular}}
\end{table}

\begin{table*}[!b]
\caption{The \textit{mAP}@$50$ results of unsupervised hashing retrieval methods. The underlined experimental results are from AGCH\cite{AGCH}, and the remaining results are reported from our own reproductions. Below, SGL stands for spectral graph learning, DN stands for distillation network, and the remaining abbreviations are consistent with the previous definitions.}
\label{tab:UHR}
\renewcommand\arraystretch{1.02}
\centering
\resizebox{1\linewidth}{!}{
\begin{tabular}{|c|c|c|c|cccc|cccc|cccc|cccc|}
\hline
\multirow{2}{*}{Task}                     & \multirow{2}{*}{Method} & \multirow{2}{*}{Category} & \multirow{2}{*}{Source} & \multicolumn{4}{c|}{MIR Flickr}                                   & \multicolumn{4}{c|}{NUS-WIDE}                                     & \multicolumn{4}{c|}{MS COCO}                                      & \multicolumn{4}{c|}{IAPR TC-12}                                    \\ \cline{5-20}
                                          &                          &                     & & 16bits         & 32bits         & 64bits         & 128bits        & 16bits         & 32bits         & 64bits         & 128bits        & 16bits         & 32bits         & 64bits         & 128bits        & 16bits         & 32bits         & 64bits         & 128bits        \\ \hline\hline
\multirow{10}{*}{I$\to$T}                      & CVH\cite{CVH}             & SGL & IJCAI11               & {\ul 0.606}    & {\ul 0.599}    & {\ul 0.596}    & {\ul 0.589}    & {\ul 0.372}    & {\ul 0.363}    & {\ul 0.404}    & {\ul 0.390}    & 0.505          & 0.509          & 0.519          & 0.510          & 0.451          & 0.457          & 0.459          & 0.450          \\
& IMH\cite{IMH}          & SGL & SIGMOD13              & {\ul 0.612}    & {\ul 0.601}    & {\ul 0.592}    & {\ul 0.579}    & {\ul 0.470}    & {\ul 0.473}    & {\ul 0.476}    & {\ul 0.459}    & 0.570          & 0.615          & 0.613          & 0.587          & 0.542          & 0.573          & 0.593          & 0.593          \\
& LSSH\cite{LSSH}           & MF & SIGIR14               & {\ul 0.584}    & {\ul 0.599}    & {\ul 0.602}    & {\ul 0.614}    & {\ul 0.481}    & {\ul 0.489}    & {\ul 0.507}    & {\ul 0.507}    & 0.652          & 0.707          & 0.746          & 0.773          & 0.571          & 0.740 & 0.763 & 0.777 \\
& CMFH\cite{CMFH}           &  MF & CVPR14                 & {\ul 0.621}    & {\ul 0.624}    & {\ul 0.625}    & {\ul 0.627}    & {\ul 0.455}    & {\ul 0.459}    & {\ul 0.465}    & {\ul 0.467}    & 0.621          & 0.669          & 0.525          & 0.562          & 0.647          & 0.688          & 0.706          & 0.718          \\
& DBRC\cite{DBRC}          &  CNN-RNN & MM17                 & {\ul 0.617}    & {\ul 0.619}    & {\ul 0.620}    & {\ul 0.621}    & {\ul 0.424}    & {\ul 0.459}    & {\ul 0.447}    & {\ul 0.447}    & 0.567          & 0.591          & 0.617          & 0.627          & 0.457          & 0.521          & 0.570          & 0.620          \\
& DJSRH\cite{DJSRH}        &  GNN & ICCV19                & {\ul 0.810}    & {\ul 0.843}    & {\ul 0.862}    & {\ul 0.876}    & {\ul 0.724}    & {\ul 0.773}    & {\ul 0.798}    & {\ul 0.817}    & 0.678          & 0.724          & 0.743          & 0.768          & 0.548          & 0.619          & 0.640          & 0.666          \\
& AGCH\cite{AGCH}           & GNN & TMM22                 & {\ul 0.865}    & {\ul 0.887}    & {\ul 0.892}    & {\ul 0.912}    & {\ul 0.809}    & {\ul 0.830}    & {\ul 0.831}    & {\ul 0.852}    & 0.741          & 0.772          & 0.789          & 0.806          & 0.661          & 0.699          & 0.705          & 0.729          \\
& DAEH\cite{DAEH}          &  DN & TCSVT22     &  0.886  &  0.895   &   0.903  &  0.909  &  0.786  &   0.799 &  0.812   &   0.820  &           0.784       &     0.796      &      0.813     &      0.832    &          0.643&  0.667    &  0.682  &    0.714       \\
& CIRH\cite{CIRH} & GNN & TKDE23 & 0.901 & 0.913& 0.929 &0.937 &0.815 & 0.836 & 0.854 & 0.862 & 0.797 & 0.819 & 0.830 & 0.849 & 0.673 & 0.712 & 0.739 & 0.746 \\
& UCMFH\cite{UCMFH}     & Transformer & IF23 & 0.923 & 0.953 & 0.964 & 0.965 & 0.849 & 0.879 & 0.894 & 0.899 & 0.892 & 0.919 & 0.938 & 0.938 & 0.694 & 0.732 & 0.716 & 0.736 \\  \hline
\multicolumn{1}{|c|}{\multirow{10}{*}{T$\to$I}} & CVH\cite{CVH}          & SGL & IJCAI11               & {\ul 0.591}    & {\ul 0.583}    & {\ul 0.576}    & {\ul 0.576}    & {\ul 0.401}    & {\ul 0.384}    & {\ul 0.442}    & {\ul 0.432}    & 0.543          & 0.553          & 0.560          & 0.542          & 0.551          & 0.583          & 0.601          & 0.602          \\
\multicolumn{1}{|c|}{}                     & IMH\cite{IMH}            & SGL & SIGMOD13              & {\ul 0.603}    & {\ul 0.595}    & {\ul 0.589}    & {\ul 0.580}    & {\ul 0.478}    & {\ul 0.483}    & {\ul 0.472}    & {\ul 0.462}    & 0.641          & 0.709          & 0.705          & 0.652          & 0.609          & 0.649          & 0.679          & 0.682          \\
\multicolumn{1}{|c|}{}                     & LSSH\cite{LSSH}          & MF & SIGIR14               & {\ul 0.637}    & {\ul 0.659}    & {\ul 0.659}    & {\ul 0.672}    & {\ul 0.577}    & {\ul 0.617}    & {\ul 0.642}    & {\ul 0.663}    & 0.612          & 0.682          & 0.742          & 0.795          & 0.543          & 0.655          & 0.705          & 0.746          \\
\multicolumn{1}{|c|}{}                     & CMFH\cite{CMFH}            & MF & CVPR14                 & {\ul 0.642}    & {\ul 0.662}    & {\ul 0.676}    & {\ul 0.685}    & {\ul 0.529}    & {\ul 0.577}    & {\ul 0.614}    & {\ul 0.645}    & 0.627          & 0.667          & 0.554          & 0.595          & 0.639          & 0.688          & 0.715          & 0.733          \\
\multicolumn{1}{|c|}{}                     & DBRC\cite{DBRC}          & CNN-RNN & MM17                 & {\ul 0.618}    & {\ul 0.622}    & {\ul 0.626}    & {\ul 0.628}    & {\ul 0.455}    & {\ul 0.459}    & {\ul 0.468}    & {\ul 0.473}    & 0.635          & 0.671          & 0.697          & 0.735          & 0.462          & 0.520          & 0.591          & 0.635          \\
\multicolumn{1}{|c|}{}                     & DJSRH\cite{DJSRH}        & GNN & ICCV19                & {\ul 0.786}    & {\ul 0.822}    & {\ul 0.835}    & {\ul 0.847}    & {\ul 0.712}    & {\ul 0.744}    & {\ul 0.771}    & {\ul 0.789}    & 0.650          & 0.753          & 0.805          & 0.823          & 0.497          & 0.606          & 0.660          & 0.678          \\
\multicolumn{1}{|c|}{}                     & AGCH\cite{AGCH}          & GNN & TMM22                 & {\ul 0.829}    & {\ul 0.849}    & {\ul 0.852}    & {\ul 0.880}    & {\ul 0.769}    & {\ul 0.780}    & {\ul 0.798}    & {\ul 0.802}    & 0.746          & 0.774          & 0.797          & 0.817          & 0.670          & 0.710          & 0.737          & 0.739          \\
                                          & DAEH\cite{DAEH}          &  DN & TCSVT22                 &  0.834   & 0.845    &  0.852   &   0.887 &  0.752  & 0.768   & 0.783    &  0.791   &    0.768       &  0.786         &     0.812      &   0.819        &    0.650      &    0.667      &  0.685        &   0.713        \\
\multicolumn{1}{|c|}{}                     & CIRH\cite{CIRH}     & GNN & TKDE23 & 0.867 & 0.885 & 0.900 & 0.901 & 0.774 & 0.803 & 0.810 & 0.817 & 0.811 & 0.847 & 0.872 & 0.895 & 0.675 & 0.718 & 0.739 & 0.753 \\
\multicolumn{1}{|c|}{}                     & UCMFH\cite{UCMFH}     & Transformer & IF23 & 0.920 & 0.947 & 0.959 & 0.961 & 0.859 & 0.884 & 0.899 & 0.903 & 0.887 & 0.921 & 0.939 & 0.942 & 0.694 & 0.715 & 0.712 & 0.728 \\ \hline
\end{tabular}}
\end{table*}
\begin{table*}[!b]
\caption{The \textit{mAP}@$500$ results of supervised hashing retrieval methods. The underlined experimental results are from GCH\cite{GCH}, and the remaining results are reported from our own reproductions. '---' means that the retrieval accuracy of the retrieval methods\cite{GCH, DSPH} is anomalous, so we no longer record their performance.}
\label{tab:SHR}
\renewcommand\arraystretch{1.02}
\centering
\resizebox{1\linewidth}{!}{
\begin{tabular}{|c|c|c|c|cccc|cccc|cccc|cccc|}
\hline
\multirow{2}{*}{Task} & \multirow{2}{*}{Method} &\multirow{2}{*}{Category}& \multirow{2}{*}{Source} & \multicolumn{4}{c|}{MIR Flickr}                                   & \multicolumn{4}{c|}{NUS-WIDE}                                     & \multicolumn{4}{c|}{MS COCO}                                      & \multicolumn{4}{c|}{IAPR TC-12}     \\ \cline{5-20}
                      &                          &                     &  & 16bits         & 32bits         & 64bits         & 128bits        & 16bits         & 32bits         & 64bits         & 128bits        & 16bits         & 32bits         & 64bits         & 128bits        & 16bits & 32bits & 64bits & 128bits \\ \hline\hline
\multirow{9}{*}{I$\to$T} & SCM\cite{SCM}           &  PL & AAAI14                & 0.886          & 0.894          & 0.900          & 0.906          & 0.646          & 0.675          & 0.681          & 0.684          & 0.567          & 0.615          & 0.643          & 0.660          & 0.310  & 0.310  & 0.306  & 0.269   \\
& DCMH\cite{DCMH}        & CNN-RNN & CVPR17                & {\ul 0.724}    & {\ul 0.731}    & {\ul 0.731}    & 0.800          & {\ul 0.568}    & {\ul 0.561}    & {\ul 0.596}    & 0.707          & {\ul 0.505}    & {\ul 0.536}    & {\ul 0.557}    & {\ul 0.566}    & 0.507  & 0.548  & 0.520  & 0.520   \\
& SSAH\cite{SSAH}          & GAN & CVPR18                & 0.903          & 0.922          & 0.925          & 0.945          & 0.691          & 0.727          & 0.728          & 0.771          & 0.632          & 0.669          & 0.668          & 0.724          & 0.617  & 0.652  & 0.671  & 0.541   \\
& MTFH\cite{MTFH}       & MF & TPAMI19                & 0.856          & 0.875          & 0.903          & 0.906          & 0.716          & 0.750          & 0.765          & 0.787          & 0.597          & 0.647          & 0.717          & 0.754          & 0.459  & 0.569  & 0.590  & 0.577   \\
& GCH\cite{GCH}          & GNN & IJCAI19               & {\ul 0.833}    & {\ul 0.857}    & {\ul 0.869}    & ---        & {\ul 0.693}    & {\ul 0.719}    & {\ul 0.753}    &    ---            & {\ul 0.648}    & {\ul 0.686}    & {\ul 0.708}    &      ---          & 0.312  & 0.305  & 0.286  & 0.311   \\
& DCHUC\cite{DCHUC}       & CNN-RNN & TKDE22                & 0.895          & 0.916          & 0.926          & 0.911          & 0.707          & 0.672          & 0.738          & 0.802          & 0.513          & 0.550          & 0.558          & 0.517          & 0.554  & 0.602  & 0.625  & 0.648   \\
& BATCH\cite{BATCH}       & MF & TKDE21                & 0.914          & 0.929          & 0.933          & 0.941          & 0.757          & 0.781          & 0.797          & 0.806          & 0.662          & 0.695          & 0.719          & 0.739          & 0.649  & 0.692  & 0.714  & 0.729   \\
& HMAH\cite{HMAH}         & GAN & TMM23                 & 0.960 & 0.965 & 0.969 & 0.971 & 0.813 & 0.825 & 0.840 & 0.849 & 0.691 & 0.732 & 0.763 & 0.790 & 0.657  & 0.682  & 0.722  & 0.743   \\
& DSPH\cite{DSPH}          &  Transformer & TCSVT24 & 0.925    & 0.940  &  0.945 & 0.952 & 0.852 & 0.905 & 0.929 &  0.940 &  0.793   & 0.815   & 0.833   &  0.847   & ---   &      ---   &      ---  & --- \\ \hline
\multirow{9}{*}{T$\to$I} & SCM\cite{SCM}        &  PL & AAAI14                & 0.883          & 0.890          & 0.897          & 0.902          & 0.667          & 0.712          & 0.724          & 0.728          & 0.608          & 0.684          & 0.781          & 0.813          & 0.486  & 0.532  & 0.424  & 0.430   \\
& DCMH\cite{DCMH}            & CNN-RNN& CVPR17                & {\ul 0.764}    & {\ul 0.749}    & {\ul 0.780}    & 0.837          & {\ul 0.558}    & {\ul 0.591}    & {\ul 0.616}    & 0.594          & {\ul 0.549}    & {\ul 0.572}    & {\ul 0.605}    & 0.611          & 0.535  & 0.570  & 0.488  & 0.481   \\
& SSAH\cite{SSAH}        &  GAN & CVPR18                & 0.896          & 0.906          & 0.915          & 0.910          & 0.658          & 0.673          & 0.666          & 0.637          & 0.583          & 0.556          & 0.664          & 0.677          & 0.617  & 0.578  & 0.520  & 0.475   \\
& MTFH\cite{MTFH}        & MF  & TPAMI19                & 0.871          & 0.884          & 0.888          & 0.892          & 0.694          & 0.735          & 0.733          & 0.736          & 0.529          & 0.649          & 0.769          & 0.804          & 0.482  & 0.553  & 0.599  & 0.631   \\
& GCH\cite{GCH}         & GNN  & IJCAI19               & {\ul 0.892}    & {\ul 0.910}    & {\ul 0.907}    &       ---         & {\ul 0.732}    & {\ul 0.766}    & {\ul 0.761}    &   ---             & {\ul 0.745}    & {\ul 0.797}    & {\ul 0.830}    &      ---         & 0.332  & 0.328  & 0.311  & 0.311   \\
& DCHUC\cite{DCHUC}        & CNN-RNN  & TKDE22                & 0.764          & 0.749          & 0.780          & 0.837          & 0.558          & 0.591          & 0.616          & 0.594          & 0.549          & 0.572          & 0.605          & 0.611          & 0.535  & 0.570  & 0.488  & 0.481   \\
& BATCH\cite{BATCH}        & MF  & TKDE21                & 0.896          & 0.906          & 0.904          & 0.911          & 0.753          & 0.767          & 0.778          & 0.787          & 0.743          & 0.796          & 0.837          & 0.866          & 0.641  & 0.688  & 0.715  & 0.734   \\
& HMAH\cite{HMAH}        &  GAN & TMM23                 & 0.915 & 0.925 & 0.938 & 0.940 & 0.783 & 0.796 & 0.814 & 0.820 & 0.800 & 0.869 & 0.904 & 0.934 & 0.680  & 0.700  & 0.732  & 0.754   \\
& DSPH\cite{DSPH}          &  Transformer & TCSVT24  &   0.897  & 0.904  & 0.911 & 0.918 & 0.859& 0.920 &0.935   &0.951   &0.792     &  0.800      &   0.819   &  0.828      &    ---  &    ---  & ---  &  ---  \\ \hline
\end{tabular}}
\end{table*}

\subsubsection{NDCG@$k$} This metric computes the cumulative gain for each query and then discounts it by the logarithm of the rank. The gain is normalized by the ideal gain for each query. Formally, it is defined as
\begin{equation}
\small
\begin{aligned}
&\ \ \ \ \ \textit{NDCG}\text{@}k=\frac{\textit{DCG}}{\textit{IDCG}},\\
\textit{DCG} = \sum_{i=1}^{k}&\frac{rel_{i}}{log_{2}(i+1)}, \ \ \textit{IDCG} = \sum_{i=1}^{|rel|_k} \frac{rel_{i}}{log_{2}(i+1)},
\end{aligned}
\end{equation}
where $rel_i$ represents the correlation degree between the $i$-th retrieved result and the query sample, and $|rel|_k$ is an ideal set consisting of the top-$k$ samples after sorting all the retrieved samples according to the real correlation. The final score is obtained by averaging the \textit{NDCG}@$k$ across all queries.

\subsection{Evaluation Result}
In this section, we present the accuracy and efficiency of several representative methods for text-image cross-modal retrieval tasks, aiming to offer insights and references for researchers. As shown in Tables \ref{tab:URR}-\ref{tab:SHR}, we compare the accuracy of various cross-modal retrieval methods using the standard metrics commonly applied in each task. Additionally, since hashing-based cross-modal retrieval methods focus on both efficiency and accuracy, their efficiency comparisons are detailed in Table \ref{tab:CRE} for a more comprehensive evaluation. From the performance comparisons, we can draw the following key observations:

\begin{itemize}
\item{Supervised cross-modal retrieval methods generally achieve higher retrieval accuracy than unsupervised ones. These results are reasonable in that the supervised methods can enhance the semantics of multi-modal features or hash codes with explicit annotations. As various deep learning techniques are introduced into cross-modal retrieval, the performance gap between unsupervised and supervised methods is gradually narrowed. This shows that the model optimization of deep neural networks does not have to rely on expensive manual annotations.}
\item{Deep cross-modal retrieval models generally outperform shallow methods on retrieval accuracy, which indicates that deep neural networks perform well in capturing the latent semantics in multi-modal data. But at the same time, it should be noted that shallow methods, such as BATCH\cite{BATCH} can achieve comparable retrieval accuracy than the complex deep methods with much less training and testing time cost. This indicates that the representation capability of deep networks has not been sufficiently exploited in existing deep methods.}
\item{In real-valued cross-modal retrieval, Transformer and VLP-based methods can achieve the most superior retrieval accuracy than traditional deep networks. Among them, VLP models can achieve the best unsupervised retrieval accuracy at present, and VLP models with supervised fine-tuning, such as CLIP4CMR \cite{CLIP4CMR}, can also exceed traditional supervised deep models.}
\item{Cross-modal hashing methods show varied performance with increasing hash code length. While some methods see improved retrieval accuracy with longer hash codes, others do not. This suggests that optimal hash code lengths differ among methods, and adding more bits beyond a certain point does not enhance performance.}
\item{As shown in Table \ref{tab:CRE}, the computational efficiency of shallow hashing retrieval methods is significantly higher than that of deep ones. Moreover, in deep hashing retrieval methods, no single network architecture has a significant inherent efficiency advantage over others. Instead, the overall design and optimization of hashing architectures are crucial.}
\end{itemize}

\begin{table}[!t]
\caption{Comparison of the training and testing time (Unit: second) for 128-bit hashing-based cross-modal retrieval methods on a single TITAN RTX GPU. '---' indicates that the retrieval accuracy of the retrieval methods\cite{GCH, DSPH} is anomalous, so we no longer record their running times.}
\label{tab:CRE}
\renewcommand\arraystretch{1.02}
\centering
\resizebox{\linewidth}{!}{
\begin{tabular}{|c|c|cc|cc|cc|cc|}
\hline
\multirow{2}{*}{Method} & \multirow{2}{*}{Category} & \multicolumn{2}{c|}{MIR Flickr} & \multicolumn{2}{c|}{NUS-WIDE} & \multicolumn{2}{c|}{MS COCO} & \multicolumn{2}{c|}{IAPR TC-12} \\ \cline{3-10}
                 & & Train   & Test  & Train   & Test   & Train   & Test  & Train   & Test  \\ \hline\hline
CVH\cite{CVH}    & SGL & 5.5     & 0.2   & 6.5     & 1.2    & 5.9     & 1.2   & 9.8     & 1.3   \\
IMH\cite{IMH}    & SGL & 31.5    & 0.4   & 31.0    & 2.7    & 32.9    & 2.7   & 50.4    & 0.6   \\
LSSH\cite{LSSH}  & MF & 48.2    & 1.9   & 50.9    & 17.4   & 49.0    & 11.5  & 328.4   & 54.5  \\
CMFH\cite{CMFH}  & MF & 732.0   & 0.9   & 748.2   & 8.1    & 782.6   & 5.8   & 153.7   & 0.3   \\
DBRC\cite{DBRC}  & CNN-RNN & 991.1   & 2.3   & 1,067.1  & 12.7   & 1,051.7  & 11.0  & 1,497.8  & 2.1   \\
DJSRH\cite{DJSRH}& GNN & 497.3   & 54.6  & 520.9   & 721.5  & 599.6   & 190.9 & 4,676.0  & 236.6 \\
AGCH\cite{AGCH}  & GNN & 327.1   & 18.1  & 343.9   & 279.0  & 386.2   & 328.1 & 748.1   & 114.9 \\
DAEH\cite{DAEH}  & DN & 678.2&135.6&981.3&123.7&498.5&75.6&884.3&156.4\\
CIRH\cite{CIRH}  & GNN & 107.6 & 1.4 & 192.9  & 10.5 & 222.4   & 8.2   & 88.7    & 1.3   \\
UCMFH\cite{UCMFH}  & Transformer & 376.5 & 35.2 & 698.1  & 93.2 & 1,543.5 & 384.7 & 457.0 & 127.6  \\  \hline
SCM\cite{SCM}    & PL & 164.8   & 0.3   & 116.3   & 2.1    & 243.9   & 1.6   & 83.9    & 0.2   \\
DCMH\cite{DCMH}  & CNN-RNN & 3,969.9  & 97.9  & 3,957.1  & 943.5  & 4,161.0  & 519.9 & 6,261.6  & 174.1 \\
SSAH\cite{SSAH}  & GAN & 5,865.8  & 116.5 & 5,558.4  & 1,719.1 & 6,954.9  & 754.7 & 3,900.0  & 216.9 \\
MTFH\cite{MTFH}& MF & 395.1     & 0.1   & 416.5    & 0.2   & 430.3    & 0.1  & 2,019.8    & 294.0   \\
GCH\cite{GCH}    & GNN & --- & --- & --- & --- & --- & --- & 33,153.8 & 151.7 \\
DCHUC\cite{DCHUC}& CNN-RNN & 461.3   & 114.3  & 1,001.8  & 1,006.9 &  1,022.3 & 644.6&  955.5  &179.9\\
BATCH\cite{BATCH}& MF & 1.5     & 2.6   & 1.5     & 14.2   & 1.7     & 11.5  & 1.0     & 2.4   \\
HMAH\cite{HMAH}  & GAN & 121.5   & 6.3   & 72.1    & 61.5   & 321.5   & 41.4  & 267.0   & 2.8   \\
DSPH\cite{DSPH}  & Transformer & 253.4 & 10.5 & 341.4&20.3&569.4&40.0&---&---\\ \hline
\end{tabular}}
\end{table}

\section{Application}
\label{Sec:Application}
Cross-modal retrieval has emerged as a powerful tool with successful applications across diverse domains. It empowers users with seamless access to pertinent information, significantly improving their ability to comprehend and represent multimedia content effectively. This versatile technique finds utility in several common application scenarios, including but not limited to:

\emph{1) E-commerce.} One common scenario in e-commerce is when users want to describe a product using text or voice input\cite{FashionBERT, EI-CLIP, MAKE}. Cross-modal retrieval systems can then provide relevant product images and videos that match the users' descriptions. This capability allows users to have a more comprehensive understanding of the product and make informed purchasing decisions.

\emph{2) Recipe search.} One exciting application of cross-modal retrieval in recipe search is the ability to search by taking a picture of a food item \cite{Recipe1, Recipe2, Recipe3}. Using image recognition technology, these retrieval systems can analyze the image and extract visual features to identify the food item accurately. Subsequently, retrieval systems can retrieve recipes associated with that specific ingredient or dish, enabling users to discover new recipes based on the ingredients they have at hand.

\emph{3) Scene text retrieval.} Text-based search allows users to input text or voice queries to find specific textual elements in scenes \cite{Scene_text1, StacMR, Scene_text2}. Additionally, cross-modal retrieval enables leveraging visual inputs to retrieve text-related information, like capturing pictures of storefronts, menus, or product labels. Image recognition and text extraction techniques identify and extract textual content from images, matching it with available text to provide relevant information and scenes.

\emph{4) Person re-identification.} Cross-modal retrieval in person re-identification allows users to search for specific individuals by providing descriptions or attributes of the person of interest\cite{ReID1, IRRA, ReID2}. By entering such descriptions, retrieval systems retrieve relevant individuals from mass surveillance videos, facilitating behavior analysis and tracking.

\emph{5) Remote sensing retrieval.} Cross-modal retrieval allows users to find relevant images by providing textual descriptions of geographic scenes or specific objects \cite{IEFT, SWAN, CCLS2T}. Users can describe land cover types, geographical features, or objects of interest, and then retrieval systems search for corresponding images. This functionality also enables users to upload their own images and obtain additional information, enhancing remote sensing systems' capabilities.

\emph{6) Medical diagnosis.} Cross-modal retrieval enables healthcare professionals to search for relevant medical images by entering textual descriptions of symptoms, diseases, or medical conditions \cite{Medical2, Medical3, Medical1}. Instead of solely relying on image-based queries, which may require specific anatomical knowledge or visual recognition skills, medical professionals can utilize their expertise in describing the clinical features or characteristics of the condition. This allows for a more intuitive and accessible search process, improving the efficiency of finding relevant medical images.

\emph{7) Sketch-based image retrieval.} In sketch-based image retrieval, users can draw a rough sketch of an object or scene, and cross-modal retrieval systems can retrieve similar images from a database \cite{CGAE, ZS-SBIR, CrossX-DFL}. It leverages the visual similarity between sketches and RGB images, enabling users to find images that match their sketches, which is particularly useful in design, art, and creative industries.

\emph{8) Video moment retrieval.} Cross-modal retrieval aids in video moment retrieval, where users can input textual queries or provide visual cues to locate specific moments or scenes within videos \cite{SVMR, MCMN, SpotVMR}. By combining text and visual information, retrieval systems can pinpoint precise moments in videos, making it easier to navigate and extract relevant segments from large video collections.

\emph{9) Retrieval-augmented generation.} In retrieval-augmented generation, cross-modal retrieval enhances content generation tasks by retrieving relevant information from a large database and using it to generate coherent and contextually accurate content \cite{RDMS, Smallcap, FLMR}. This technique integrates retrieved data into the generation process, improving the relevance and quality of the generated content, and is beneficial for applications such as automated content creation, question answering, and personalized recommendations.

\section{Discussion and Outlook}
\label{Sec:Discussion and Outlook}
Over the last two decades, cross-modal retrieval has made remarkable advancements and has been extensively adopted in search engines. While it effectively fulfills users' needs for retrieving diverse modalities, challenges remain in managing the complexity of heterogeneous data. As cross-modal retrieval continues to progress, it has the potential to revolutionize multimedia information retrieval and enable more efficient and comprehensive access to various data sources. Several promising research directions include:

\emph{1) Efficient cross-modal semantic modeling.} The quality of cross-modal semantic modeling directly determines the cross-modal retrieval performance. The use of vision-language pre-training models \cite{Oscar, Uniter, Unicoder, SOHO, ALIGN, CLIP} has proven to be effective in capturing complex cross-modal semantics, enabling advanced retrieval tasks. However, the process of pre-training these large models can be computationally intensive, requiring significant storage and energy resources \cite{CVLPLR}. One approach to enhancing efficiency is to draw inspiration from the human brain's information processing mechanisms \cite{Brain}. By studying how the brain processes and integrates information from different modalities, researchers can devise novel methods to streamline cross-modal semantic modeling. Another promising avenue is to combine data-driven multi-modal foundation models with human-generated knowledge \cite{Prior}. Human cognition offers valuable semantic information that can streamline training by reducing unnecessary computations. By incorporating human-generated annotations or semantic priors into the modeling framework, cross-modal retrieval systems can benefit from the high-level understanding provided by humans, reducing the reliance on large-scale data for training.

\emph{2) Uncertain multi-modal data modeling.} Uncertain multi-modal data modeling is a critical challenge in cross-modal retrieval, as existing models often rely on assumptions or prior knowledge that may not accurately reflect real-world scenarios. Assumptions such as one-to-one correspondence, similarity relationships, or perfect semantic annotations between multi-modal data can be limiting in practical scenarios where these assumptions may not hold. Issues like imbalanced multi-modal data \cite{LTCMH}, incomplete correspondence \cite{DAVAE}, noisy annotations \cite{MRL}, and misalignment \cite{DECL} can significantly impact the performance of cross-modal retrieval models. To address these limitations, it is necessary to explore more effective techniques for modeling uncertain multi-modal data. One promising direction is the utilization of probabilistic models and Bayesian approaches \cite{UPB}, which can capture and quantify uncertainty in the cross-modal data representations and correlations. Moreover, the use of self-supervised or weakly-supervised learning methods \cite{Bert, MAE} can help alleviate the reliance on accurate annotations, as they can learn from the inherent structure and correlation present in the data itself. Techniques such as self-supervised representation learning, co-training, or multi-instance learning can be employed to exploit the underlying relationships within the uncertain multi-modal data. Furthermore, adversarial learning approaches \cite{AL} can be applied to handle data misalignment and distribution discrepancies across modalities. By learning domain-invariant representations or applying domain adaptation techniques, cross-modal retrieval models can effectively align and bridge the gap between different modalities, even in the presence of uncertain or misaligned data.

\emph{3) Multi-modal distribution adaptation.} Multi-modal distribution adaptation is a critical aspect of cross-modal retrieval, as the distribution of multi-modal data can vary significantly due to factors such as time, location, scenes, and other sources of variability \cite{Transfer1}. These variations can have a negative impact on the performance of cross-modal retrieval systems.  To overcome these challenges, it is essential for cross-modal retrieval systems to demonstrate robustness and adaptability. One effective approach to addressing these challenges is through the utilization of cross-domain transfer learning techniques. By leveraging knowledge from related domains or auxiliary tasks, cross-modal retrieval systems can learn to adapt and generalize across different distributions. Transfer learning methods \cite{Transfer2}, such as domain adaptation \cite{DA}, domain generalization \cite{DG}, or adversarial learning \cite{AL}, can be employed to bridge the distribution gaps between modalities and improve the overall retrieval performance.

\emph{4) Continuous retrieval model updating.} Continuous retrieval model updating is a vital aspect of cross-modal retrieval to ensure that retrieval systems stay up-to-date with the dynamic nature of multi-modal data. As new data is generated or removed from the retrieval database, it becomes necessary to promptly update the retrieval model parameters to reflect these changes. This allows retrieval systems to adapt and maintain high-quality retrieval performance over time. One approach for continuous retrieval model updating is online learning \cite{OL}, where the model is updated incrementally as new data arrives. Online learning algorithms enable the efficient processing of large-scale data by updating the model parameters on the fly, avoiding the need for repeated model training from scratch. This allows retrieval systems to adapt quickly to changes in the data distribution and maintain high retrieval performance. Another approach is the use of adaptive learning algorithms \cite{TAL} that dynamically adjust the model parameters based on the relevance and significance of new data. These algorithms can prioritize the update process to focus on the most informative and influential data samples, thereby optimizing the efficiency of retrieval model updating. Furthermore, techniques like transfer learning \cite{Transfer2} and knowledge distillation \cite{KD_survey} can be utilized to transfer knowledge from existing retrieval models to updated models. This can help preserve the learned representations and prevent catastrophic forgetting when incorporating new data into retrieval systems.

\emph{5) Interactive and user-centric retrieval.} Interactive and user-centric retrieval is a crucial aspect of cross-modal retrieval systems, recognizing the diverse goals and preferences of users and ensuring their satisfaction \cite{IUCR1, IUCR2}. Users often have different intentions or expectations when issuing a query, and their requirements can evolve over time. To address these variations and provide personalized results, cross-modal retrieval systems need to incorporate interactive mechanisms and user-centric strategies. One key element of interactive retrieval is user feedback. Users can provide feedback on retrieved results, indicating the relevance or preference of specific items \cite{IUCR1}. Feedback-driven learning approaches, such as relevance feedback or active learning, can be employed to improve the accuracy and relevance of retrieved results. In addition to feedback, users may modify their queries during the retrieval process \cite{IUCR2}. Query modification techniques, such as query expansion or reformulation, can be employed to refine and optimize the query based on user intent. By understanding the user's information needs and dynamically adjusting the query, cross-modal retrieval systems can improve the relevance and coverage of retrieved results. To support interactive and user-centric retrieval, cross-modal retrieval systems should employ interactive interfaces and visualization tools. These interfaces should provide intuitive ways for users to provide feedback, modify queries, and specify preferences. Visualizations can help users explore and navigate the retrieved results, facilitating better understanding and decision-making.

\emph{6) Responsible cross-modal retrieval.} As cross-modal retrieval systems become integral to search engines and multimedia platforms, addressing responsible concerns like transparency, trustworthiness, model fairness, and data bias is crucial, especially with the rise of AI-related regulations. Specifically, transparency is crucial for building user trust and accountability in cross-modal retrieval systems. Achieving this requires clear explanations of the decision-making processes involved in retrieval operations. Implementing explainable AI (XAI) techniques \cite{XAI} can clarify how models interpret user queries and return multi-modal retrieved results. For example, providing insights into the features influencing retrieval outcomes helps users understand why certain results are presented, fostering trust in the system. Additionally, allowing users to query the models reasoning enables them to assess the reliability of the retrieved information, promoting a more responsible approach to retrieval. Trustworthiness is contingent upon the robustness and prudence of retrieval systems, encompassing their resistance to adversarial attacks and their ability to prevent excessive mining of private information. Despite some advancements in this area\cite{CMLA, PIP}, it remains insufficiently explored. As adversarial techniques become more diverse and the emphasis on personal privacy intensifies, it is urgent to develop trusted cross-modal retrieval systems. Addressing model bias (i.e., ensuring model fairness\cite{Fair}) is important for improving user experience. However, most current research focuses on generative tasks\cite{Fair1, Fair2}. In cross-modal retrieval, it is essential to develop fairness-aware models that can detect and mitigate biases arising from training data or external factors, preventing biased results that could reinforce stereotypes or marginalize certain user groups. Moreover, data bias resulting from the under-representation of specific modalities, languages, or cultures \cite{Bias1, Bias2} can disproportionately affect certain user groups. This issue underscores the need for not only more robust model learning strategies but also the development of diverse datasets and de-biasing algorithms for bias detection and correction.

\emph{7) Distributed cross-modal retrieval.} Given the increasing scale and complexity of multi-modal data, the effective storage and management of distributed multi-modal data become crucial for cross-modal retrieval systems. Distributed cross-modal retrieval offers several advantages, including reduced data transmission, accelerated retrieval speed, and enhanced privacy protection for multi-modal data. These improvements ultimately enhance the efficiency and scalability of retrieval systems, making them capable of handling large-scale data with better performance. To implement distributed cross-modal retrieval, various techniques can be employed. Distributed computing frameworks, such as Apache Hadoop \cite{Apache1} or Apache Spark \cite{Apache2}, can facilitate parallel processing and indexing of multi-modal data across distributed nodes. Cloud computing platforms offer scalable infrastructure and storage capabilities, enabling efficient retrieval and management of large-scale multi-modal datasets. Additionally, federated learning \cite{FL} emerges as a promising approach for distributed retrieval scenarios, allowing retrieval models to be trained collaboratively across multiple local sites while preserving data privacy and security. Although there have been some initial studies\cite{FedCMR, PT-FUCH, PEPFCH, FedCAFE} in this area, it is still in the early stages of federated cross-modal retrieval.

\section{Conclusion}
\label{Sec:Conclusion}
Cross-modal retrieval addresses the growing need for accessing and utilizing diverse multi-modal data. The evolution of research in this field has improved the accuracy, stability, and scalability of retrieval systems. The paper presents a comprehensive taxonomy, reviews numerous papers, and provides insights into cross-modal retrieval methods and architectures. It also offers guidance on dataset selection and performance evaluation metrics. The paper explores opportunities, challenges, and future research directions, contributing to the understanding and development of cross-modal retrieval. Further exploration and innovation in this field are encouraged.

\bibliographystyle{IEEEtran}
\bibliography{bare_jrnl_new_sample4}

\vspace{-10mm}

\begin{IEEEbiography}[{\includegraphics[width=1in,height=1.25in,clip,keepaspectratio]{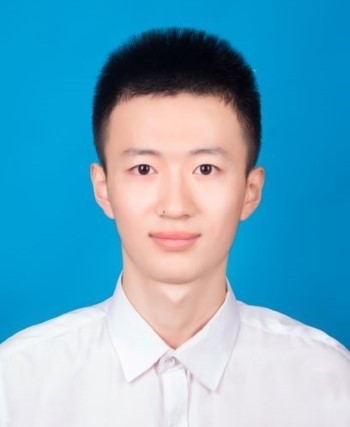}}]{Tianshi Wang} is currently a PhD student with the School of Information Science and Engineering, Shandong Normal University, China. He is also a visiting student at the School of Electronic and Information Engineering, Tongji University, China. His research interests include the area of large-scale multimedia content analysis and retrieval.
\end{IEEEbiography}

\vspace{-10mm}

\begin{IEEEbiography}[{\includegraphics[width=1in,height=1.25in,clip,keepaspectratio]{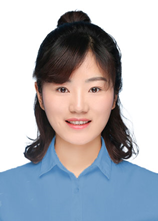}}]{Fengling Li} is currently a PhD student with Australian Artificial Intelligence Institute, Faculty of Engineering and Information Technology, University of Technology Sydney. Her current research interest mainly focuses on the multi-modal analysis and retrieval.
\end{IEEEbiography}

\vspace{-3mm}

\begin{IEEEbiography}[{\includegraphics[width=1in,height=1.25in,clip,keepaspectratio]{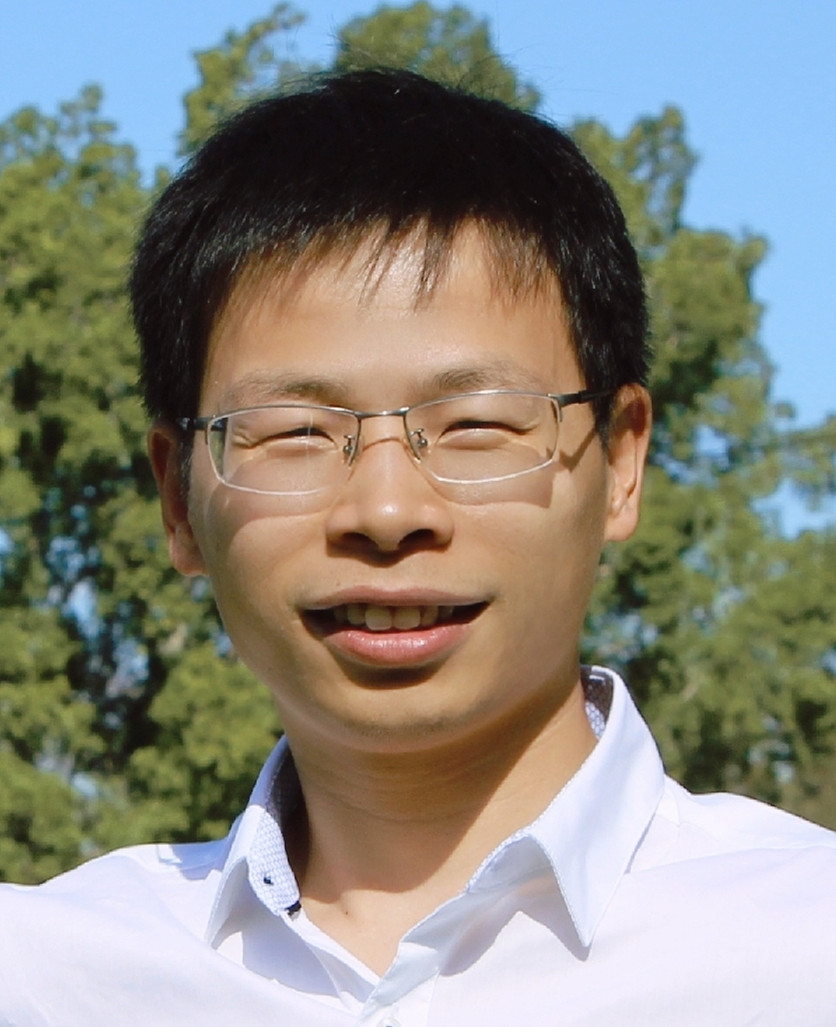}}]{Lei Zhu} is currently a professor with the School of Computer Science and Technology, Tongji University. He received his B.Eng. and Ph.D. degrees from Wuhan University of Technology in 2009 and Huazhong University Science and Technology in 2015, respectively. He was a Research Fellow at the University of Queensland (2016-2017). His research interests are in the area of large-scale multimedia content analysis and retrieval. Zhu has co-/authored more than 100 peer-reviewed papers, such as ACM SIGIR, ACM MM, IEEE TPAMI, IEEE TIP, IEEE TKDE, and ACM TOIS. His publications have attracted more than 9,200 Google citations. At present, he serves as the Associate Editor of IEEE TBD, ACM TOMM, and Information Sciences. He has served as the Area Chair of ACM MM/IEEE ICME, Senior Program Committee for SIGIR/CIKM/AAAI. He won ACM SIGIR 2019 Best Paper Honorable Mention Award, ADMA 2020 Best Paper Award, ChinaMM 2022 Best Student Paper Award, ACM China SIGMM Rising Star Award, Shandong Provincial Entrepreneurship Award for Returned Students, and Shandong Provincial AI Outstanding Youth Award.
\end{IEEEbiography}

\vspace{-3mm}

\begin{IEEEbiography}[{\includegraphics[width=1in,height=1.25in,clip,keepaspectratio]{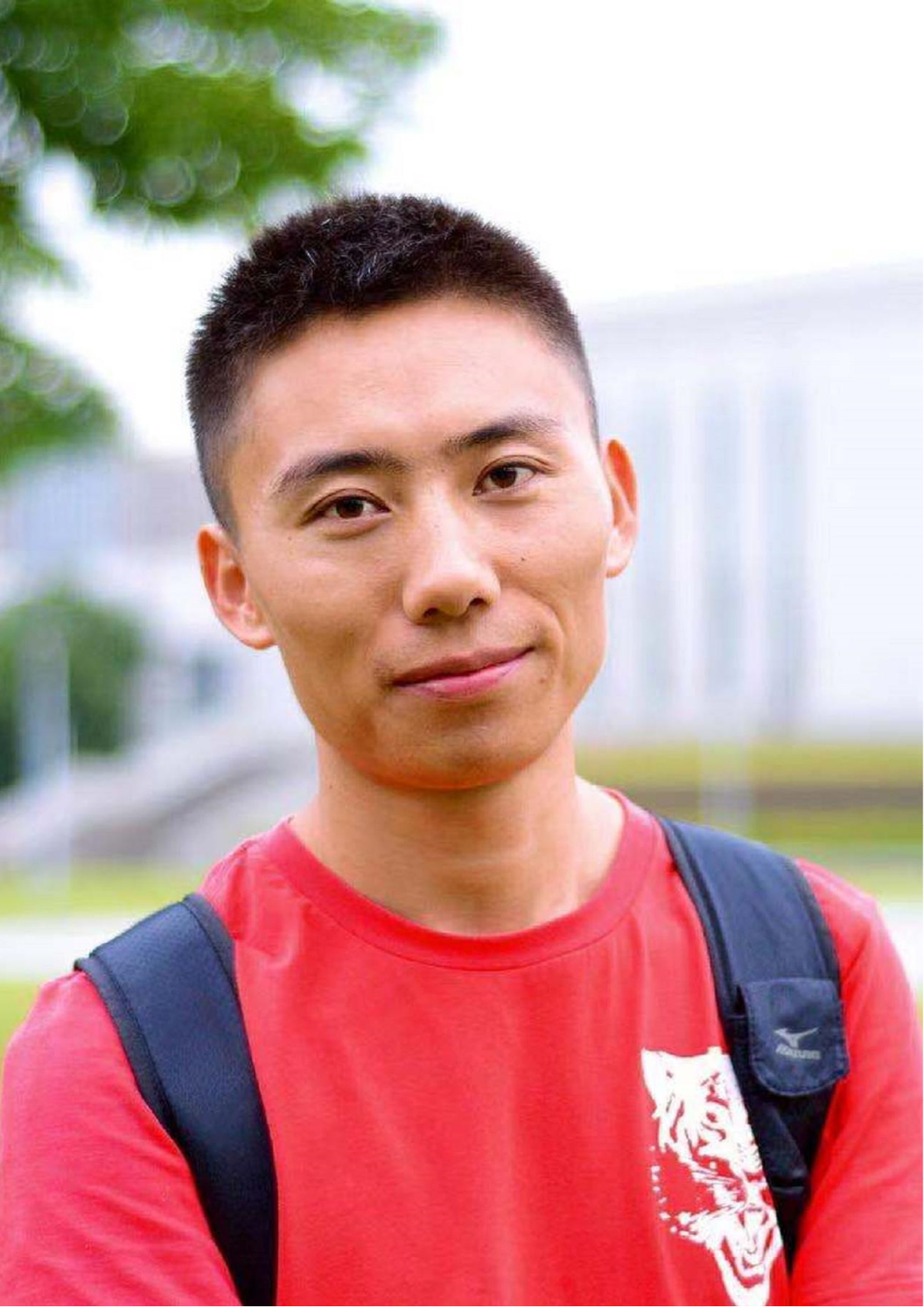}}]{Jingjing Li} received the M.Sc. and Ph.D. degrees in computer science from the University of Electronic Science and Technology of China in 2013 and 2017, respectively. He is currently a full Professor with the School of Computer Science and Engineering, University of Electronic Science and Technology of China. His current research interests include computer vision, machine learning and multimedia analysis, especially transfer learning, domain adaptation and cross-modal retrieval. Dr. Jingjing Li has published over 70 peer-reviewed papers on top-ranking journals and conferences, including IEEE TPAMI, TIP, TKDE, CVPR and NeurIPS. He has long served as a reviewer/PC/SPC/AC for TPAMI, TIP, TOIS, AAAI, CVPR, WACV and ACM MM. He won the Excellent Doctoral Thesis Award of The Chinese Institute of Electronics, and The Excellent Young Scholar of Wu Wen Jun AI Science \& Technology Award.
\end{IEEEbiography}

\vspace{-3mm}

\begin{IEEEbiography}[{\includegraphics[width=1in,height=1.25in,clip,keepaspectratio]{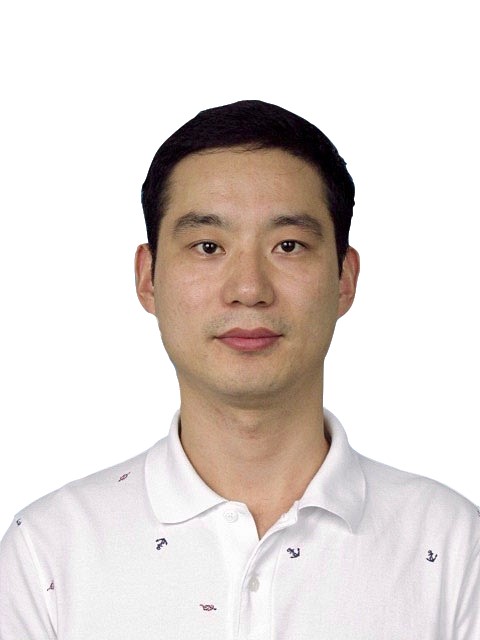}}]{Zheng Zhang} received his Ph.D. degree from Harbin Institute of Technology, China, in 2018. Dr. Zhang was a Postdoctoral Research Fellow at The University of Queensland, Australia. He is currently with Harbin Institute of Technology, Shenzhen, China. He has published over 150 technical papers at prestigious journals and conferences. He serves an Editorial Board Member for the IEEE Trans. on Affective Computing (T-AC), IEEE Journal of Biomedical and Health Informatics (J-BHI), Information Fusion (INFFUS), Information Processing \& Management journal, and also serves/served as the AC/SPC member for several top conferences, like CVPR and ACM Multimedia. His research interests mainly focus on multimedia content analysis and understanding.
\end{IEEEbiography}

\vspace{-3mm}

\begin{IEEEbiography}[{\includegraphics[width=1in,height=1.25in,clip,keepaspectratio]{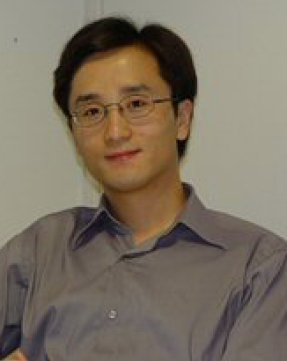}}]{Heng Tao Shen} obtained his BSc with first-class Honours and Ph.D. from the Department of Computer Science, the National University of Singapore in 2000 and 2004 respectively. His research interests mainly include Multimedia Search, Computer Vision and Artificial Intelligence. He is a Member of Academia Europaea, Fellow of ACM, IEEE and OSA.
\end{IEEEbiography}

\vfill

\end{document}